\documentclass[12pt,preprint]{aastex631}
\usepackage{float}

\graphicspath{ {./figures/} }

\newcommand{\um}{$\mu$m}

\newcommand{\Msun}{M$_{\odot}$}

\newcommand{\kms}{km~s$^{-1}$}

\newcommand{\nthpnt}{N$_{2}$H$^{+}$}
\newcommand{\nthp}{N$_{2}$H$^{+}$\,(1--0)}

\newcommand{\hcopnt}{HCO$^{+}$}
\newcommand{\hcop}{HCO$^{+}$\,(1--0)}
\newcommand{\siont}{\mbox{SiO}}
\newcommand{\sio}{\mbox{SiO~(2--1)}}
\newcommand{\htcopnt}{H$^{13}$CO$^{+}$}
\newcommand{\htcop}{H$^{13}$CO$^{+}$\,(1--0)}

\newcommand{\tctfs}{$^{13}$C$^{34}$S\,(2--1)}
\newcommand{\tcsnt}{$^{13}$CS}
\newcommand{\tcs}{$^{13}$CS~(2--1)}

\newcommand{\chtcn}{CH$_{3}$CN\,(5$_1$--4$_1$)}
\newcommand{\hctn}{HC$_{3}$N\,(10--9)}
\newcommand{\hctnnt}{HC$_{3}$N}
\newcommand{\hncnt}{HNC}
	
\newcommand{\hcnnt}{HCN}
	
\newcommand{\hncofznt}{HNCO}
\newcommand{\cchnt}{C$_{2}$H} 	   
\newcommand{\hntcnt}{HN$^{13}$C}
\newcommand{\hnc}{HNC~(1--0)}
\newcommand{\hctccn}{HC$^{13}$CCN\,(10--9)}	
\newcommand{\hcn}{HCN (1--0)}
\newcommand{\hncofo}{HNCO\,($4_{1,3}$--$3_{1,2}$)}	
\newcommand{\hncofz}{HNCO\,($4_{0,4}$--$3_{0,3}$)}	
\newcommand{\cch}{C$_{2}$H (N=1--0;~J=3/2--1/2;~F=2--1)} 	   
\newcommand{\hntc}{HN$^{13}$C\,(1--0)}

\newcommand{\tastar}{T$^{*}_{A}$}

\def\addtodot#1.#2\relax{#1\rlap{.}^{\dotadd}#2}

\newcommand{\ahcop}{$A_{HCO^+}$}
\newcommand{\ahnc}{$A_{HNC}$}
\newcommand{\ahtcop}{$A_{H^{13}CO^+}$}

\usepackage[section]{placeins}
\begin{document}

\shorttitle{Rapidly Collapsing, Dense Molecular Clumps}
\shortauthors{Jackson et al.}

\title{Dense Molecular Clumps with Large Blue Asymmetries: Evidence for Collapse}

\author[0000-0002-3466-6164]{James M. Jackson}
\affiliation{Green Bank Observatory,155 Observatory Road, Green Bank, WV 24944, USA}
\affiliation{USRA SOFIA Science Center, NASA Ames Research Center, Moffett Field, CA 94045, USA}
\affiliation{School of Mathematical and Physical Sciences, University of Newcastle, University Drive, Callaghan NSW 2308, Australia}
\affiliation{Current Address: 802/21 Cadigal Ave., Pyrmont NSW 2009, Australia}

\author{J. Scott Whitaker}
\affiliation{Physics Department, Boston University, 590 Commonwealth Ave., Boston, MA 02215, USA}

\author{Edward Chambers}
\affiliation{USRA SOFIA Science Center, NASA Ames Research Center, Moffett Field, CA 94045, USA}
\affiliation{Space Science Institute, 4765 Walnut St, Suite B
Boulder, CO 80301, USA}

\author{Robert Simon}
\affiliation{I. Physikalisches Institut, Universität zu Köln, Zülpicher Str. 77, 50937 Köln, Germany}

\author{Cristian Guevara}
\affiliation{Instituto de Astronomía, Universidad Católica del Norte, Av. Angamos 0610, 1270398 Antofagasta, Chile}

\author[0000-0002-4173-2852]{David Allingham}
\affiliation{School of Mathematical and Physical Sciences, University of Newcastle, University Drive, Callaghan NSW 2308, Australia}

\author{Philippa Patterson}
\affiliation{School of Mathematical and Physical Sciences, University of Newcastle, University Drive, Callaghan NSW 2308, Australia}

\author[0000-0002-0831-4886]{Nicholas Killerby-Smith}
\affiliation{School of Mathematical and Physical Sciences, University of Newcastle, University Drive, Callaghan NSW 2308, Australia}
\affiliation{Research School of Astronomy and Astrophysics, Australian National University, Canberra 2611, ACT, Australia}

\author{Jacob Askew}
\affiliation{School of Mathematical and Physical Sciences, University of Newcastle, University Drive, Callaghan NSW 2308, Australia}
\affiliation{Centre for Astrophysics and Supercomputing, Swinburne University of Technology, P.O. Box 218, Hawthorn, Victoria 3122, Australia}
\affiliation{OzGrav: The Australian Research Council Centre of Excellence for Gravitational Wave Discovery, Hawthorn VIC 3122, Australia}

\author[0000-0002-7125-7685]{Patricio Sanhueza}
\affiliation{Department of Astronomy, School of Science, The University of Tokyo, 7-3-1 Hongo, Bunkyo, Tokyo 113-0033, Japan}

\author{Ian W. Stephens} 
\affiliation{Department of Earth, Environment, and Physics, Worcester State University, 486 Chandler Street, Worcester MA  01602, USA}
\affiliation{Center for Astrophysics, Harvard \& Smithsonian, 60 Garden St., Cambridge, MA 02138, USA}

\author[0000-0002-1730-8832]{Anika Schmiedeke}
\affiliation{Green Bank Observatory,155 Observatory Road, Green Bank, WV 24944, USA}

\author {Robert Loughnane}
\affiliation{Instituto de Radioastronom\'{i}a y Astrof\'{i}sica, Universidad Nacional Aut\'{o}noma de M\'{e}xico, Apdo. Postal 3-72, Morelia, Michoac\'{a}n, 58089, M\'{e}xico}

\begin{abstract}
An analysis of the Millimetre Astronomy Legacy Team 90 GHz (MALT90) survey has produced a sample of 27 candidate dense molecular clumps with large collapse motions, as revealed by large ``blue'' asymmetrical line profiles of the optically thick \hcop\, line.
New, more sensitive molecular line observations of this sample, conducted with the Mopra 22-m telescope, confirm the blue asymmetries in the \hcop\, line profiles, with large, positive values of the asymmetry parameter $A$ ($\bar{A}_{HCO^+} = 0.69\pm0.01$), and positive, but smaller asymmetries in the \hcn\, and \hnc\, lines: ($\bar{A}_{HCN} = 0.35\pm0.01$ and $\bar{A}_{HNC} = 0.28\pm0.01$), as expected for a less optically thick tracer in collapsing clumps. The small, positive mean asymmetry parameters for \cch\, and \htcop, $\bar{A}_{C_2H} = 0.15\pm0.02$ and $\bar{A}_{H^{13}CO^+} = 0.18\pm0.03$, likely indicate slightly optically thick emission for at least some clumps.  The hyperfine ratios for \nthp\, are in their optically thin, LTE, values, but for \hcn\ they are not; the $F=1 \to 1$ hyperfine line shows abnormally weak intensities. A simple two-component model shows that self-absorption of the background $F = 1 \to 1$ hyperfine line by the main $F = 2 \to 1$ hyperfine line of a cold, foreground, redshifted cloud can reproduce the observed \hcn\, hyperfine intensities and match the \hcn\, and \hcop\, line profiles.  All of these results are consistent with self-absorption of the optically thick lines on the red side of the profile, as expected for collapsing clumps.  A simple two-cloud model suggests that this sample represents dense clumps with extreme collapse velocities, $V_{inf} \sim 2.4$ \kms.
\end{abstract}

\section{Introduction}

One common technique to detect collapse motions in star-forming molecular clumps is to search for asymmetries in the line profiles of optically thick molecular lines (e.g., \citealt{Evans1999, Evans2003}).  In collapsing clumps with higher excitation temperatures toward the center and lower excitation temperatures in the exterior, the spectral profiles for optically thick lines are often asymmetric, with brighter emission on the higher-frequency ``blue" side of the line profile than on the lower-frequency ``red'' side.  Such excitation temperature gradients can arise from internal heating by star formation, or alternatively, by higher densities toward the center, leading to thermalized excitation in the interior and subthermal excitation in the periphery.  

This so-called ``blue-red asymmetry'' or ``blue asymmetry'' arises from the fact that, in a collapsing cloud, the optically thick lines probe different regions in the cloud for the blue and red velocities projected along the line of sight.  At redshifted velocities, the lower-excitation, outer regions on the near side of the cloud are first encountered along the line of sight, while at blueshifted velocities, the higher excitation inner regions towards the center of the cloud are first encountered along the line of sight. For an optically thick line, these regions obscure any line emission at the same velocity from further along the line of sight.  Thus, since the line brightness for an optically thick line is higher for regions with higher excitation temperatures, the blue side of the profile is brighter than the red side because it probes regions with higher excitation temperatures.  In the earliest reports of this effect toward low-mass star forming regions, the line profiles are typically double-peaked, with the blue peak having a higher intensity than the red peak \citep{SnellLoren1977,Walker1986,Mardones1997}. Along the same line of reasoning, an expanding cloud would have a red asymmetry, as the blueshifted velocities would be in the outer regions on the near side of the cloud.  For reviews on blue asymmetries and their use as a collapse tracer in low mass star-forming regions, see \cite{Evans1999} and \cite{Evans2003}.  

Simulations of high-mass star-forming regions with realistic, complicated filamentary structures and flows, however, predict more complicated line profiles, typically with only a slight blue asymmetry and no distinct double-peaked profiles (e.g., \citealt{Smith2012,Smith2013}).
For high-mass star-forming regions, therefore, the classic asymmetric, double-peaked profile is rare, although it has been detected toward a few sources \citep{SunGao2009,Liu2013,Contreras2018,Morii2025}. Thus, statistical studies of optically thick tracers like \hcopnt\, in large samples are usually necessary to infer collapse \citep{Fuller2005, Wu2007, SunGao2009, Chen2010, Lopez2010, Reiter2011, Rygl2013, Campbell2016, Pillai2023}.

The Millimetre Astronomy Legacy Team (MALT90) survey used the 22-m Mopra single dish telescope near Coonabarrabran, NSW, Australia to characterize the molecular line emission with frequencies near 90 GHz from a large sample of $\sim 3,000$ dense molecular clumps \citep{Foster2011, Jackson2013, Rathborne2016}, first identified by their 870 \um\, continuum emission by the APEX Telescope Large Area Survey of the Galaxy (ATLASGAL)  \citep{Schuller2009}.  An analysis of the MALT90 data toward $\sim1,000$ of these molecular clumps \citep{Jackson2019} showed statistical evidence for blue asymmetries in the optically thick \hcop\, line with respect to the systemic velocity established by the optically thin \nthp\, and \htcop\, lines.  This result can be interpreted as indicating overall collapse motions in dense molecular clumps.  

To characterize the degree of the blue asymmetry, \cite{Jackson2019} devised an {\it asymmetry parameter} $A$, defined as $A = \frac{I_{blue} - I_{red}}{I_{blue} 
+ I_{red}}$, where $I_{blue}$ is the integrated line flux to velocities more negative than the systemic velocity, and $I_{red}$ the integrated line flux to more positive velocities.  $A$ quantifies the fraction of excess flux in the line profile lying to more negative (``blue'') radial velocities of the systemic velocity, established by Gaussian fits to an optically thin tracer such as \nthp\, or \htcop. If all of the line flux is blueward of the systemic velocity, $A = 1$, and if all the flux is redward of the systemic velocity, $A = -1$.  Perfectly symmetric line profiles have $A=0$.  (If a line shows absorption features against the continuum on the red side of the line profile, the absorbing flux below the continuum baseline is counted as negative, so that for inverse P Cygni profiles, $A$ can have a value greater than one.)  The average \hcop\, asymmetry parameter for the entire ensemble of MALT90 targets is $\bar{A}_{HCO^+} = 0.083\pm0.010$ \citep{Jackson2019}, consistent with model predictions \citep{Smith2012, Smith2013} and indicating overall collapse motions for dense molecular clumps.

Here we examine a subsample of 27 candidate clumps with large collapse motions selected from the MALT90 survey.  These sources were selected to have large values of $A$, specifically $A>0.8$, for the \hcop\, line, using \nthp\, as the optically thin reference line. Given the relatively poor signal to noise of the MALT90 survey, however, these values are uncertain.  This study presents new, more sensitive Mopra data in order to verify whether the blue asymmetries are real, to quantify the asymmetries better, and to determine the best candidates of collapsing clumps for follow up studies. 

\begin{figure}
    \centering
    \includegraphics[width=\linewidth,height=0.9\textheight,keepaspectratio]{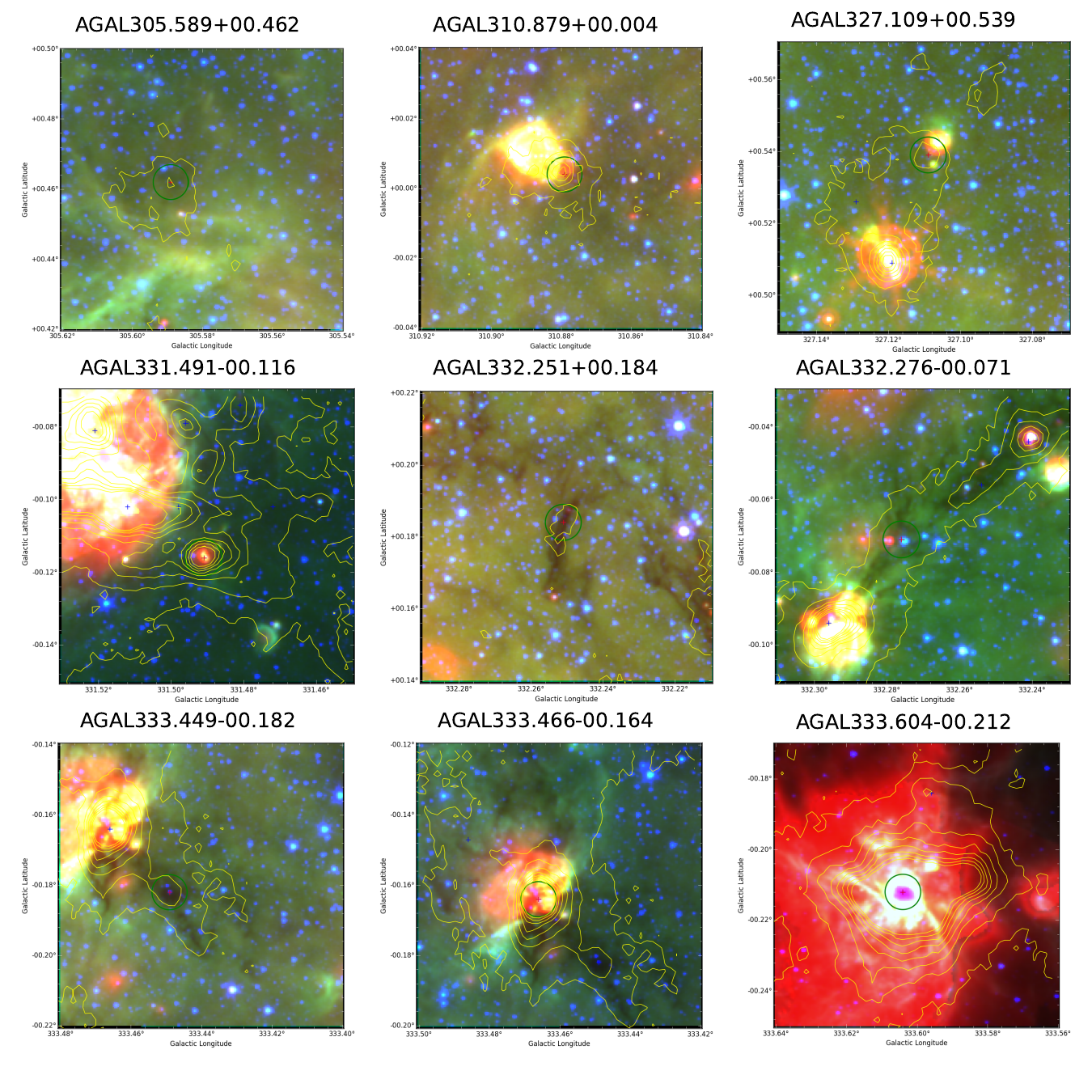} 
    \caption{{\it Spitzer} mid- (GLIMPSE) and far-infrared (MIPSGAL) images of the selected sources, with 3.6 \um\, in blue, 8.0 \um\, in green, and 24 \um\, in red.  Red crosses indicate the source position. The green circle indicates the size of the Mopra beam and the observed position.  Yellow contours indicate ATLASGAL 870 \um\, flux with levels at 0.1 Jy/beam.}
    \label{fig:fig0_spitzer}
\end{figure}

\begin{figure}
    \centering
    \includegraphics[width=\linewidth,height=0.9\textheight,keepaspectratio]{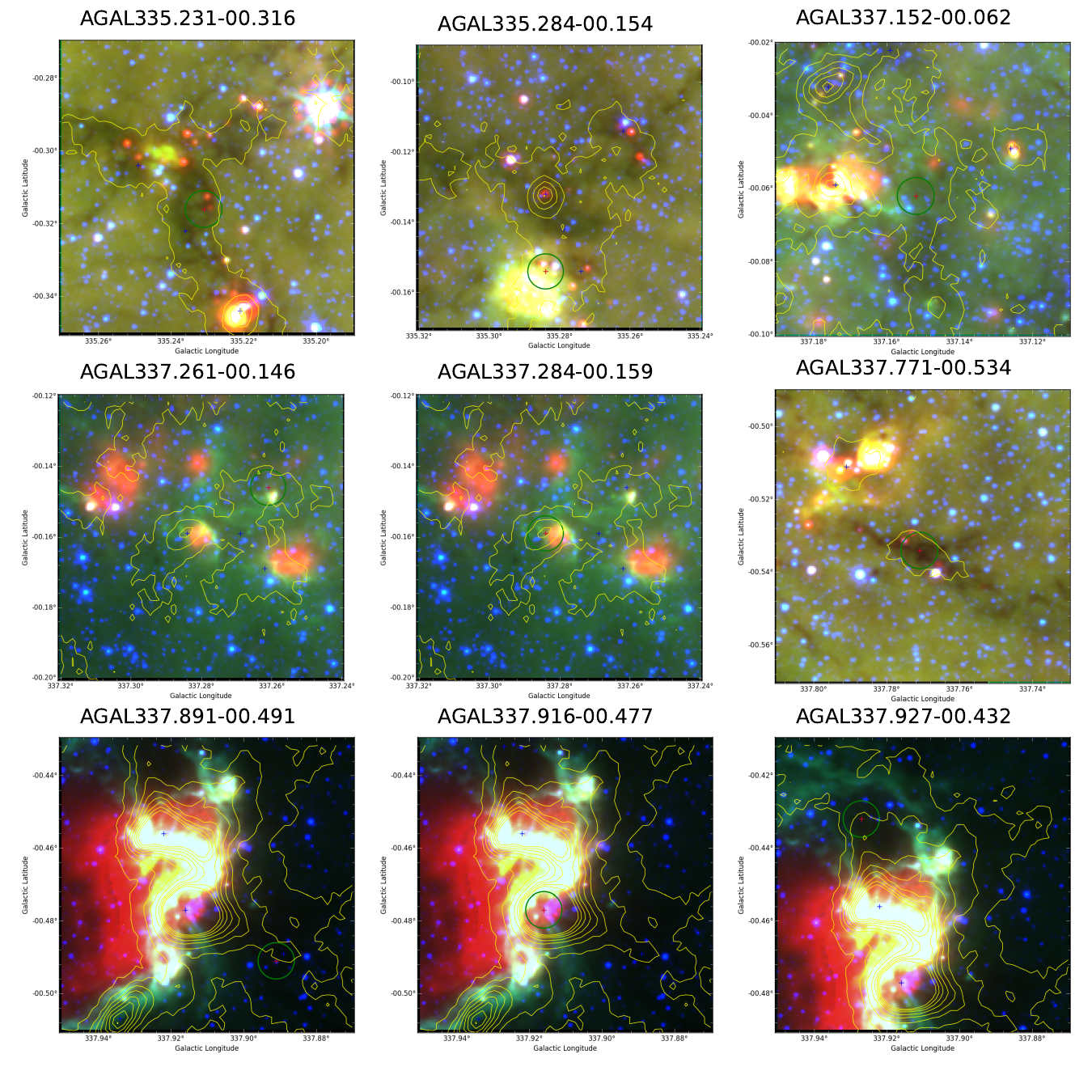} 
    \caption{{\it Spitzer} mid- (GLIMPSE) and far-infrared (MIPSGAL) images of the selected sources, with 3.6 \um\, in blue, 8.0 \um\, in green, and 24 \um\, in red.  Red crosses indicate the source position. The green circle indicates the size of the Mopra beam and the observed position.  Yellow contours indicate ATLASGAL 870 \um\, flux with levels at 0.1 Jy/beam.}
    \label{fig:fig1_spitzer}
\end{figure}

\begin{figure}
    \centering
    \includegraphics[width=\linewidth,height=0.9\textheight,keepaspectratio]{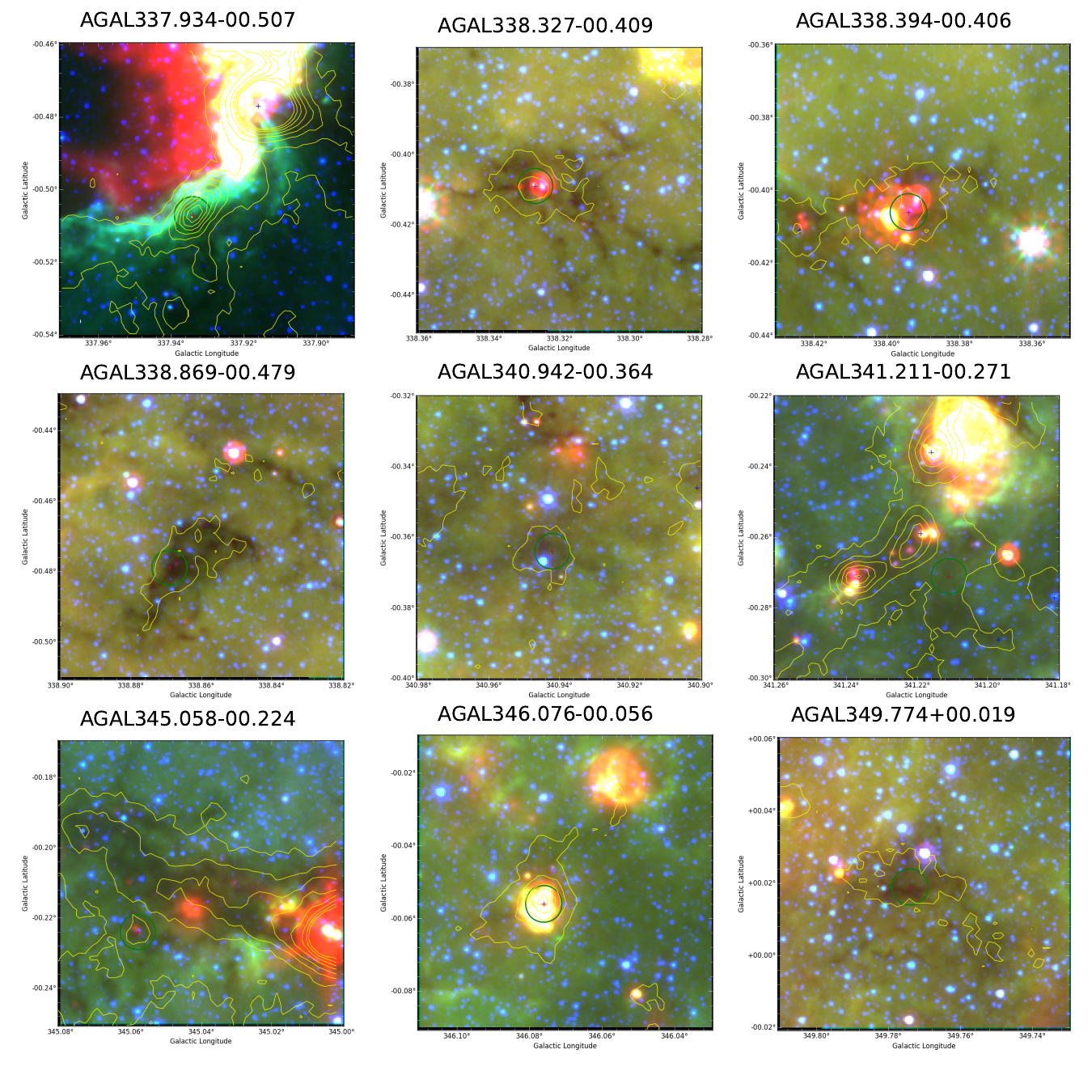} 
    \caption{{\it Spitzer} mid- (GLIMPSE) and far-infrared (MIPSGAL) images of the selected sources, with 3.6 \um\, in blue, 8.0 \um\, in green, and 24 \um\, in red.  Red crosses indicate the source position. The green circle indicates the size of the Mopra beam and the observed position.  Yellow contours indicate ATLASGAL 870 \um\, flux with levels at 0.1 Jy/beam. }
    \label{fig:fig2_spitzer}
\end{figure}

Table \ref{tab:parameters} lists the 27 candidate clumps and their properties.  Since MALT90 sources were selected from the ATLASGAL survey \citep{Schuller2009, Contreras2013}, the source names are AGAL$ll.lll$$\pm$$bb.bbb$ where $l$ and $b$ are the Galactic longitude and latitude, respectively, in degrees.  Distances to the sources are taken from \cite{Whitaker2017} and source classifications, dust temperatures, and mass column densities from \cite{Guzman2015}.  The effective radii are calculated from the effective angular radius presented in \cite{Guzman2015} and the distances from \cite{Whitaker2017}, using $r = \theta D$.
Following \cite{Contreras2017}, dust masses are calculated from $M_{dust}=\langle N\rangle\pi r_{eff}^2$.  Figures \ref{fig:fig0_spitzer} to \ref{fig:fig2_spitzer} show the mid-infrared images for these candidate collapsing clumps from the {\it Spitzer} GLIMPSE \citep{Benjamin2003} and MIPSGAL \citep{Carey2009} surveys, along with 870 \um\, continuum emission from ATLASGAL \citep{Schuller2009}.

\section{Observations}

 The sample of 27 candidate collapsing clumps was observed with the 22-m Mopra telescope near Coonabarabran, NSW, Australia in several molecular lines in the 86-93 GHz frequency range using the MOPS backend. Observations were conducted from 7 May 2018 to 13 September 2018. The receiver and MOPS backend set up was identical to that used for the MALT90 Survey \citep{Jackson2013}.  Table \ref{tab:restfreqs} gives the spectral lines observed and their adopted rest frequencies.  The spectral resolution for these observations is 0.11 \kms, and the angular resolution is 38$''$.
 
 Instead of mapping the source as was done in MALT90, only a single position centered on the ATLASGAL source coordinates was observed.  This approach resulted in substantially longer integration times: 30 minutes per position in this study as compared with 30 seconds in MALT90.  These longer integration times resulted in a much better noise performance.  In units of \tastar, the antenna temperature corrected to the top of the atmosphere, the rms noise for these observations has an average value of 0.034 K and a median value of 0.026 K.  This is a substantial improvement over MALT90, which had typical noise values near 0.25 K \citep{Jackson2013}.

\section{Results}

\begin{figure}[H]
    \centering
    \includegraphics[width=\linewidth,height=0.9\textheight,keepaspectratio]{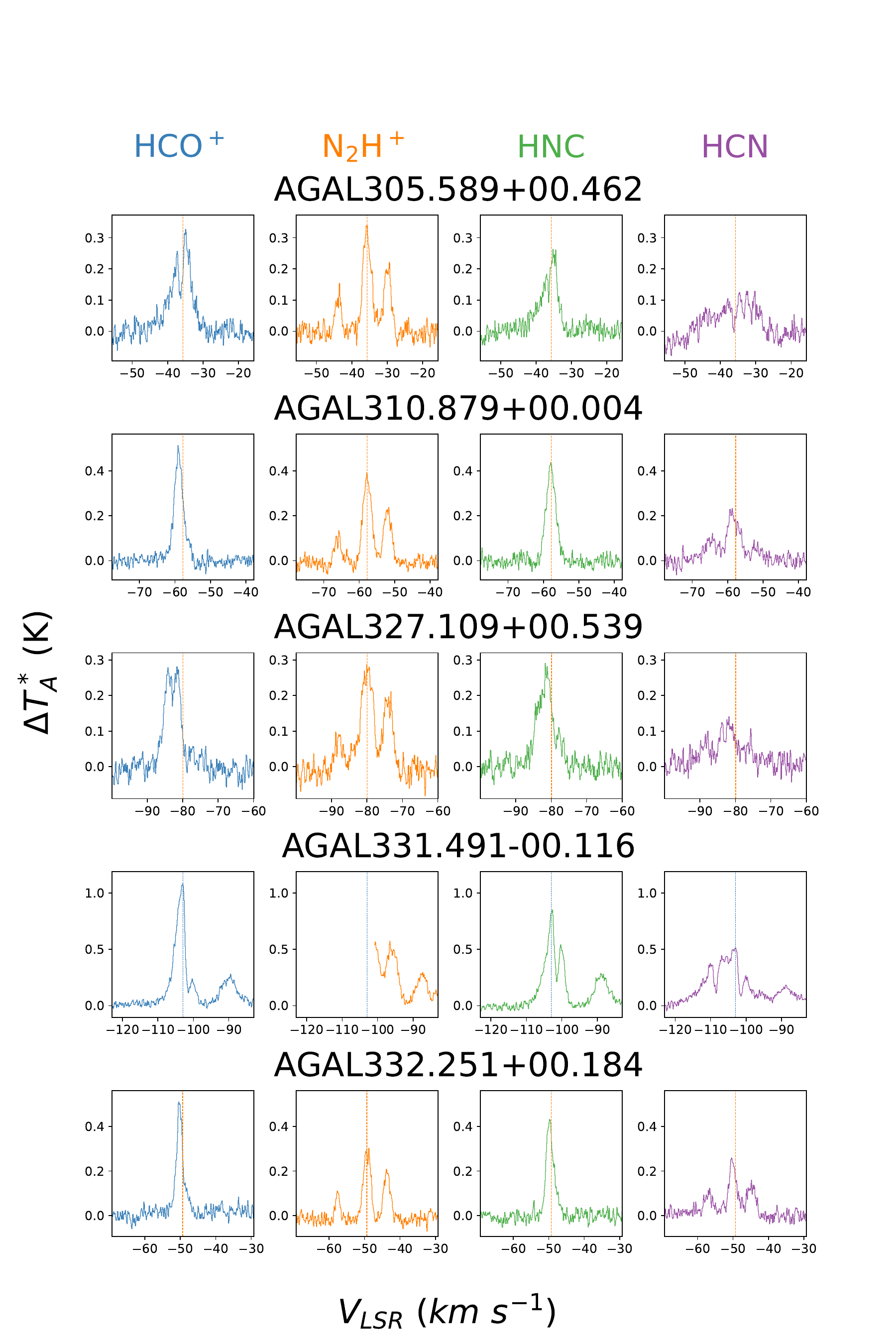} 
    \caption{ Mopra 3 mm spectra of the \hcop, \nthp, \hcn, and \hnc\, lines.  The vertical lines show the fitted velocity of the \nthp\, line in orange, if available, otherwise the velocity of the peak \hcop\, intensity in cyan.}
    \label{fig:fig0_bright}
\end{figure}

\begin{figure}[H]
    \centering
    \includegraphics[width=\linewidth,height=0.9\textheight,keepaspectratio]{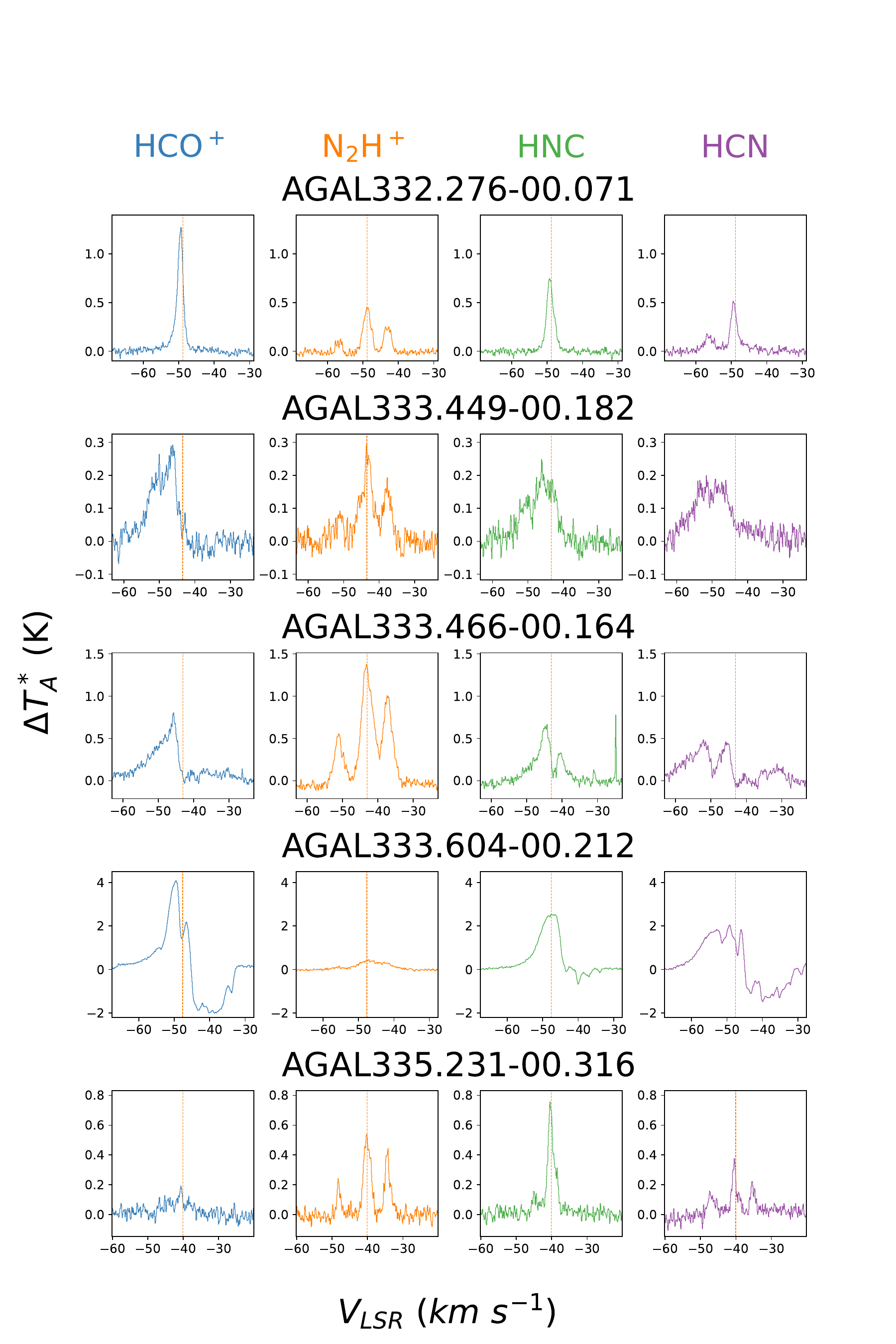} 
    \caption{ Mopra 3 mm spectra of the \hcop, \nthp, \hcn, and \hnc\, lines. Same as Fig, \ref{fig:fig0_bright}, but for different sources.}
    \label{fig:fig1_bright}
\end{figure}

\begin{figure}[H]
    \centering
    \includegraphics[width=\linewidth,height=0.9\textheight,keepaspectratio]{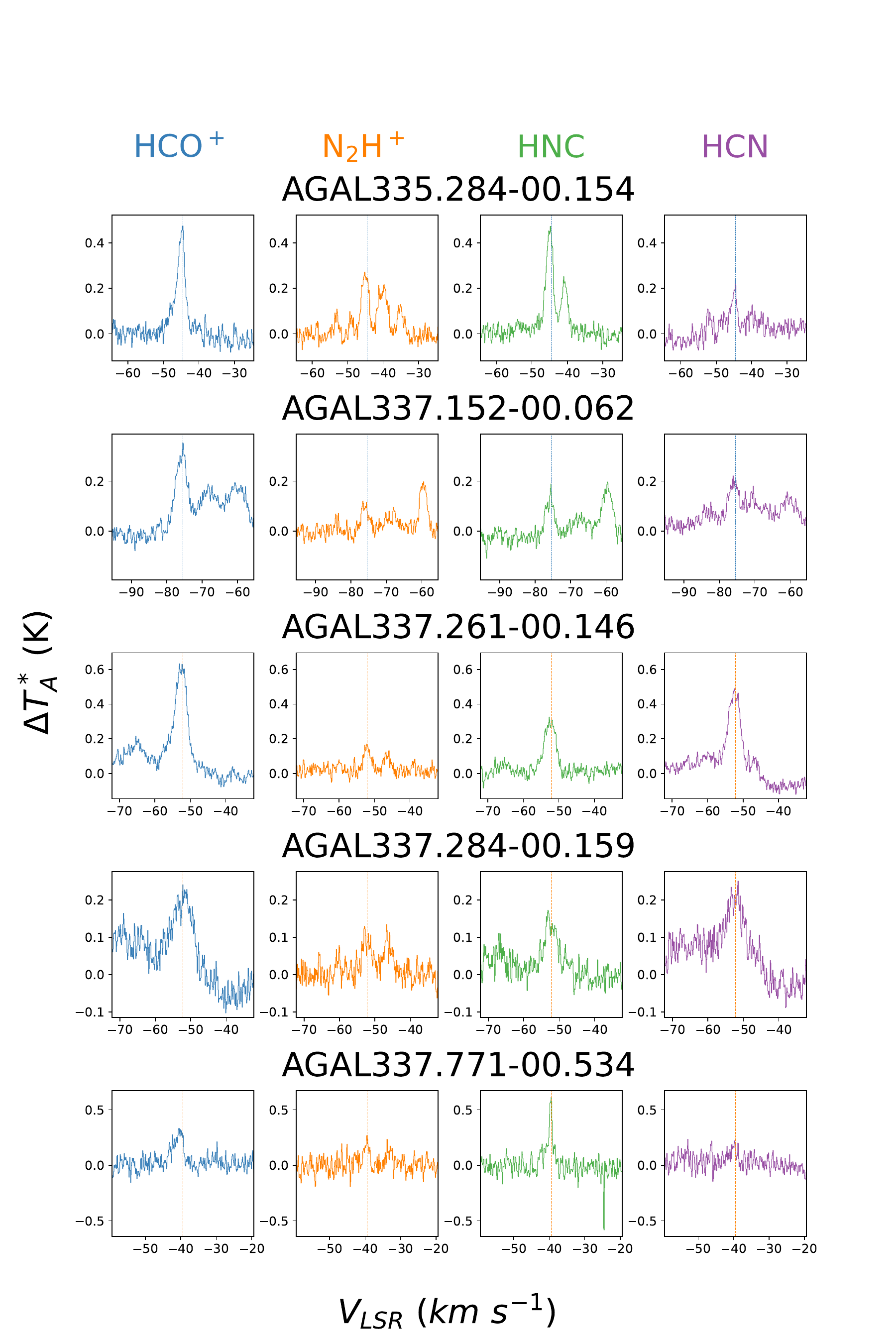} 
    \caption{ Mopra 3 mm spectra of the \hcop, \nthp, \hcn, and \hnc\, lines. Same as Fig, \ref{fig:fig0_bright}, but for different sources.}
    \label{fig:fig2_bright}
\end{figure}

\begin{figure}[H]
    \centering
    \includegraphics[width=\linewidth,height=0.9\textheight,keepaspectratio]{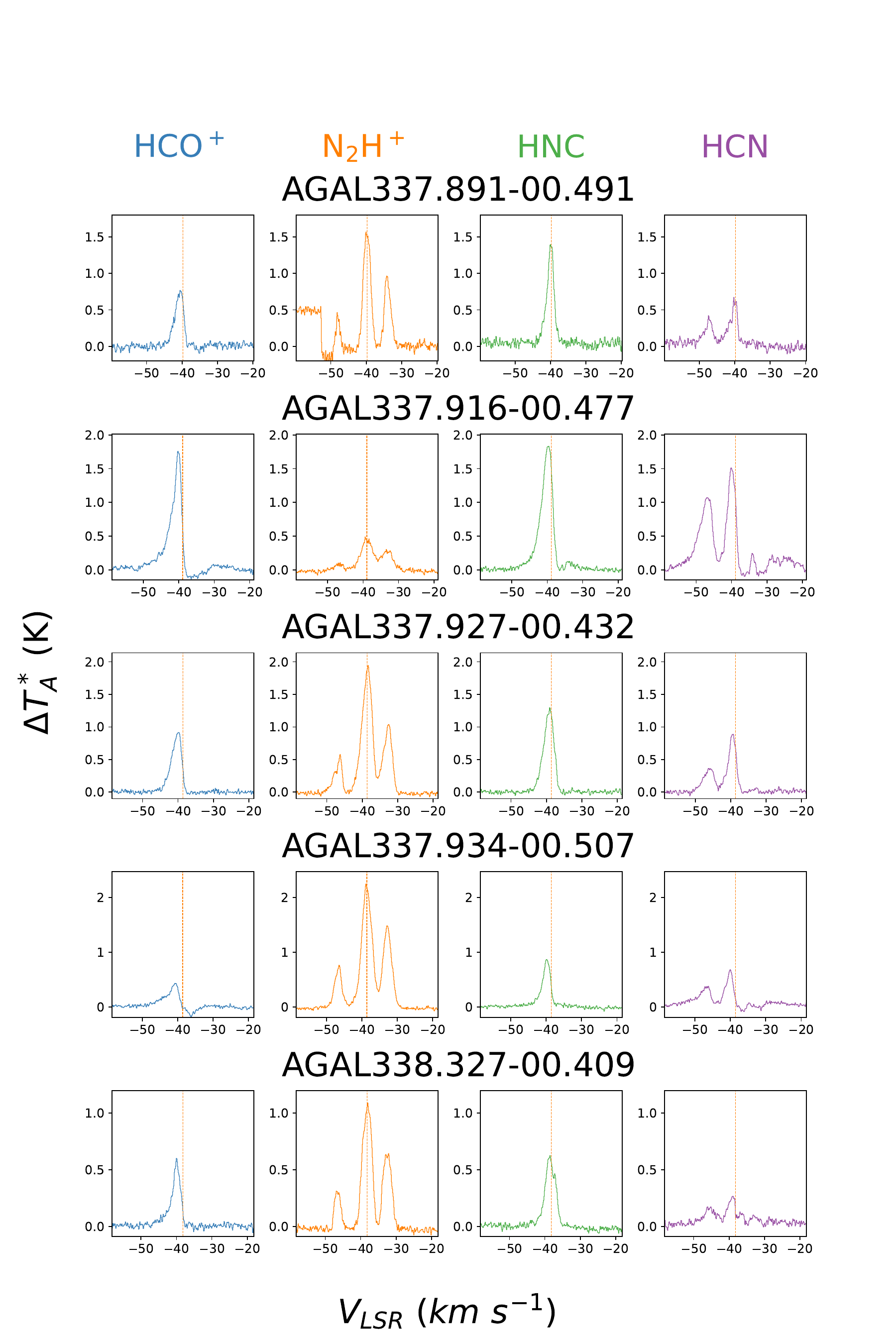} 
    \caption{ Mopra 3 mm spectra of the \hcop, \nthp, \hcn, \hnc, and \htcop\, lines. Same as Fig, \ref{fig:fig0_bright}, but for different sources.}
    \label{fig:fig3_bright}
\end{figure}

\begin{figure}[H]
    \centering
    \includegraphics[width=\linewidth,height=0.9\textheight,keepaspectratio]{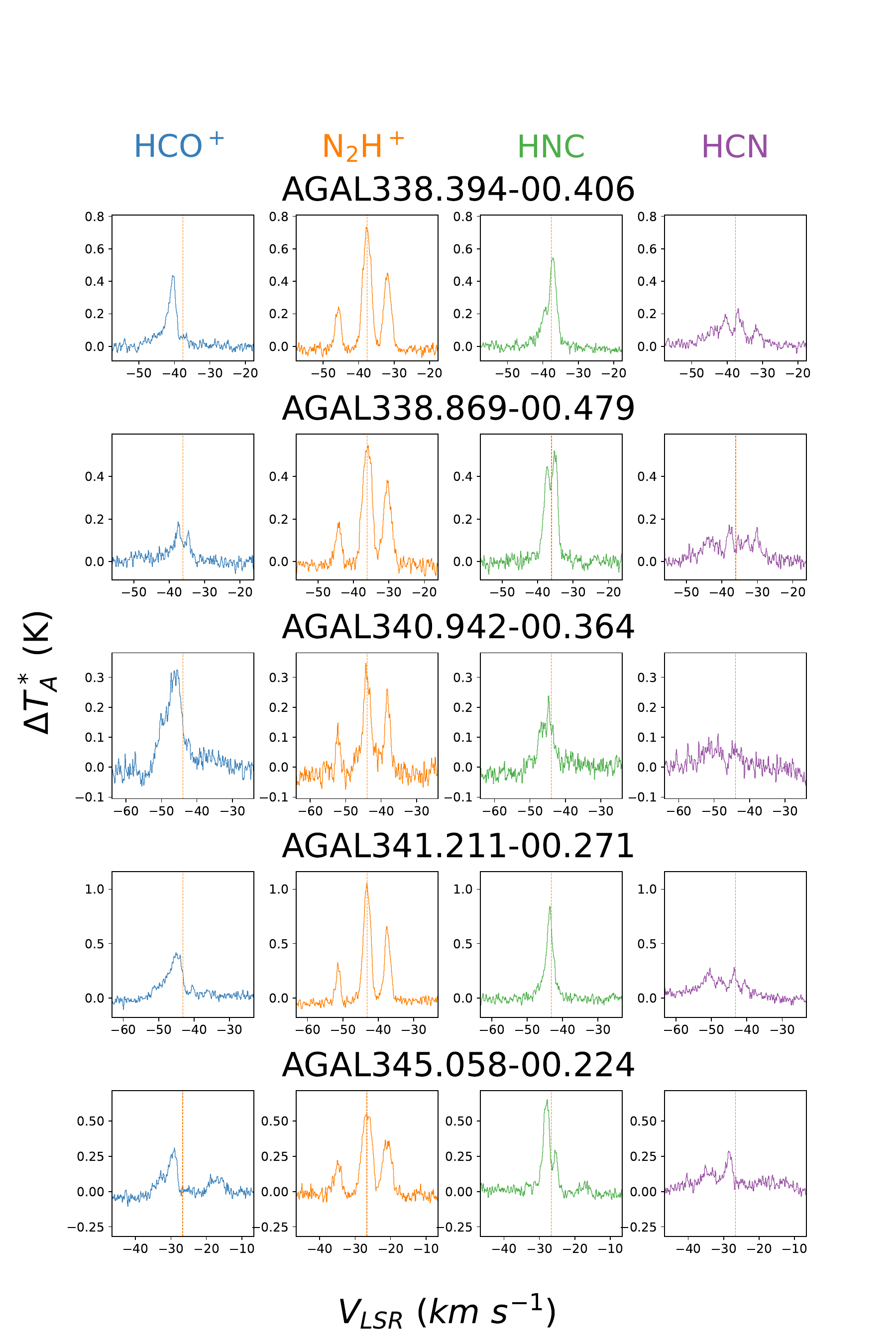} 
    \caption{ Mopra 3 mm spectra of the \hcop, \nthp, \hcn, \hnc, and \htcop\, lines. Same as Fig, \ref{fig:fig0_bright}, but for different sources.}
    \label{fig:fig4_bright}
\end{figure}

\begin{figure}[H]
    \centering
    \includegraphics[width=\linewidth,height=0.9\textheight,keepaspectratio]{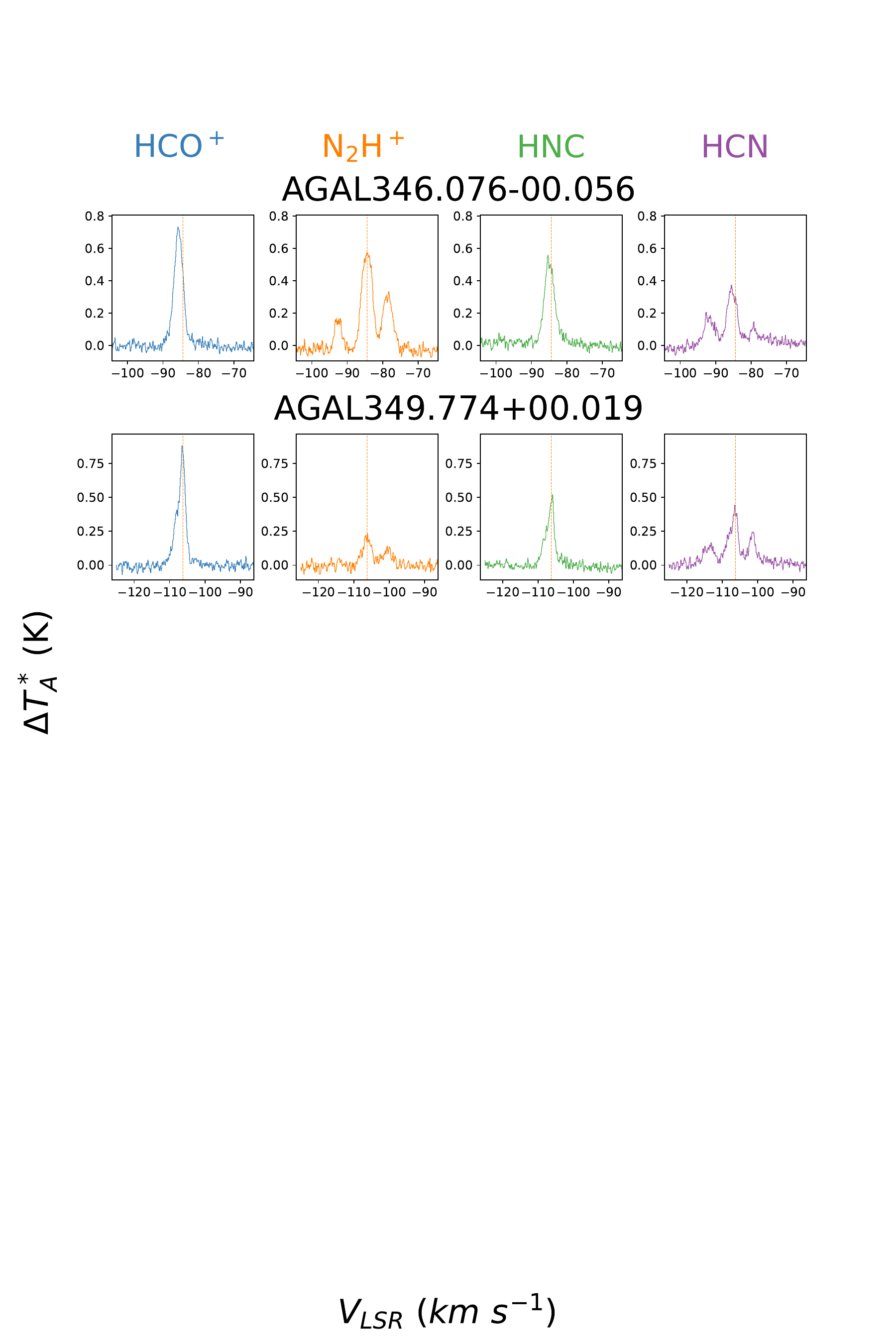} 
    \caption{ Mopra 3 mm spectra of the \hcop, \nthp, \hcn, \hnc, and \htcop\, lines. Same as Fig, \ref{fig:fig0_bright}, but for different sources.}
    \label{fig:fig5_bright}
\end{figure}

\begin{figure}[H]
    \centering
    \includegraphics[width=\linewidth,height=0.9\textheight,keepaspectratio]{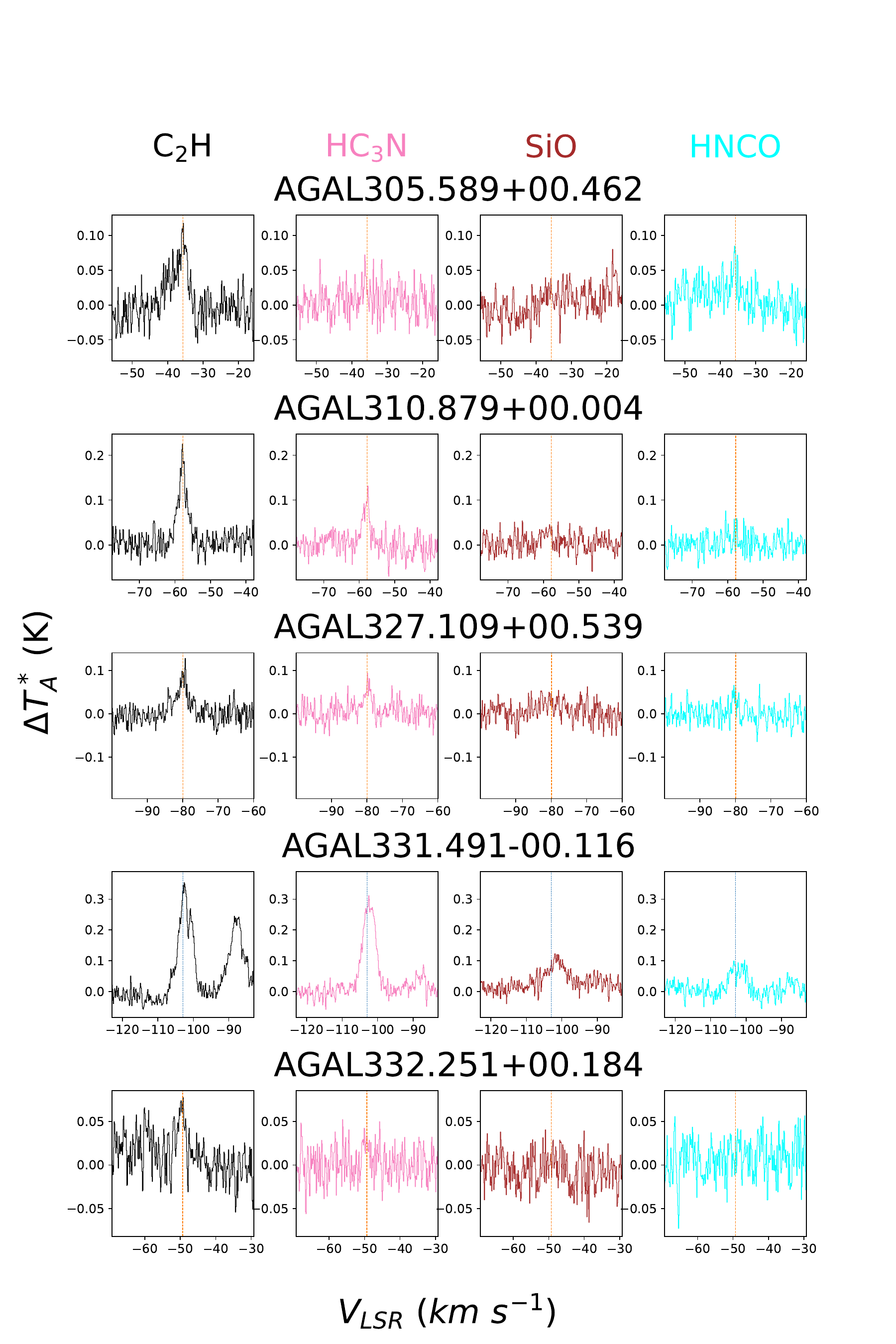} 
    \caption{ Mopra 3 mm spectra of the \cch, \hctn\, \hncofz\, and \sio\, lines.  The vertical lines show the fitted velocity of the \nthp\, line in orange, if available, otherwise the velocity of the peak \hcop\, intensity in cyan.}
    \label{fig:fig0_faint}
\end{figure}

\begin{figure}[H]
    \centering
    \includegraphics[width=\linewidth,height=0.9\textheight,keepaspectratio]{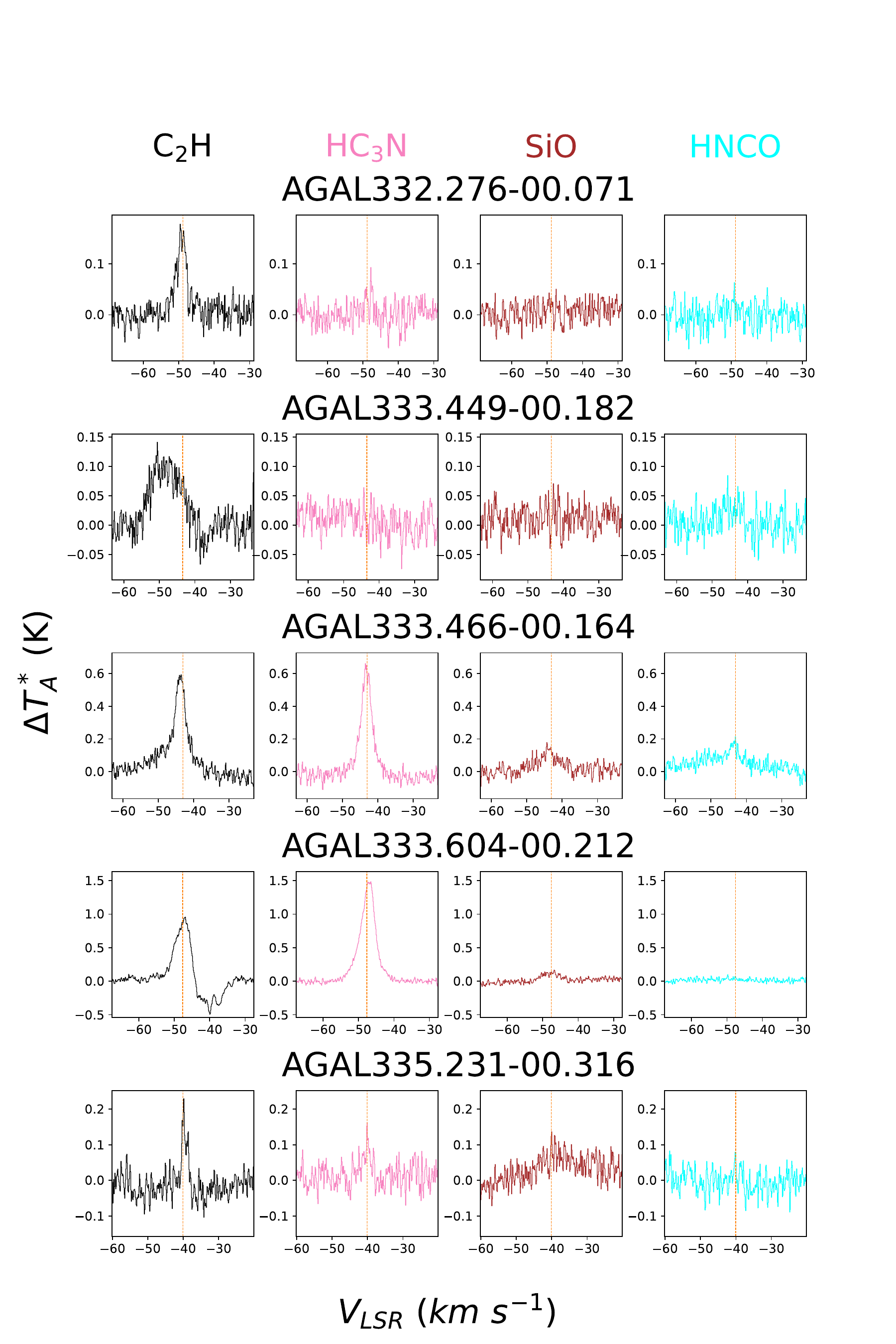} 
    \caption{ Mopra 3 mm spectra of the \cch, \hctn\, \hncofz\, and \sio\, lines. Same as Fig, \ref{fig:fig0_faint}, but for different sources.}
    \label{fig:fig1_faint}
\end{figure}

\begin{figure}[H]
    \centering
    \includegraphics[width=\linewidth,height=0.9\textheight,keepaspectratio]{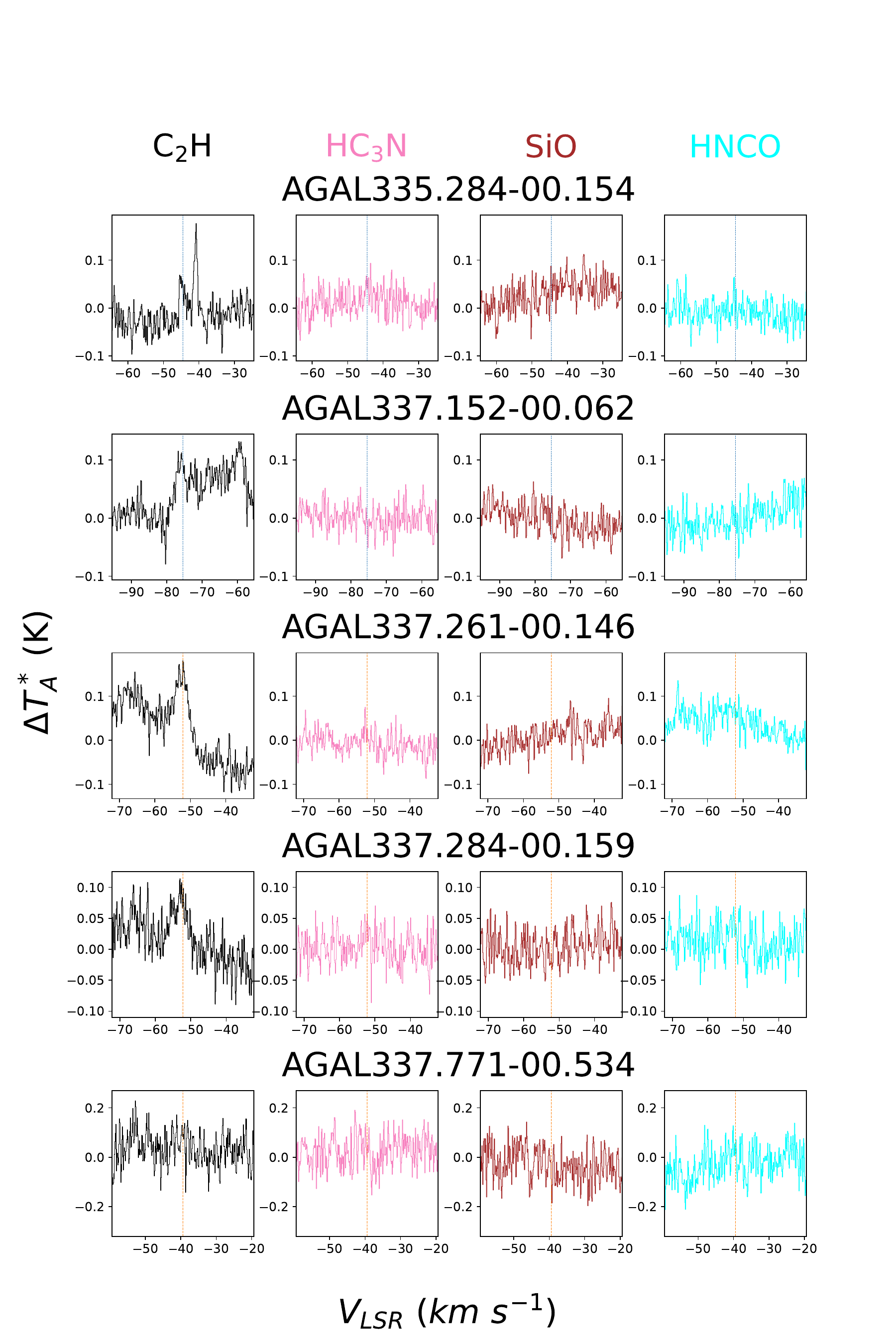} 
    \caption{ Mopra 3 mm spectra of the \cch, \hctn\, \hncofz\, and \sio\, lines. Same as Fig, \ref{fig:fig0_faint}, but for different sources.}
    \label{fig:fig2_faint}
\end{figure}

\begin{figure}[H]
    \centering
    \includegraphics[width=\linewidth,height=0.9\textheight,keepaspectratio]{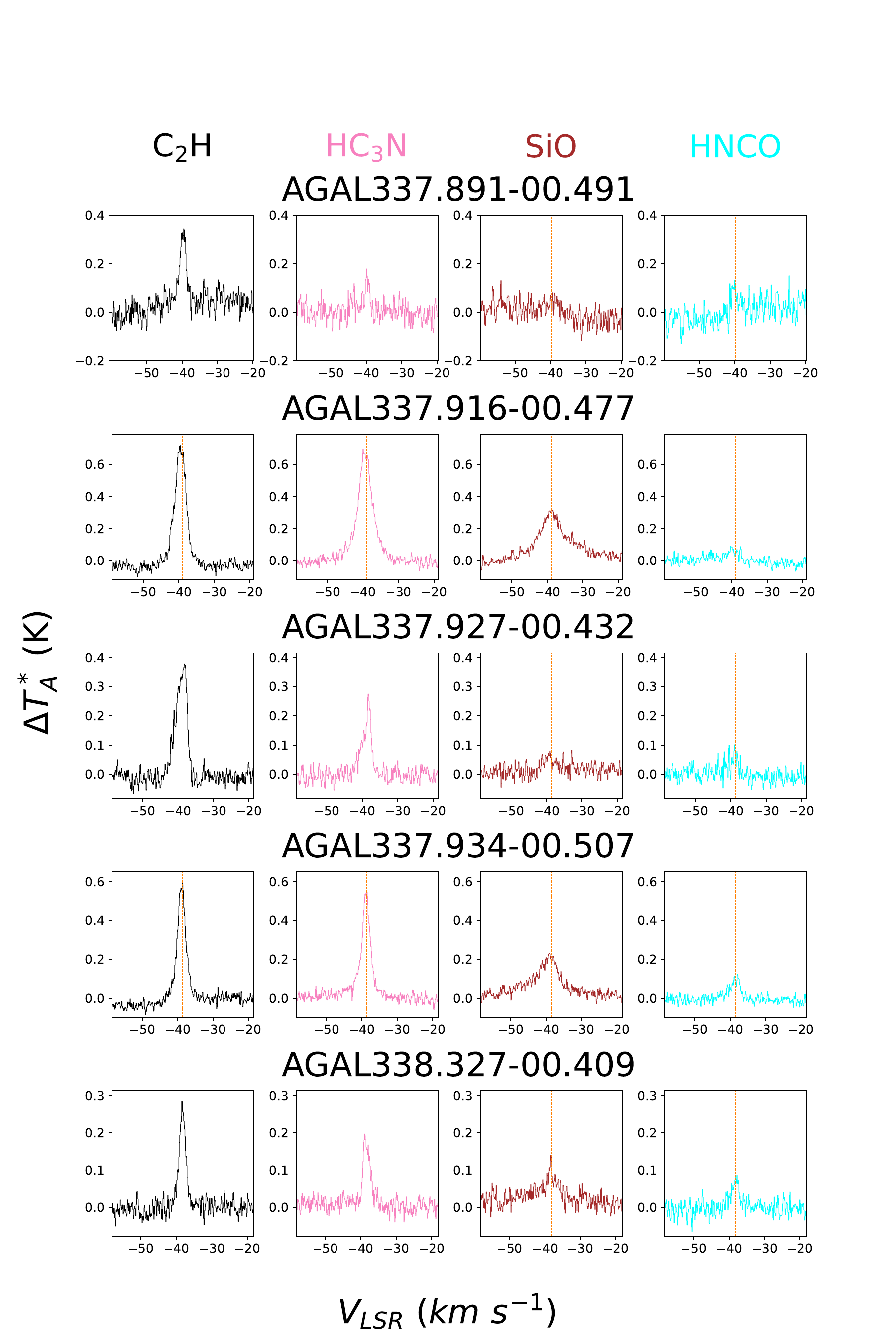} 
    \caption{ Mopra 3 mm spectra of the \cch, \hctn\, \hncofz\, and \sio\, lines. Same as Fig, \ref{fig:fig0_faint}, but for different sources.}
    \label{fig:fig3_faint}
\end{figure}

\begin{figure}[H]
    \centering
    \includegraphics[width=\linewidth,height=0.9\textheight,keepaspectratio]{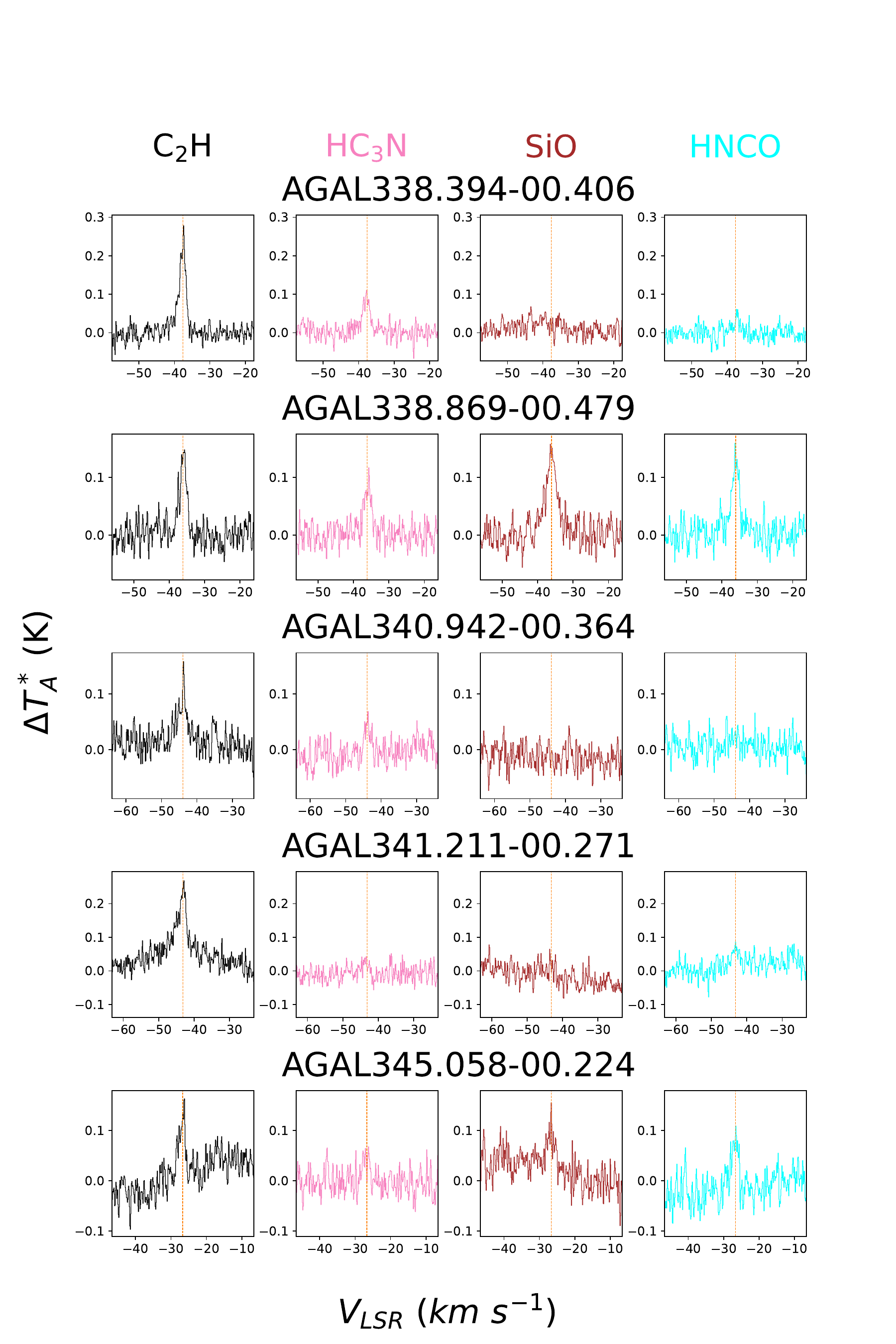} 
    \caption{ Mopra 3 mm spectra of the \cch, \hctn\, \hncofz\, and \sio\, lines. Same as Fig, \ref{fig:fig0_faint}, but for different sources.}
    \label{fig:fig4_faint}
\end{figure}

\begin{figure}[H]
    \centering
    \includegraphics[width=\linewidth,height=0.9\textheight,keepaspectratio]{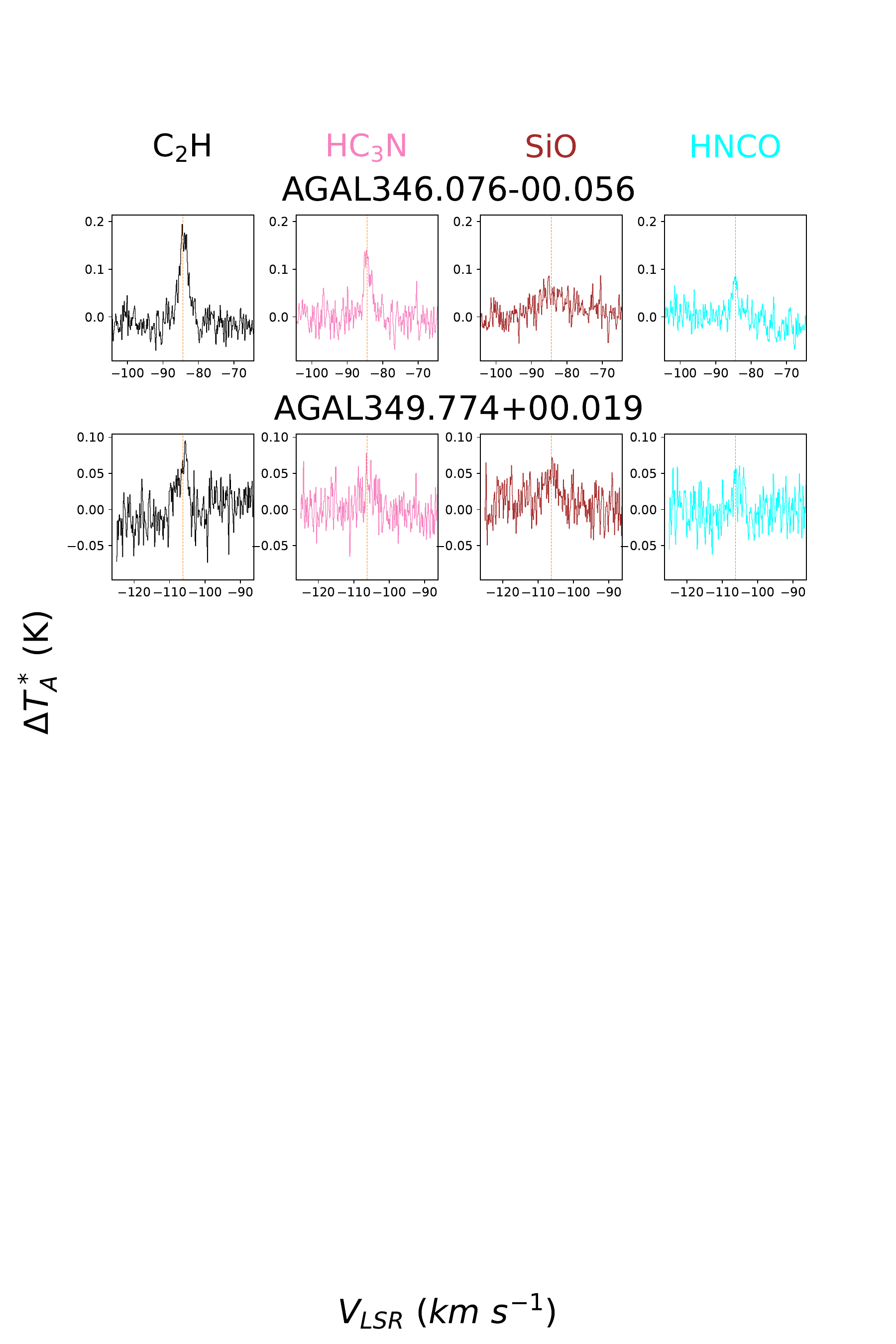} 
    \caption{ Mopra 3 mm spectra of the \cch, \hctn\, \hncofz\, and \sio\, lines. Same as Fig, \ref{fig:fig0_faint}, but for different sources.}
    \label{fig:fig5_faint}
\end{figure}

\begin{figure}[H]
    \centering
    \includegraphics[width=\linewidth,height=0.9\textheight,keepaspectratio]{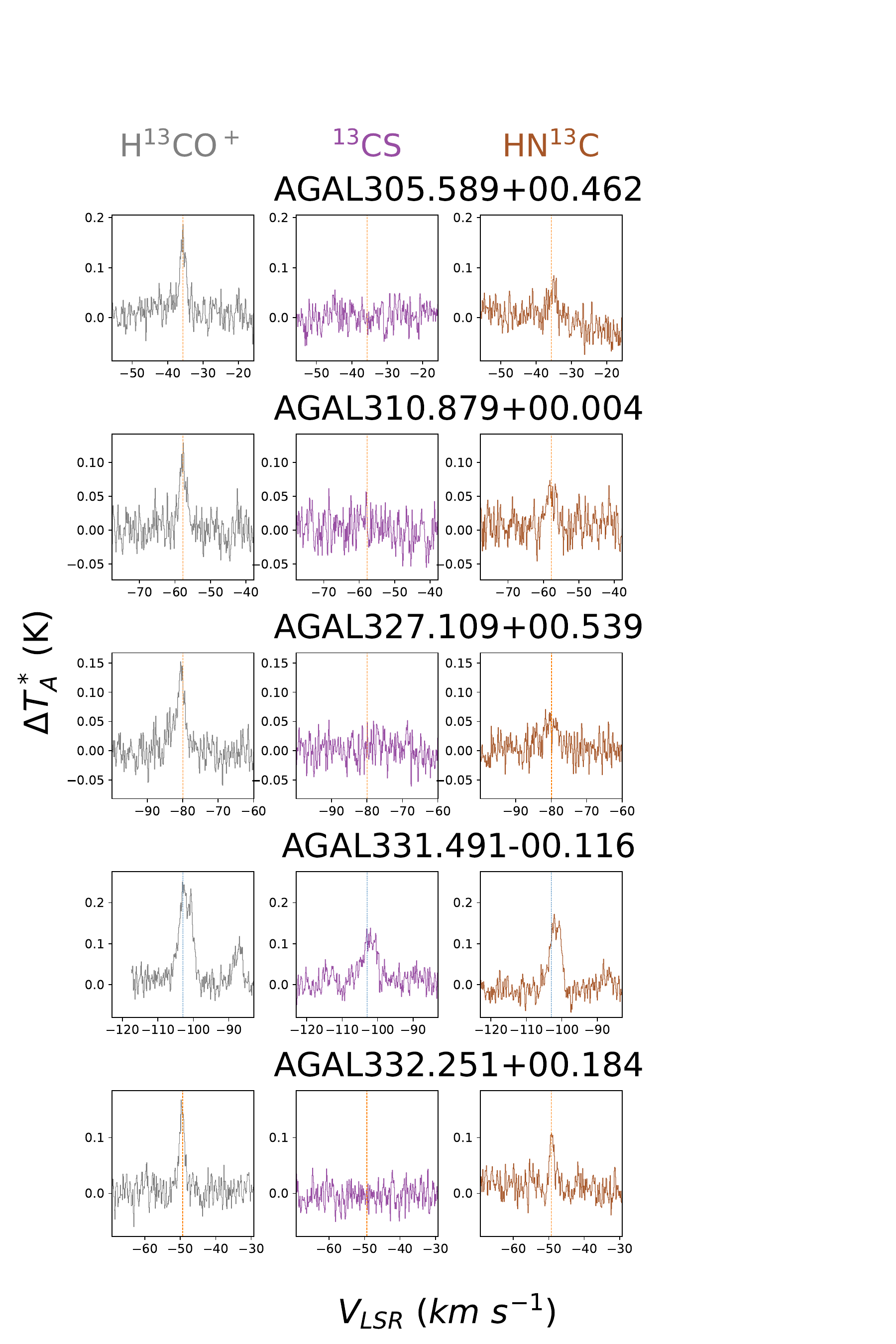} 
    \caption{ Mopra 3 mm spectra of the \htcop, \hntc, and \tcs\, lines.  The vertical lines show the fitted velocity of the \nthp\, line in orange, if available, otherwise the velocity of the peak \hcop\, intensity in cyan.}
    \label{fig:fig0_isotopologue}
\end{figure}

\begin{figure}[H]
    \centering
    \includegraphics[width=\linewidth,height=0.9\textheight,keepaspectratio]{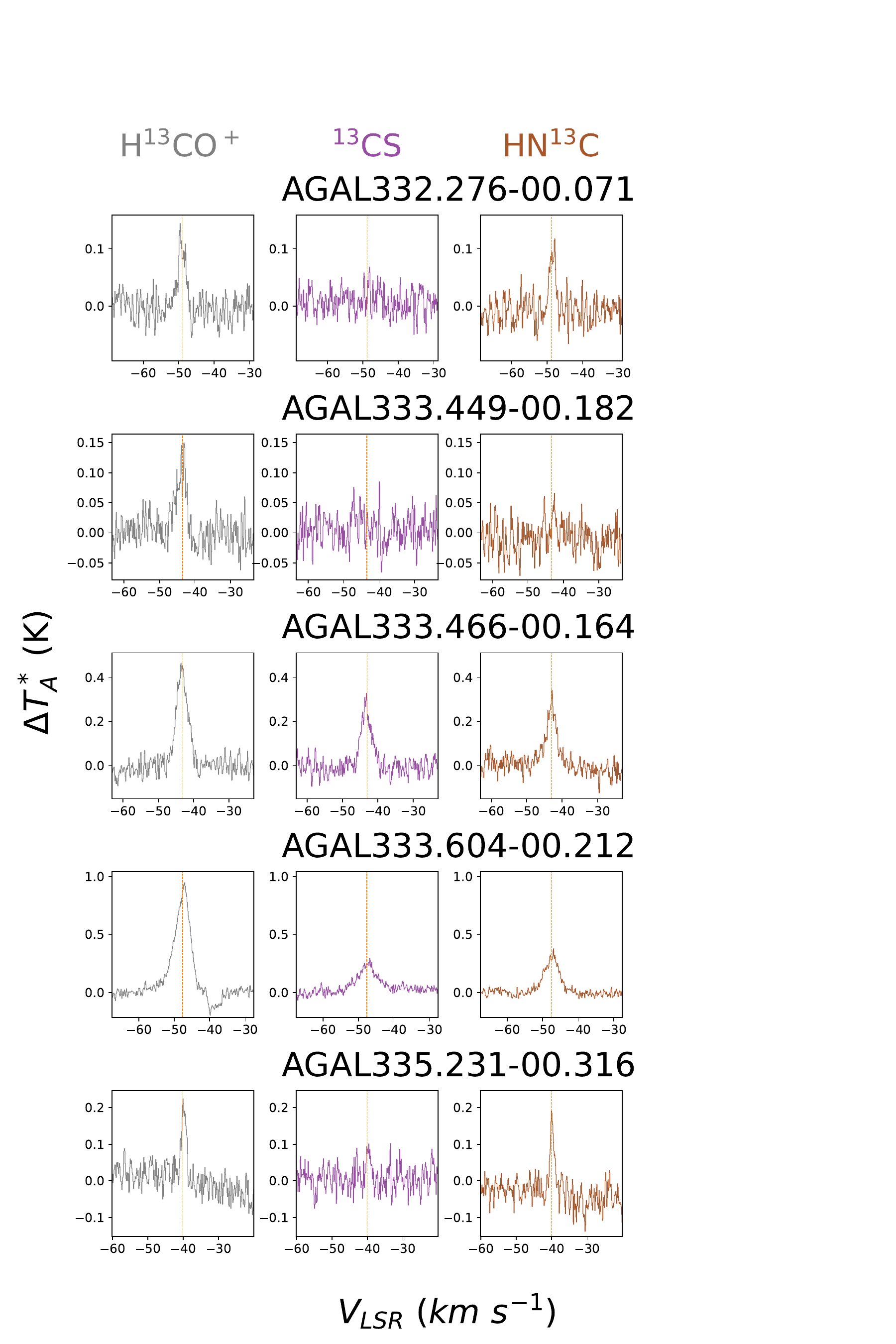} 
    \caption{ Mopra 3 mm spectra of the \htcop, \hntc, and \tcs\, lines. Same as Fig, \ref{fig:fig0_isotopologue}, but for different sources.}
    \label{fig:fig1_isotopologue}
\end{figure}

\begin{figure}[H]
    \centering
    \includegraphics[width=\linewidth,height=0.9\textheight,keepaspectratio]{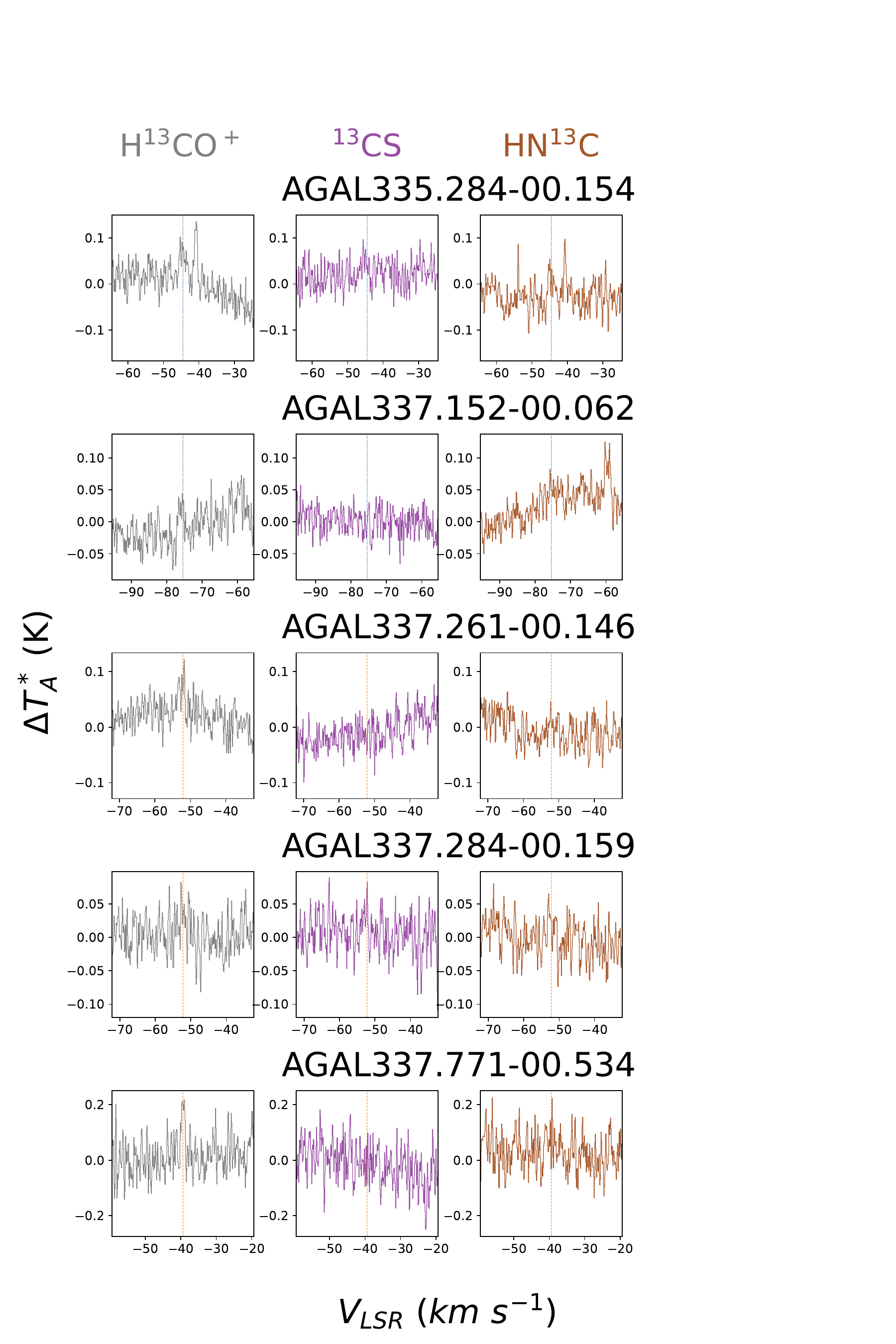} 
    \caption{ Mopra 3 mm spectra of the \htcop, \hntc, and \tcs\, lines. Same as Fig, \ref{fig:fig0_isotopologue}, but for different sources.}
    \label{fig:fig2_isotopologue}
\end{figure}

\begin{figure}[H]
    \centering
    \includegraphics[width=\linewidth,height=0.9\textheight,keepaspectratio]{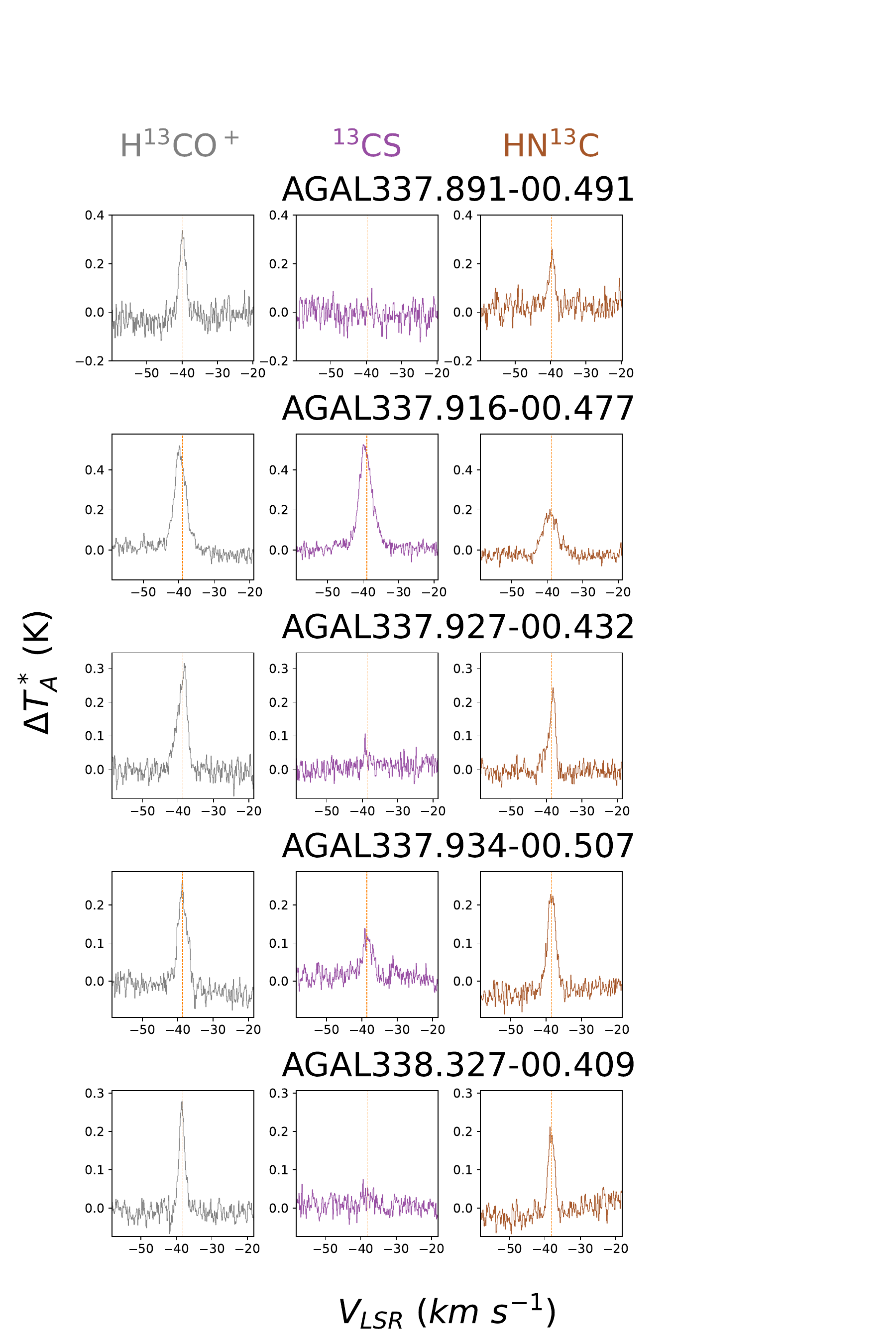} 
    \caption{ Mopra 3 mm spectra of the \htcop, \hntc, and \tcs\, lines. Same as Fig, \ref{fig:fig0_isotopologue}, but for different sources.}
    \label{fig:fig3_isotopologue}
\end{figure}

\begin{figure}[H]
    \centering
    \includegraphics[width=\linewidth,height=0.9\textheight,keepaspectratio]{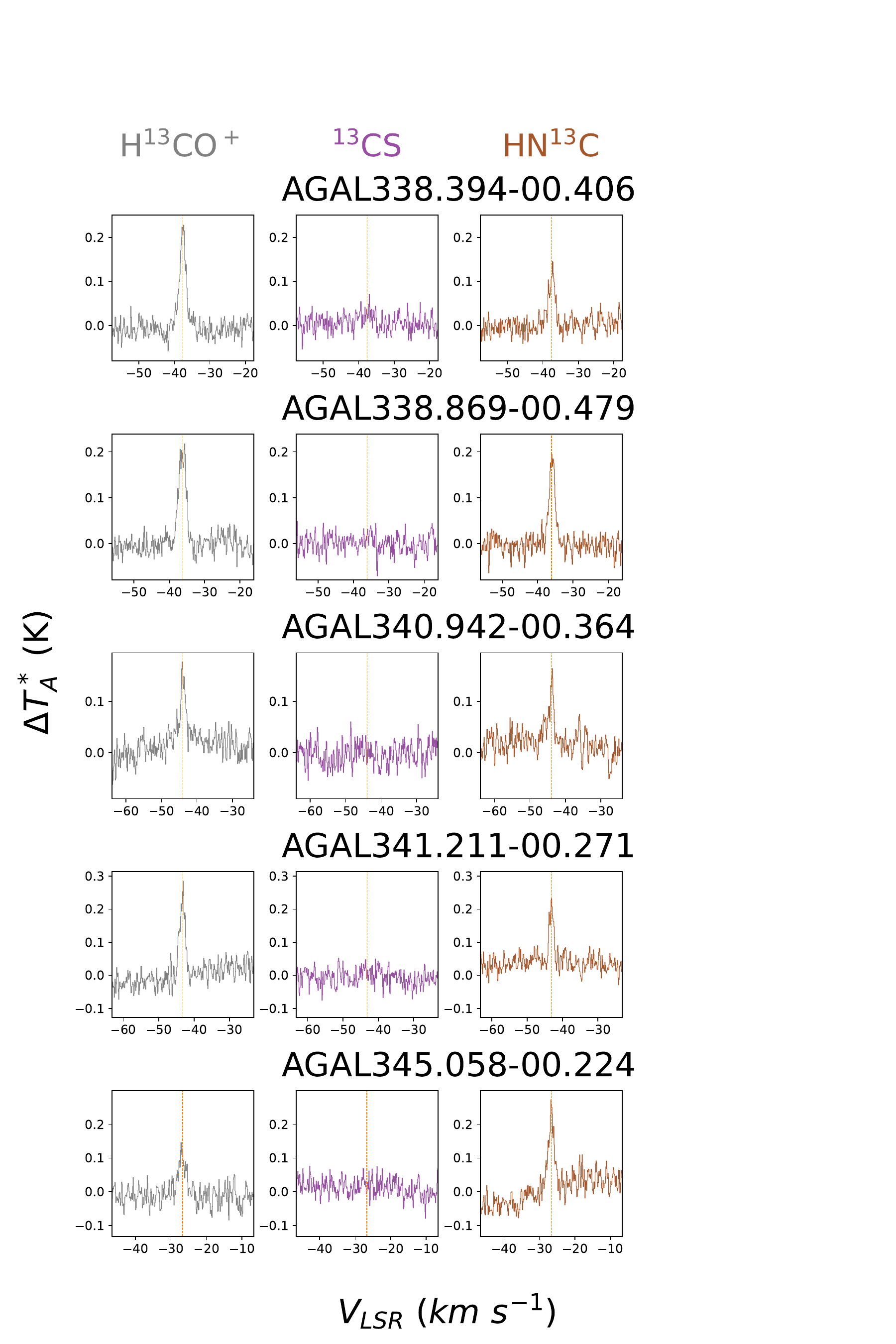} 
    \caption{ Mopra 3 mm spectra of the \htcop, \hntc, and \tcs\, lines. Same as Fig, \ref{fig:fig0_isotopologue}, but for different sources.}
    \label{fig:fig4_isotopologue}
\end{figure}

\begin{figure}[H]
    \centering
    \includegraphics[width=\linewidth,height=0.9\textheight,keepaspectratio]{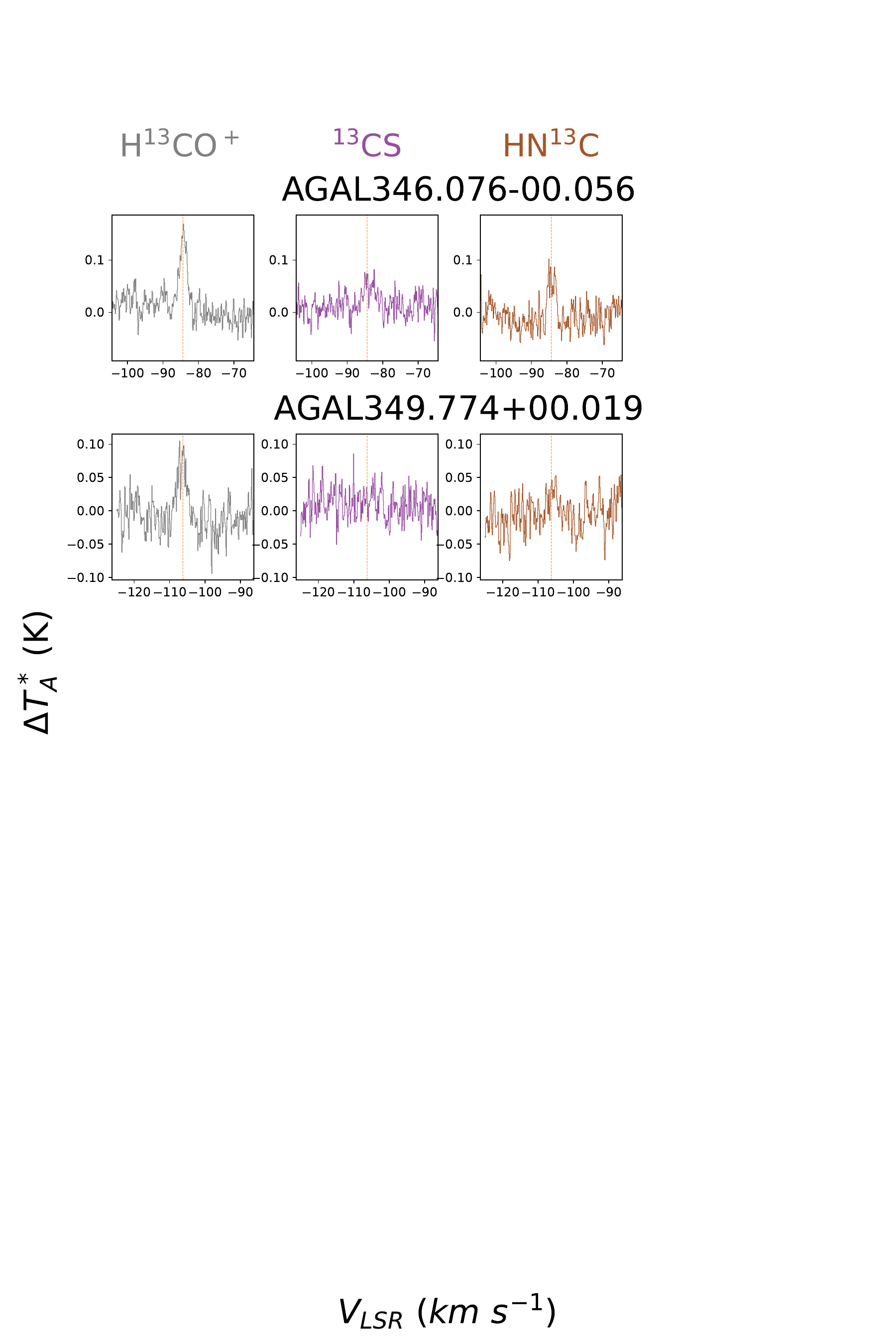} 
    \caption{ Mopra 3 mm spectra of the \htcop, \hntc, and \tcs\, lines.Same as Fig, \ref{fig:fig0_isotopologue}, but for different sources.}
    \label{fig:fig5_isotopologue}
\end{figure}

Spectra were collected for 16 different spectral lines.  We present the data for the 11 brightest molecular lines that were routinely detected. For clarity, we group the lines into three categories for display. The ``bright'' lines [\hcop, \hnc, \hcn, and \nthp] are  shown  in Figures \ref{fig:fig0_bright} to \ref{fig:fig5_bright}. The ``faint'' lines [\cch, \hctn, \hncofz, and \sio] are shown in Figures \ref{fig:fig0_faint} to \ref{fig:fig5_faint}.  The ``isotopologue'' lines [\htcop, \hntc, and \tcs]  are  shown in Figures \ref{fig:fig0_isotopologue} to \ref{fig:fig5_isotopologue}. 

The \nthp\, line was used as the optically thin reference line to determine the LSR radial velocities for each source.  Based on their \nthp\, profiles, three sources were excluded from further analysis because velocities from \nthpnt\, could not be reliably determined.  The LSR velocity of AGAL331.491$-$00.116 was so negative that the entire \nthp\, line hyperfine structure did not fit into the available passband, and, moreover, in many other lines, its line profile shows two velocity components.  The velocity profile for AGAL337.152$-$00.062 shows emission extending over 20 \kms, likely due to the blending of three or more velocity components.  Finally, AGAL335.284$-$00.154 appears to have two velocity components, and the \nthp\, hyperfine lines for the two components are blended, so that a velocity determination using \nthp\, is problematic.

For the remaining 24 sources, we determined the rest velocities by fitting the \nthp\, hyperfine profile, modeled as three Gaussians in the optically thin (5:3:1) amplitude ratio for the $F_1=2 \to 1$ (``main''), $F_1=1 \to 1$ (``right''), and $F_1=0 \to 1$ (``left'') hyperfine components, respectively, as discussed in Section 4.2 below.  The modeled velocity offsets from that of the main hyperfine component were 0.00 \kms, 5.74 \kms,  and $-8.20$ \kms\, for the $F_1=2 \to 1$, $F_1=1 \to 1$, and $F_1=0 \to 1$ hyperfine components, respectively \citep{Pagani2009}. Section 4.2 below shows that the optically thin assumption for the \nthp\, hyperfine components is well justified.
The LSR velocities from \nthp\, are well determined, typically with errors $\leq 0.03$ \kms, or $\sim$30\% of a spectral channel. 
Table \ref{tab:nthp_fit} presents the results of the \nthp\, velocity fits. These fitted velocities are shown as vertical orange lines in Figs. \ref{fig:fig0_bright} to \ref{fig:fig5_isotopologue}.  For the three sources where the \nthp\, velocity was unavailable, we display a vertical blue line at the velocity of the peak \hcop\, intensity instead.

\section{Discussion}

\subsection{Preponderance of Blue Asymmetric Profiles}

Having established the LSR velocities, we then determined asymmetry parameters $A$ for eleven different lines: \hcop, \hcn, \hnc, \cch, \htcop, \tcs, \chtcn, \nthp, \hncofz, \hctn, \sio, \tcs, and \hntc.   The integrals $I_{blue}$ and $I_{red}$ were limited to regions of 40 spectral channels ($\approx 4.4$ \kms) to either side of the systemic velocity $V_{LSR}$. We chose this value as a good compromise that captures most of the line emission without introducing noise from signal-free regions of the spectrum or unwanted contamination from outflow emission (e.g., \citealt{Richer1999, Arce2006}).  For reference, the mean velocity dispersion $\sigma_V$ in the optically thin \nthp\, line is 1.22 \kms. The spectral channel that contained the systemic velocity $V_{LSR}$ was excluded from both integrations. 

Errors in the asymmetry parameter arise from two factors: (1) $\sigma_{fit}$, the error arising from uncertainties to the velocity fit to the optically thin reference line and (2) $\sigma_{int}$,  the statistical error arising from determining the integrals $I_{blue}$ and $I_{red}$ with noisy data. To model the first error, we conducted Monte Carlo simulations to find the value of $A$ by varying $V_{LSR}$ around the nominal fitted value with random Gaussian errors with a dispersion given by the uncertainty in the velocity fit.  The value of $A$ was calculated for each trial, and the dispersion of the ensemble of the resulting $A$ values in the simulation was taken as the error $\sigma_{fit}$.  To determine the second error, $\sigma_{int}$, the formal error in the integrals was calculated assuming the RMS error $\sigma_T$ for each spectrum describes standard, independent, Gaussian noise fluctuations in the amplitudes for each channel.  If $N$ is the number of channels summed in each integral $I_{blue}$ and $I_{red}$, then the error for $A$ due to the noise in the spectrum over the integration range is given by $\sigma_{int} = \frac{2\sigma_T \sqrt{N \left( I_{\text{blue}}^2 + I_{\text{red}}^2 \right)}}{(I_{\text{blue}} + I_{\text{red}})^2}$.  Finally, we assume that the fit errors $\sigma_{fit}$ and the integration errors $\sigma_{int}$ are independent and add in quadrature.  The total error then is $\sigma_{tot} = \sqrt{\sigma_{fit}^2 + \sigma_{int}^2}$.  
In practice, since the velocities are well determined, $\sigma_{int}$ typically dominates the total error.

Table \ref{tab:Asymm_all} shows the values of the asymmetry parameters for each source for the five brightest lines: \hcop, \hcn, \hnc, \cch, and \htcop.  The \hcop\, asymmetry parameter $A_{HCO^+}$ is typically positive, with a mean asymmetry for the sample of $\bar{A}_{HCO^+} = 0.69 \pm 0.01$. This positive value indicates a preponderance of blue asymmetries in the \hcop\, line for this sample, confirming the selection method and expected in collapsing clumps for a line with large optical depth.  Large optical depths in the \hcop\, in this sample are evident by small ratios in the \hcopnt/\htcopnt line ratios and the fact that \htcop\, often peaks at local minima of the \hcop\, line profiles. 

The average value of $\bar{A}_{HCO^+} = 0.69$ is smaller than the original cutoff value of $A_{HCO^+} = 0.8$ used to select these sources.  However, the asymmetry parameters derived from the much noisier MALT90 data are occasionally unreliable. For example, the source AGAL305.589+00.462, originally thought to have a large blue asymmetry based on the MALT90 data, is instead found with the better sensitivity here to have a negative value for $A_{HCO^+}$, indicating a red rather than a blue asymmetry.  Nevertheless, the values of $A_{HCO^+}$ for this sample are in general remarkably large, with a mean value  
over 8 times larger than the value of $\bar{A}_{HCO^+} = 0.083\pm0.010$ for the entire sample of clumps observed in MALT90 \citep{Jackson2019}.  Moreover, 92\% of the 24 clumps have significantly positive asymmetry parameters $A_{HCO^+}$ greater than 3 times the uncertainty on the individual measurement.  This study therefore confirms that this sample of candidate collapsing clumps indeed has very large blue asymmetries in their \hcop\, line profiles, which can be interpreted as arising from large collapse motions.

We searched for differences in \ahcop\, as a function of the clumps' evolutionary stage, but find no significant trends, confirming the result of \cite{Pillai2023}.  The mean \hcopnt\, asymmetry parameter is slightly larger for the Protostar category $\bar{A}_{HCO^+} = 0.76\pm0.01$  than the Quiescent category $\bar{A}_{HCO^+} = 0.69\pm0.004$, but not significantly so.  With such small sample sizes in each category, it is difficult to detect such effects, should they exist, with high confidence. 

For collapsing clumps the asymmetry parameters are expected to decrease as the optical depth decreases.  
This decrease in the asymmetry parameter results from the fact that optically thinner species trace a larger volume of the clump, so that the flux differences between the red and blue parts of the profile become smaller.  Indeed, for optically thin emission for a spherically symmetric cloud, the line profiles should be perfectly symmetric, since emission from every molecule can be detected in both the blue- and red-shifted parts of the collapsing clump, and by symmetry, the blue- and red-shifted velocities exactly mirror each other on the front and back sides. Thus, for spherically symmetric collapse, the asymmetry parameter should vanish for optically thin lines.

Table \ref{tab:Asymm_all} shows asymmetry parameters for four other lines: \hcn, \hnc, \cch\, and \htcop.  Since \hcn\, and \hnc\, in dense clumps are both typically optically thick, but less optically thick than \hcop\, \citep{Cunningham2003, Jones2003, Sanhueza2012, Hoq2013, JimenezDonaire2017}, for collapsing clumps the asymmetry parameters for \hcn, and \hnc\, are expected to be positive, but smaller than that for \hcop.  Since \cch, and \htcop\, are expected to be much optically thinner than \hcop, their asymmetries are expected to be smaller still and approach zero in the optically thin limit.

\begin{figure}
    \centering
    \includegraphics[width=0.49\linewidth,keepaspectratio]{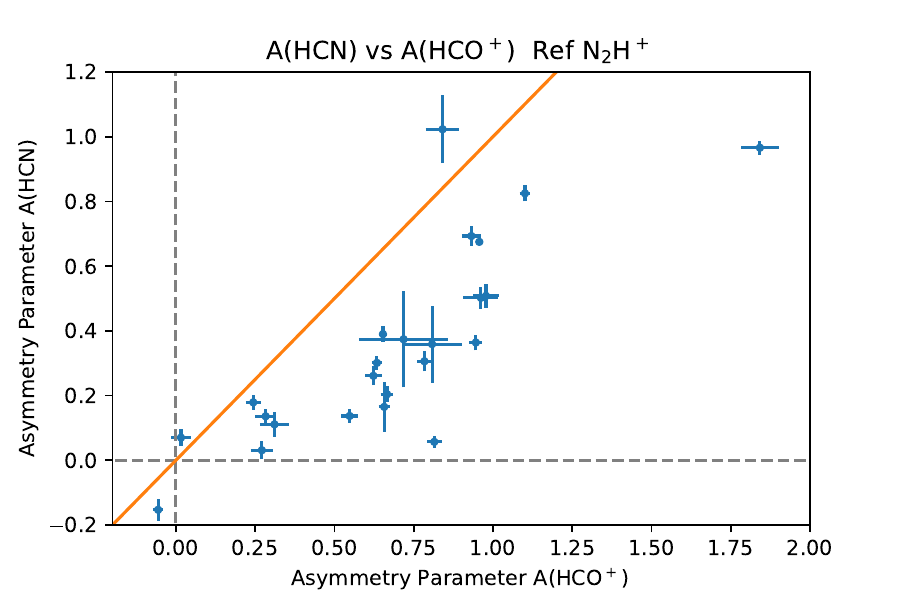} 
    \includegraphics[width=0.49\linewidth,keepaspectratio]{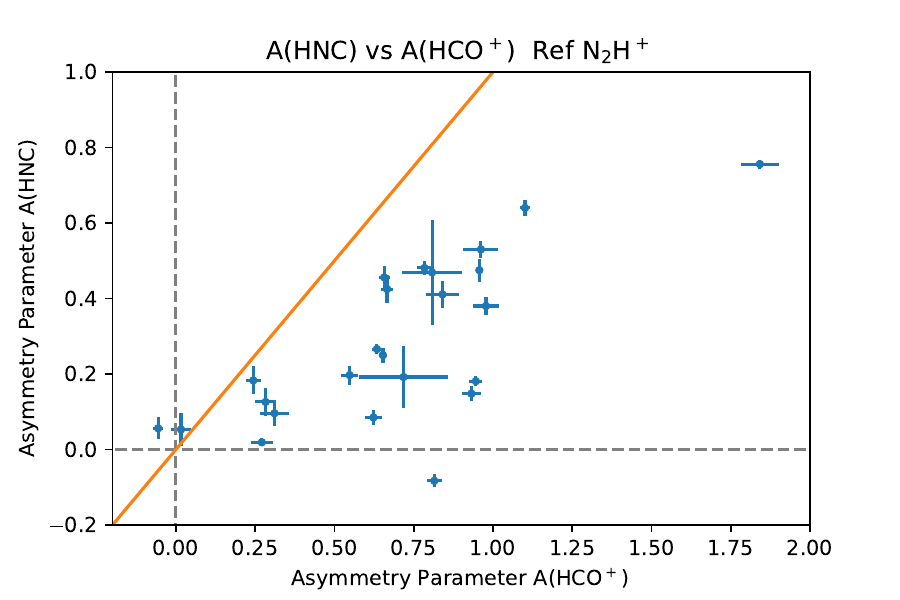} 
    \par\medskip
    \includegraphics[width=0.49\linewidth,keepaspectratio]{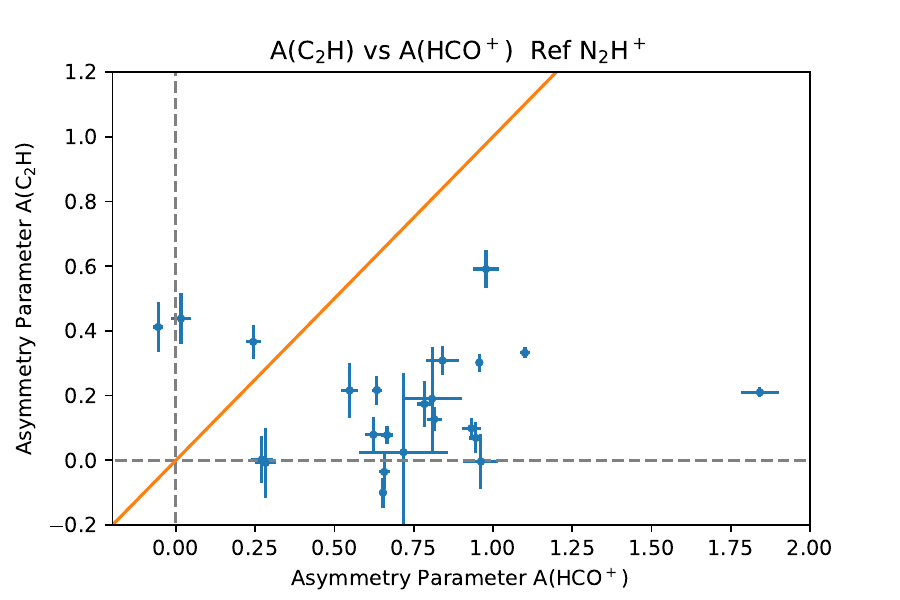}
    \includegraphics[width=0.49\linewidth,keepaspectratio]{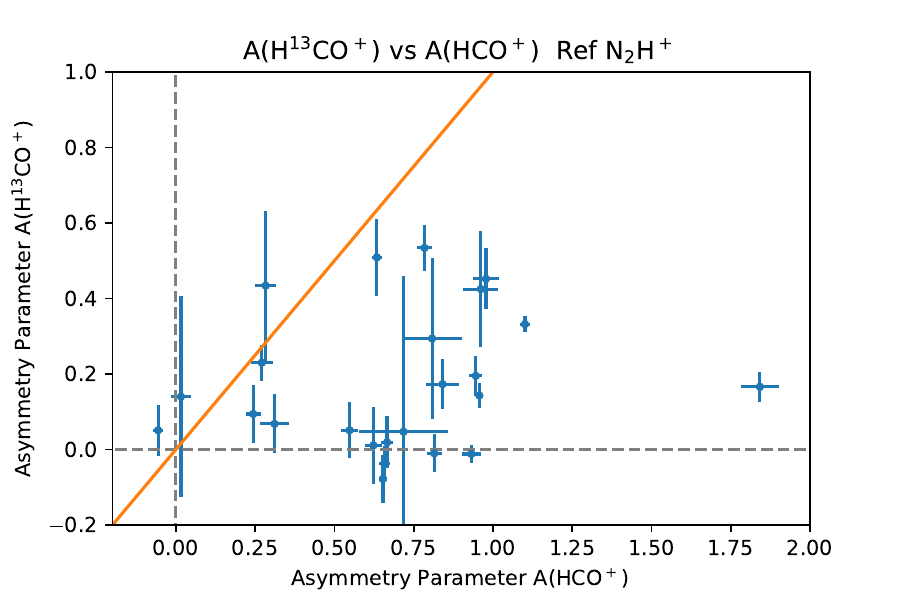}
    \caption{(Top Left) The asymmetry parameter A for \hcn\, vs. the asymmetry parameter $A$ for \hcop, using \nthp\, as the reference line. (Top right) The asymmetry parameter A for \hnc\, vs. the asymmetry parameter $A$ for \hcop, using \nthp\, as the reference line. (Bottom left) The asymmetry parameter A for \cch, vs. the asymmetry parameter $A$ for \hcop, using \nthp\, as the reference line. (Bottom right) The asymmetry parameter A for \htcop\, vs. the asymmetry parameter $A$ for \hcop, using \nthp\, as the reference line.  For all plots, the orange diagonal line represents equal values of $A$ for both lines.  The dotted gray lines indicate values of $A=0$ for both lines.  The plotted uncertainties are $\sigma_{tot}$, as described in the text. All of these lines typically have positive values for $A$.  The largest values of $A$ occur for \hcopnt.  As the optical depth decreases, the values of $A$ also decrease.} 
    \label{fig:Asymm-all}
\end{figure}

Figure \ref{fig:Asymm-all} compares the asymmetry parameters for \hcop\, with those of \hcn, \hnc, \cch\, and \htcop.  The vertical and horizontal dashed lines indicate $A$ = 0.  Clearly, the values of $A$ for all of these lines tend to be positive, and cluster in the upper right quadrant of the plot, to the right of the diagonal line indicating equal values of $A$. 
This location for most of the data points on the plot indicates that the $A$ values for \hcop\, are typically larger than those of the \hcn, \hnc, \cch, and \htcop.  Moreover, as expected, the mean asymmetry parameter decreases with decreasing optical depth.  \hcn\, and \hnc, the lines expected to have the highest optical depth after \hcop, typically have positive asymmetry parameters, but smaller than those of \hcop.  The mean asymmetry parameter for \hcn\, is $\bar{A}_{HCN} = 0.35\pm0.01$ about 2.0 times smaller than that for \hcop, and for \hnc, $\bar{A}_{HNC} = 0.28\pm0.01$, about 2.5 smaller than that for \hcop. \footnote{The hyperfine structure of \hnc\, leads to an additional uncertainty in deriving the asymmetry parameter, as the different hyperfine components can saturate at different column densities and skew the symmetry of the line profile.  However, the \hnc\, hyperfine splitting is small ($\Delta V = +0.20$ and $-0.27$ \kms), and even for optical depths as large as $\tau_0 = 10$, for the typical observed linewidths in this sample the effect is slightly to decrease $A$ by small amounts, less than half of the typical errors.} These smaller asymmetry parameters for \hnc\, and \hcn\, are consistent with collapse motions causing the asymmetry, but from lines with smaller, but still somewhat optically thick, optical depths.  \cch\, should have smaller optical depths still, and its mean asymmetry parameter, $\bar{A}_{C_2H} = 0.15\pm0.02$ is even smaller, a factor of 4.6 smaller than that of \hcop.  

Finally, Table \ref{tab:Asymm_all} presents asymmetry parameters for \htcop.  Because the abundance ratio for \htcopnt/\hcopnt\ is small \citep[$\sim 0.017$ to 0.050;][]{Stark1981}, \htcop\, is usually thought to be optically thin.  Hence, if \htcop\, is indeed optically thin, in a collapsing clump its asymmetry parameter would be zero. In fact, for the MALT90 sample, the mean asymmetry parameter for \htcop\, was found to be zero, $\bar{A}_{H^{13}CO^+}=0.000\pm0.014$ \citep{Jackson2019}.  This new sample of candidate collapsing clumps, however, might be expected to have very large optical depths \citep[which has been seen in other MALT90 clumps;][]{Stephens2015}, and thus, even \htcop\, may be slightly optically thick and therefore exhibit small, positive asymmetry parameters in some sources.  The mean value for the asymmetry parameter is $\bar{A}_{H^{13}CO^+}=0.18 \pm 0.03$, and the mean is positive at the $\sim 6\sigma$ level.  

Figure \ref{fig:Asymm-all} plots \ahcop\, vs. \ahtcop\, in the lower right panel.
This plot indicates that the \htcop\, line may be marginally optically thick for many sources in this sample, as indicated by positive values for \ahtcop.  Nevertheless, the smaller values of \ahtcop\, compared with \ahcop\, $A_{HCN}$ and \ahnc\, conforms to expectations for an optically thinner tracer of collapsing clumps.

Since \htcop\, is often thought to be optically thin, it is sometimes used as an optically thin reference line to establish the systemic velocity. Table \ref{tab:A_hcop_compare_ref}  compares \ahcop\, values derived using \nthp\, and \htcop\, as the reference lines. The agreement between the values is usually quite good, but the overall mean for \ahcop\, is slightly larger when \nthp\, is used as the reference line (0.69 compared with 0.60).  This difference can be explained if, as suggested above by the positive values for \ahtcop, the \htcop\, line is sometimes optically thick for sources in this sample.  Unusually large column densities leading to high \htcop\, optical depths are certainly plausible, as this sample may well represent clumps with extreme properties.  Another possibility is that blue outflow wings or a fainter second velocity component slightly blueward of the main component may lead to blue asymmetries in certain sources.  Such blue wings, however, are only obvious in two sources:  AGAL332.276$-$00.071 and AGAL337.927$-$00.432.

All of the asymmetry parameters conform to expectations of collapsing clumps.  The most optically thick \hcop\, line has the largest positive $A$ values, followed by the less optically thick \hcn\, and \hnc\, lines with smaller, yet still positive $A$ values, and finally, with the smallest values of the asymmetry parameters for the optically thinnest (but possibly slightly optically thick) \cch\ and \htcop\, lines.

Table \ref{tab:Asymm_mean} shows the mean values of the asymmetry parameters for all 11 lines analyzed for this study.  The relatively bright \nthp\, line has a mean asymmetry parameter, $\bar{A}_{N_2H^+} = -0.04\pm0.01$, which is slightly negative, although it should be zero for an optically thin tracer.  \nthp\, however, has hyperfine structure, and emission from the $F_1 = 1 \to 1$ hyperfine line with a velocity offset of 5.74 \kms\, from the main $F_1 = 2 \to 1$ hyperfine line can contaminate the asymmetry calculation with excess red flux.  The remaining five lines, \hncofz, \hctn, \sio, \tcs, and \hntc, are typically faint, and the lack of signal often leads to large errors in the asymmetry parameters.  To account for this effect, Table \ref{tab:Asymm_mean} gives two sets of values, one set where all values, including the noisy ones, are included, and a second set where sources with faint signals (those with $I = I_{blue} + I_{red} < 3\sigma$) are excluded.  For these five lines, if all values are included, all of the mean asymmetry parameters are consistent with zero, as expected for optically thin emission from a collapsing clump.  If sources with faint signals are excluded the means are typically consistent within 3$\sigma$ of zero, with the exception of \hncofz, which has a positive value of $\bar{A}$ at the 4$\sigma$ level, perhaps indicating somewhat optically thick emission.

\subsection{Hyperfine Intensity Ratios}

The intensity ratios of the \nthp\, and \hcn\, nuclear quadrupole hyperfine lines provide additional information about the emitting gas.  In this section, we analyze these hyperfine line ratios and show that the observed \nthp\, hyperfine lines are consistent with optically thin emission in local thermodynamic equilibrium (LTE), whereas the \hcn\, hyperfine lines are not in LTE, and likely heavily self-absorbed at redshifted velocities.  This self-absorption is modeledin Section 4.4 and found to be consistent with a collapsing clump.

The electric quadrupole interaction in molecules containing nuclei with non-zero spins splits the rotational energy states into hyperfine states.  These hyperfine energy levels are labeled by the angular momentum $\vec{F}$, which is the sum of the the nuclear spin $\vec{I}$ and the rotational angular momentum $\vec{J}$, $\vec{F} = \vec{I} + \vec{J}$. Since the nitrogen nucleus has a spin $I=1$, the rotational levels $J$ for linear nitrogen-bearing molecules are each split into three hyperfine states, with $F = J-1, J,$ and $J+1$, except for $J=0$, which remains unsplit. (Actually, for \nthpnt\, the relevant quantum number is $F_1$, which represents the coupling between the rotational angular momentum $\vec{J}$ and the spin of the outermost nitrogen nucleus $\vec{I_1}$.)  The $J=0$ rotational energy level has a single, unsplit energy level with $F=1$. The $J = 1$ rotational transition, however, is split into three hyperfine levels, $F=2$, $F=1$, and $F=0$.  The statistical weight $g_F$ of each of these $F$ states is $g_F = 2F+1$.  If the hyperfine lines are in LTE, their optical depths are proportional to their statistical weights.

\begin{equation}{\tau(F=2 \to 1):\tau(F=1 \to 1):\tau(F=0 \to 1) = 5:3:1}\end{equation}

Hereafter, we will write $\tau_{21} = \tau(F=2 \to 1)$, $\tau_{11} = \tau(F = 1 \to 1)$, and $\tau_{01} = \tau(F = 0 \to 1)$.

The equation of radiative transfer
\begin{equation}{\Delta T_B = (T_{ex} -T_{bg}) (1-e^{-\tau})}\end{equation}
then gives, for the LTE (equal excitation temperatures for all transitions) hyperfine line ratios
\begin{equation}{R_{12} \equiv \frac{\Delta T_B(F=1 \to 1)}{\Delta T_B(F=2 \to 1)} = \frac{1-e^{-\tau_{11}}}{1-e^{-\tau_{21}}} = \frac{1-e^{-0.6\tau_{21}}}{1-e^{-\tau_{21}}}}\end{equation}
and
\begin{equation}{R_{02} \equiv \frac{\Delta T_B(F=0 \to 1)}{\Delta T_B(F=2 \to 1)} = \frac{1-e^{-\tau_{01}}}{1-e^{-\tau_{21}}} = \frac{1-e^{-0.2\tau_{21}}}{1-e^{-\tau_{21}}}}\end{equation}

For small optical depths, $R_{12} = 0.6$ and $R_{02}=0.2$, while for large optical depths, $R_{12} = R_{02} = 1$.

\begin{figure}
    \centering
    \includegraphics[width=\linewidth,height=0.9\textheight,keepaspectratio]{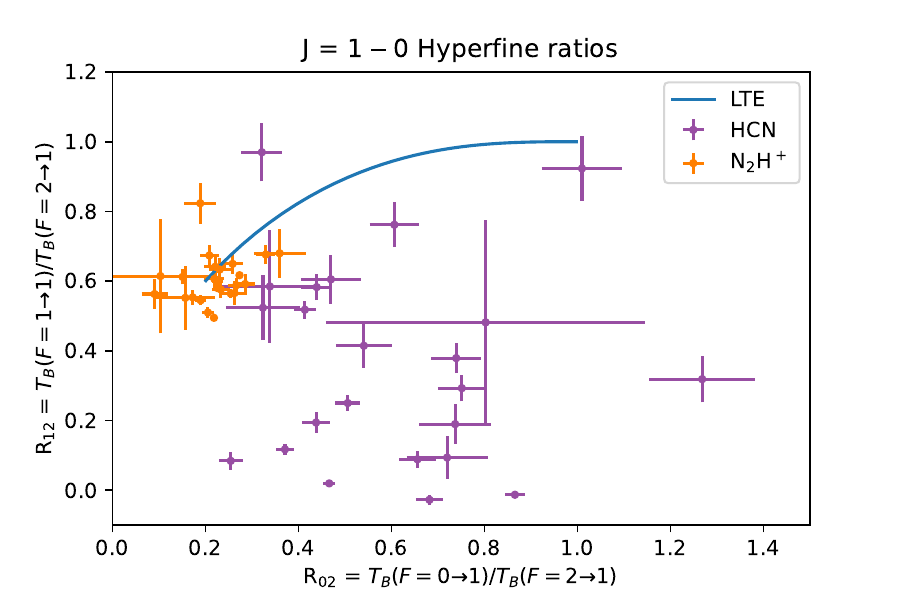} 
    \caption{ The observed hyperfine intensity ratios and their uncertainties for \hcn\, and \nthp. The solid line represents the locus of LTE ratios for various optical depths, with optically thin ratios (0.2,0.6) at the lower left, and optically thick ratios (1.0, 1.0) at the upper right.}
    \label{fig:HCN-nthp-hf-ratios}
\end{figure}

Figure \ref{fig:HCN-nthp-hf-ratios} plots the observed \nthp\, and \hcn\, hyperfine ratios observed towards the clumps in our sample.  Also included in Fig. \ref{fig:HCN-nthp-hf-ratios} is the LTE ratio from Equations 3 and 4, shown in blue. The lower left end of this curve is the optically thin limit $(R_{02},R_{12}) = (0.2, 0.6)$, and the upper right end is the optically thick limit $(R_{02},R_{12}) = (1.0, 1.0)$.

The \nthp\, hyperfine ratios, in orange, cluster near the lower left end of the LTE curve, consistent with optically thin \nthp\, hyperfine lines in LTE. In no case do we observe any evidence for optically thick \nthp\, hyperfine lines, such as flat-topped line profiles.  This result justifies our use of the optically thin 5:3:1 hyperfine line amplitude ratios when we determine the systemic velocities from the \nthp\, lines.

The hyperfine line ratios for \hcn, however, show a different behavior.  These ratios do not match the expectations from LTE.  Specifically, the $R_{12}$ values are much smaller than expected for every source but one.  In other words, the $F=1 \to 1$ lines are unusually faint, much fainter than their expected LTE values.  

Non-LTE intensity ratios of the \hcn\, hyperfine lines have been observed for decades \citep{Gottlieb1975, Baudry1980, Walmsley1982, Snell1983, Sandell1983, Cernicharo1984, Loughnane2012}.  Toward 
warm molecular clouds, most studies find $R_{12} < 0.6$ and $R_{02} > 0.2$.  Toward cold dark clouds, however, the satellite lines are often brighter than the main line, with $R_{12} > 1$
and $R_{02} > 1$.  The ratios found toward this sample, however, are extreme.  About half of the sources have $R_{12} \leq 0.3$, while $R_{02}$ ranges from $\approx 0.2$ to 1, a regime that to our knowledge has never been observed before.

Previous attempts to explain the \hcn\, hyperfine intensity anomalies \citep{KwokScoville1975,Guilloteau1981, GonzalezAlfonso1993,Loughnane2012,Mullins2016,Goicoechea2022} usually invoke high optical depths and hyperfine line overlaps in higher $J$ lines to produce non-LTE excitatision.  Alternatively, for the dark clouds, \cite{Walmsley1982} suggest that self-absorption by an optically thick, foreground, colder cloud is likely.

Because the clumps in this sample show evidence for large blue asymmetries, probably due to a cold, redshifted, foreground component from a collapsing clump's exterior and a warm, blueshifted, background component from the collapsing clump's interior, the unusual \hcn\, hyperfine intensities observed here may well result from self-absorption, as suggested by \cite{Walmsley1982} for cold dark clouds.  In this case, though, the self-absorption is unusual and extreme.  For many of the observed \hcn\, spectra, the $F=1 \to 1$ hyperfine line is almost entirely suppressed.  Since this $F = 1 \to 1$ hyperfine line is redshifted by 4.8 \kms\, with respect to the main $F = 2 \to 1$ hyperfine line, we speculate that the reason for its low intensity is self-absorption by the main $F = 2 \to 1$ line of redshifted, foreground, optically thick, low excitation temperature gas in the exterior of the collapsing clump. In Section 4.4 below, a simple two-cloud model demonstrates the plausibility of this hypothesis.

\subsection{Infall Velocity}

Interpreting the blue asymmetric profiles as infall requires that the outer portions of the clump be sufficiently redshifted with respect to the clump interior to produce the observed asymmetries.
At the same time, the size scale over which the collapse occurs must be large enough to maintain an excitation temperature gradient, for example, through interior heating by star formation, yet small enough to be contained within a single dense clump, so that the clump's exterior has a smaller excitation temperature than the clump's interior. 

To determine whether the speeds and size scales are reasonable, we estimate the free-fall velocity:
$v = \sqrt{(2GM/r)} = 0.093$ \kms\, $\sqrt{[M(M_\odot)/r(pc)]}$.  For the average mass of clumps this sample, $M = 4500$ \Msun, and for a radius of 1 pc, the free fall velocity is 6.2 \kms.  Thus, for a $\sim$ 1 pc size scale comparable to that of dense molecular clumps, free-fall can produce sufficiently redshifted velocities to account for the observed self-absorption.  Infall speeds and size scales that match those required to explain the observed profiles as due to collapse are therefore reasonable.

Whether clumps collapse ``quickly'' at time scales comparable to the free-fall time \citep{Sakai2025, Morii2025} or ``slowly'' at time scales several times the free-fall time \citep{KrumholzTan2007, Fuller2005} remains an open question.  

Toward sources with a strong continuum source, we sometimes detect an inverse P Cygni profile in both the \hcn\, and \hcop\, lines (for example: AGAL333.604$-$00.212, AGAL337.891$-$00.491, AGAL337.916$-$00.471, and AGAL337.934$-$00.507).  For these sources the presence of a redshifted absorbing gas with a velocity offset of a few \kms\, is unambiguous.

For the remaining sources, we have attempted to estimate the infall velocities  by fitting the \hcopnt\, line profiles to the ``Hill5'' model of \cite{deVriesMyers2005}.  Unfortunately, applying Hill5 fits to the data was unsuccessful, as the model fits to the data were poor. The line profiles produced by the Hill5 model tend to be double-peaked, but our line profiles rarely have this shape.  Because the self-absorption toward these sources is extreme, we usually detect little or no emission from the self-absorbed gas on the red side of the profile.  Typically, the only surviving emission is on the blue side of the profile.  The Hill5 model tried to model this emission but failed to capture the self-absorbed component, even when we forced the systemic velocity to match the \nthp\, velocities.   The Hill5 model of spherical collapse with a uniform infall speed and a simple linear increase of excitation temperature as a function of radius may well fail to provide an adequate model of the clumps observed here.  

Although attempts to model the collapse with a Hill5 model were unsuccessful, we can, however, successfully model the line profiles arising from a collapse scenario with a simpler model than Hill5.  In the next section, we show that a simple two-cloud model supports the idea that the clumps in this sample are collapsing, and that their infall velocities are indeed large, $V_{inf} \geq 2.4$ \kms.

\subsection{Simple Two-Cloud Model}

We hypothesize that a collapse scenario in which a colder, redshifted, foreground component that absorbs a warmer, blueshifted, background component can explain the blue asymmetric line profiles, the observed asymmetry parameters, and the suppressed $F= 1 \to 1$ \hcn\, hyperfine line intensities.  To test the plausibility of this hypothesis, we have made a simple ``two-cloud'' model of a collapsing clump as consisting of just two velocity components, a warm background component representing the clump interior (Cloud 1), and a cold foreground component representing the clump exterior (Cloud 2).  We then examine the radiative transfer of line emission through this two-cloud system.  Each cloud is parameterized by an excitation temperature $T_{ex}$, an LSR velocity $V$, a velocity dispersion $\sigma$, and an optical depth at the velocity of the line center $\tau_0$. By varying the parameters of both clouds, we can fit the resulting spectra to determine whether the model can match the observed line profiles, asymmetry parameters, and \hcn\, hyperfine intensity ratios.  

For a very optically thick line, a two-cloud model is apt because in a collapsing cloud, the $\tau=1$ surface on the redshifted side of the line profile arises in the cold exterior of the clump, but on the blueshifted side it arises in the warm interior of the clump.  The rest of the cloud, however, is mostly obscured.  Hence, we focus on modeling the two most optically thick line tracers, \hcn\, and \hcop.  This simple model is meant simply to demonstrate the effects of self-absorption and the plausibility of a collapse interpretation.

First, we generate the expected line profiles for the \hcn\, line in the two-cloud model.
Both the foreground and background \hcn\, hyperfine lines are modeled as Gaussians with the appropriate velocity spacing (-7.063 \kms, 0.000 \kms, and 4.843 \kms) with respect to the main hyperfine line and optical depth ratios (1:5:3) for the $F = 0 \to 1$, $F = 2 \to 1$, and $F = 1 \to 1$ hyperfine lines, respectively.   We assume a continuum background with brightness temperature $T_{bg}$ shining through both clouds.
The emerging line brightness, with the continuum baseline subtracted, then consists of the sum of three terms:
a term representing the background continuum and its attenuation by both clouds, as well as the continuum baseline subtraction:
\begin{equation}
{T_C = T_{bg}[e^{-\{\tau_1(V)+\tau_2(V)\}} - 1]}
\end{equation}
a term representing the line self-emission from Cloud 1, attenuated by Cloud 2:
\begin{equation}
{T_{L1} = T_{ex,1}[1 - e^{-\tau_1(V)}]e^{-\tau_2(V)}}
\end{equation}
and a term representing the line self-emission from Cloud 2:
\begin{equation}
{T_{L2} = T_{ex,2}[1 - e^{-\tau_2(V)}]}.
\end{equation}

The total brightness temperature with the continuum baseline subtracted is then given by
\begin{equation}
{\Delta T_B =T_C +T_{L1} + T_{L2}}. 
\end{equation}

\begin{figure}
    \centering
    \includegraphics[width=\linewidth,height=0.9\textheight,keepaspectratio]{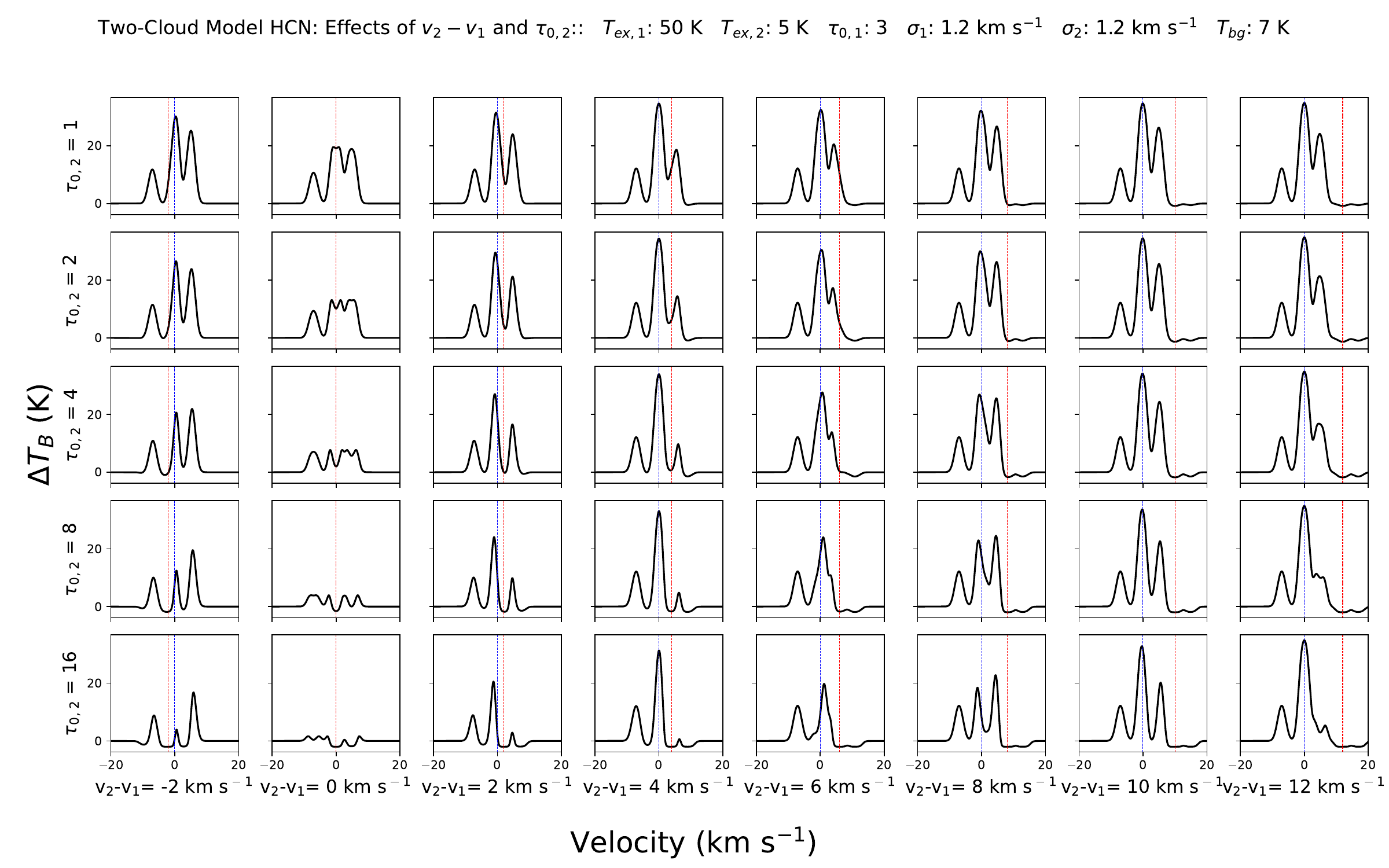} 
    \caption{Simulated line profiles of \hcn\, in a simple two-cloud model.  The blue vertical lines represent the velocity of the warm background cloud and the red vertical lines the velocity of the cold foreground cloud.  The profiles show a grid of various foreground cloud optical depths and velocity offsets, with foreground cloud optical depth ranging from 1 to 16, varying by factors of 2 (y direction) and offset velocities $V_2 - V_1$ ranging from $-2$ to 12 \kms, in steps of 2 \kms\, (x-direction).  Parameters for this particular model are shown in the title.  The rightmost $F = 1 \to 1$ hyperfine line in the background cloud is suppressed by self-absorption by the main $F=2\to 1$ hyperfine line in the foreground cloud for large foreground optical depths and velocity offsets of $\sim 2$ to 8 \kms, the lower middle panels in the plot.}
    \label{fig:two_cloud_hcn}
\end{figure}

Representative theoretical \hcnnt\, line profiles for this simple two-cloud model for various optical depths and relative velocities are shown in Fig. \ref{fig:two_cloud_hcn}.  This Figure shows the self-absorbed line profiles on a grid of relative velocities $V_2-V_1$ on the x-axis and foreground cloud optical depth values $\tau_{0,2}$ on the y-axis. The two-cloud model often generates line profiles similar to those observed.  Specifically, for large optical depths, and velocity offsets near the $F = 1 \to 1$ hyperfine spacing of 4.7 \kms\, (the lower-middle panels), the faint $F = 1 \to 1$ hyperfine line intensities of the modeled line profiles match those in the typical observed line profiles.

\begin{figure}
    \centering
    \includegraphics[width=\linewidth,height=0.9\textheight,keepaspectratio]{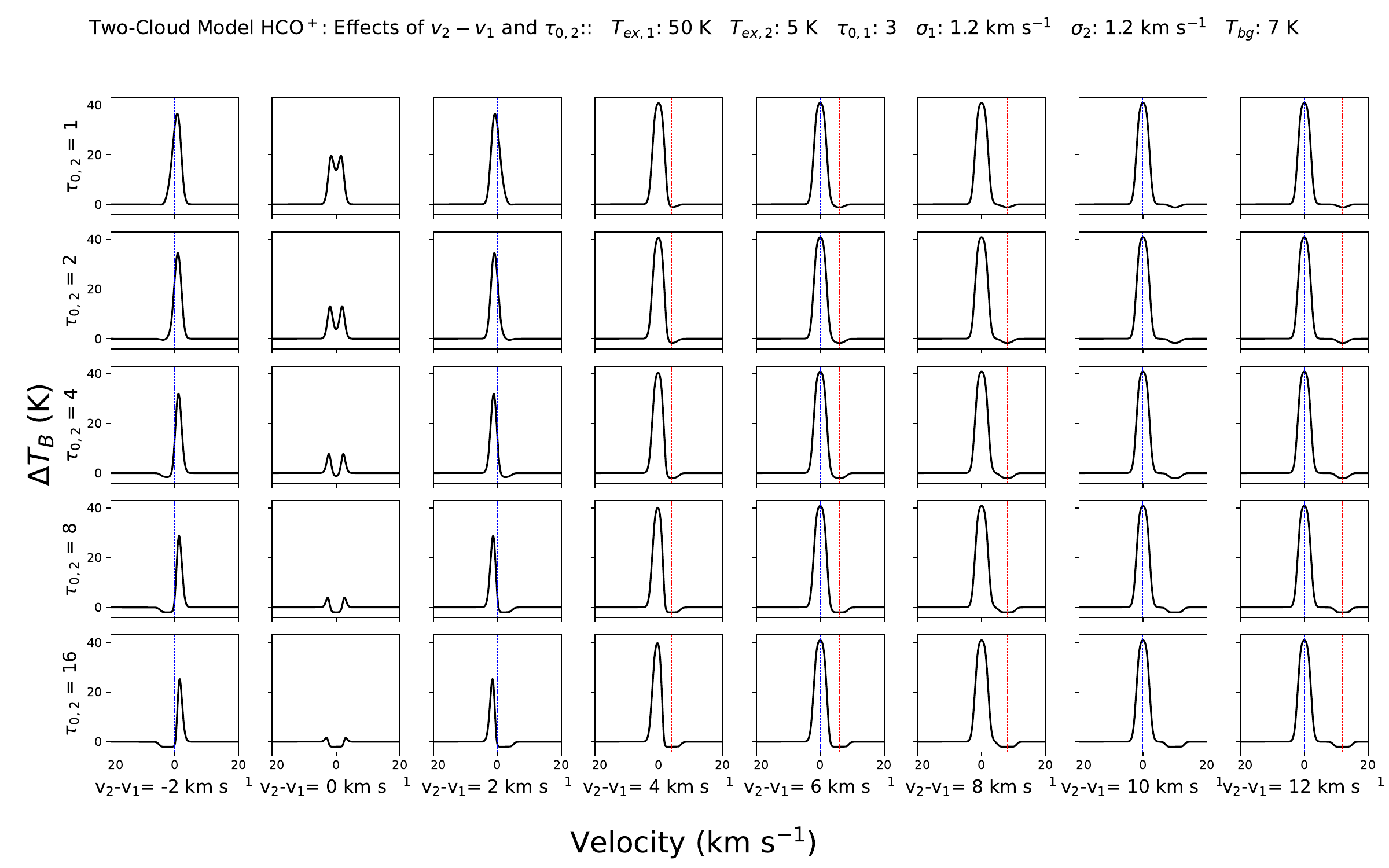} 
    \caption{Simulated line profiles of \hcop\, in a simple two-cloud model.  The blue vertical lines represent the velocity of the warm background cloud and the red vertical lines the velocity of the cold foreground cloud.  The profiles show a grid of various foreground cloud optical depths and velocity offsets, with foreground cloud optical depth ranging from 1 to 16, varying by factors of 2 (y direction) and offset velocities $V_2 - V_1$ ranging from $-2$ to 12 \kms, in steps of 2 \kms\, (x direction).  Parameters for this particular model are shown in the title. The line profiles show skewed, blue asymmetries for large foreground optical depths and velocity offsets of $\sim 2$ to 8 \kms, the lower middle panels in the plot.}
    \label{fig:two_cloud_hcop}
\end{figure}

Next, we model the \hcop\, line, following the same procedure with a single Gaussian profile for the foreground and background clouds.  Representative theoretical \hcop\, line profiles are shown in Fig. \ref{fig:two_cloud_hcop}. Again, the two-cloud model can generate line profiles similar to those observed, especially the inverse P Cygni line profiles with redshifted absorption against the continuum, again in the lower-middle panels. 

The two-cloud model can be extended to fit both the \hcn\, and \hcop\, lines simultaneously. 
For the simultaneous fits, we assume that both lines arise from two clouds with the same parameters $(V, \sigma, T_{ex})$ for each line.  
For each source, we also fit the ratio of the optical depths of the two lines: $\tau_{0}({\rm HCO^+)}/\tau_{0}({\rm HCN})$.  
Finally, we assume a common beam filling factor $\phi$ for both lines and a main beam efficiency $\eta = 0.49$ for the Mopra telescope \citep{Ladd2005}.  
We first find the best fit to the \hcn\, profile, and then use these parameters as the initial guess for the simultaneous fits.  
If absorption against the continuum was absent at $\Delta T_A^{*}$ levels $< 0.1$ K in both spectra, we restricted $T_{bg}$ to values between 2.5 and 3.0 K.  
If absorption against the continuum was detected at $\Delta T_A^{*}$ levels $> 0.1$ K, $T_{bg}$ values up to 100 K were allowed.

\begin{figure}
    \centering
    \includegraphics[width=\linewidth,height=0.9\textheight,keepaspectratio]{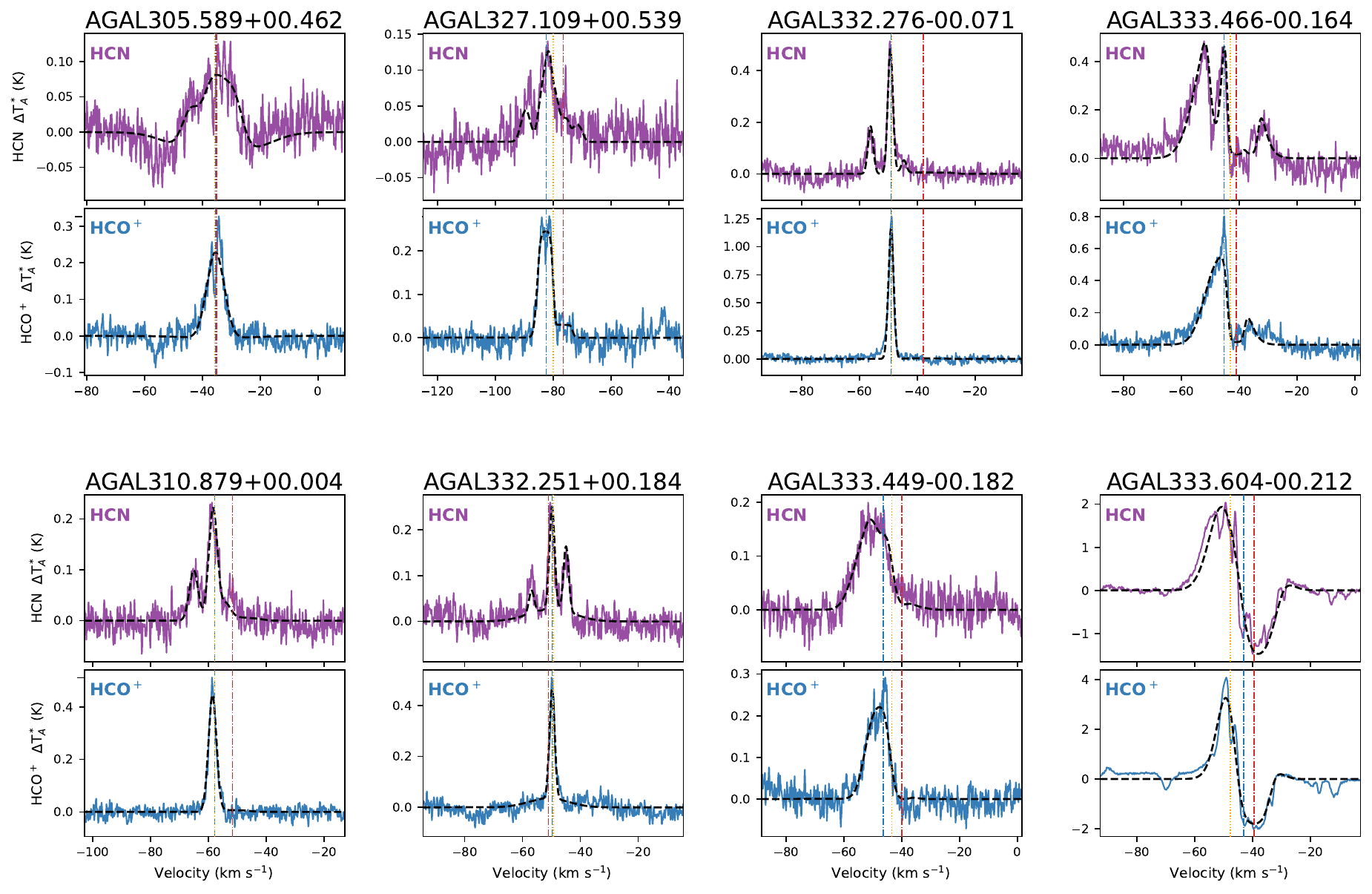} 
    \caption{Simultaneous fits of the two-cloud model to both the \hcn\, and \hcop\, lines.  The fit parameters are given in Table \ref{tab:two-cloud}.  The solid colored lines are the data and the dashed black lines the best fit.  The dotted vertical lines represent the fitted velocity of the background cloud, $V_1$ (blue line), the fitted velocity of the foreground cloud, $V_2$ (red), and the systemic velocity from the \nthp\, line (orange).}
    \label{fig:two_cloud_joint_fit1}
\end{figure}

\begin{figure}
    \centering
    \includegraphics[width=\linewidth,height=0.9\textheight,keepaspectratio]{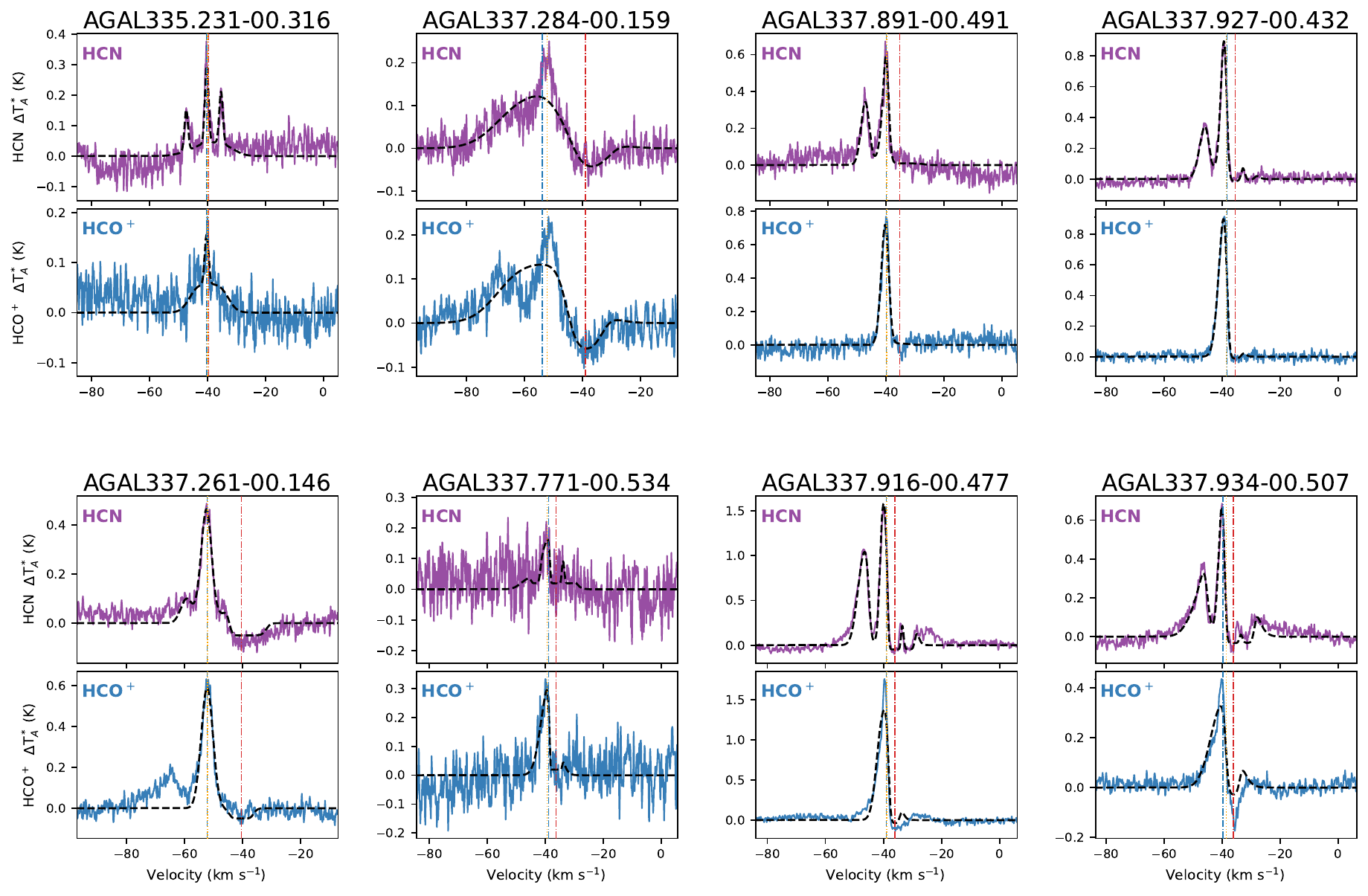} 
    \caption{Same as Fig. \ref{fig:two_cloud_joint_fit1} but for different sources. }
    \label{fig:two_cloud_joint_fit2}
\end{figure}

\begin{figure}
    \centering
    \includegraphics[width=\linewidth,height=0.9\textheight,keepaspectratio]{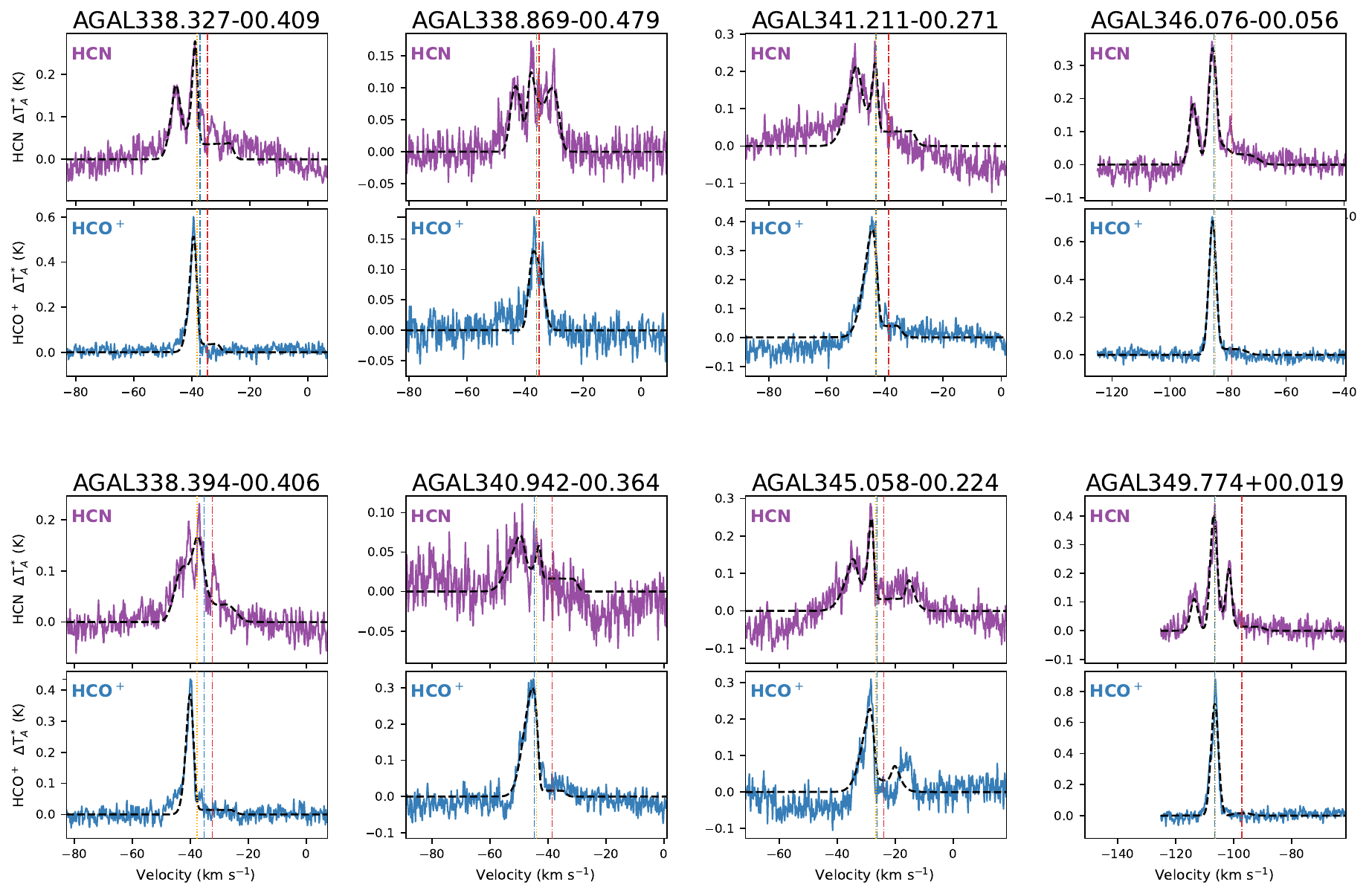} 
    \caption{Same as Fig. \ref{fig:two_cloud_joint_fit1} but for different sources. }
    \label{fig:two_cloud_joint_fit3}
\end{figure}

Table \ref{tab:two-cloud} and Figures \ref{fig:two_cloud_joint_fit1} to \ref{fig:two_cloud_joint_fit3} show the results of the simultaneous two-cloud model best fits.  In general, the joint two-cloud model provides satisfactory fits to the line profiles for both lines.   

\begin{figure}
    \centering
    \includegraphics[width=\linewidth,height=0.9\textheight,keepaspectratio]{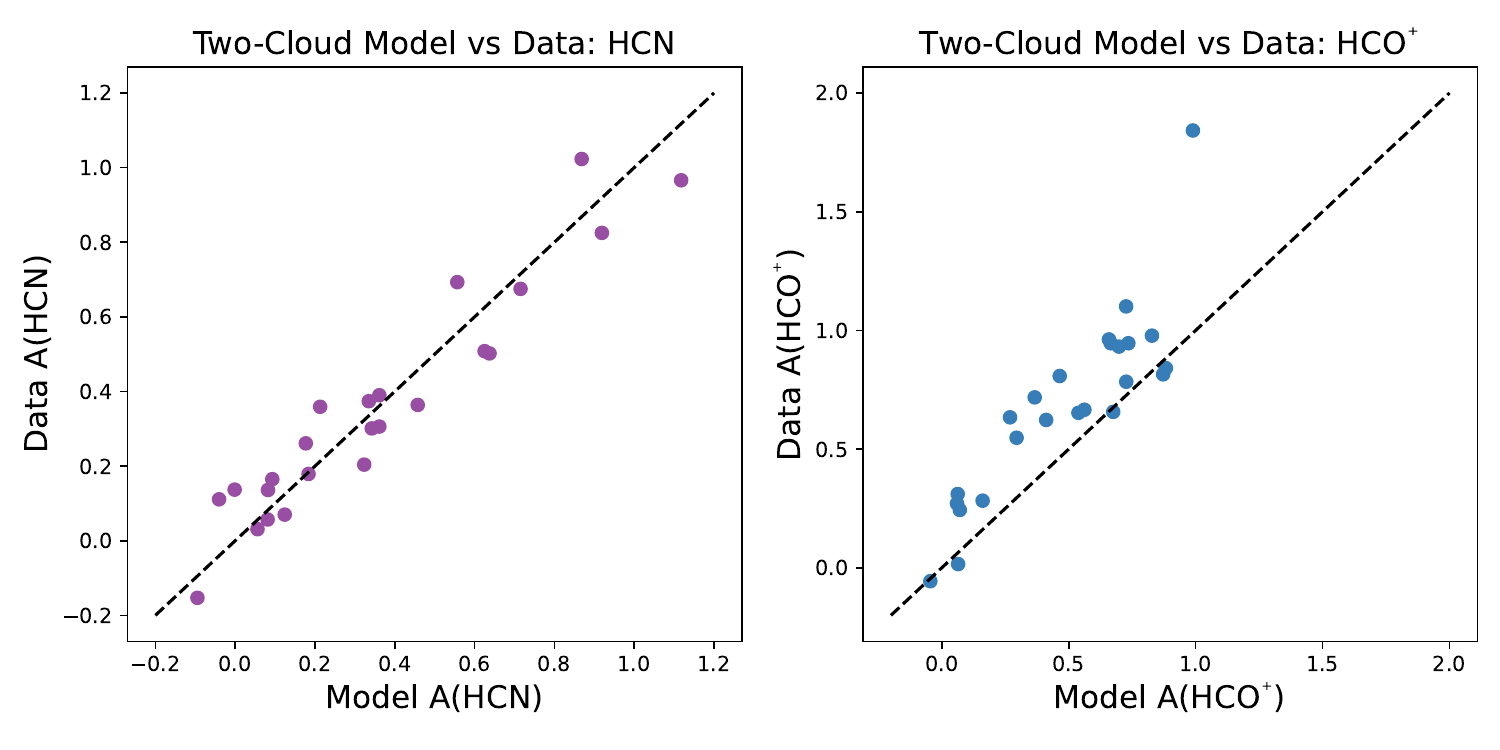} 
    \caption{The measured vs. the modeled asymmetry parameters from the joint \hcn\, and \hcop\, two-cloud fits to the data.   }
   \label{fig:two_cloud_asymmetry} 
\end{figure}

In addition to fitting the line shapes, the two-cloud model fits also reproduce the observed asymmetry parameters reasonably well.  Table \ref{tab:two-cloud-asym} compares the observed asymmetry parameters to those derived from the best fit two-cloud models.  Figure \ref{fig:two_cloud_asymmetry} plots the measured values of $A$ vs. te modeled values of $A$ for both \hcnnt\, and \hcopnt.  For \hcnnt, the model predicts the measured value of $A$ very well.  The dispersion $\Delta A(HCN)$, the modeled value of $A(HCN)$ minus the measured value, has a mean of $ 0.00 \pm 0.02$.  For \hcopnt, the model tends slightly to underestimate the asymmetry parameter, with a mean $\Delta A(HCO^+)$ of $0.20 \pm 0.04$.  For both mean values, the dispersion is comparable to the errors on the measured values.

In addition to providing satisfactory fits to the line profiles and predictions for the asymmetry parameter, the two-cloud model also predicts the suppression of the $F = 1 \to 1$ hyperfine line due to absorption of the background right $F = 1 \to 1$ hyperfine line by the main $F = 2 \to 1$ hyperfine line of the cold foreground cloud.  We ran a series of two-cloud models with various parameter values, and then fit the resulting two-component \hcn\, profile assuming a single velocity component with three hyperfine Gaussians at the same spacing as the \hcn\, hyperfine lines, but with independent, unconstrained amplitudes.  

\begin{figure}
    \centering
    \includegraphics[width=\linewidth,height=0.9\textheight,keepaspectratio]{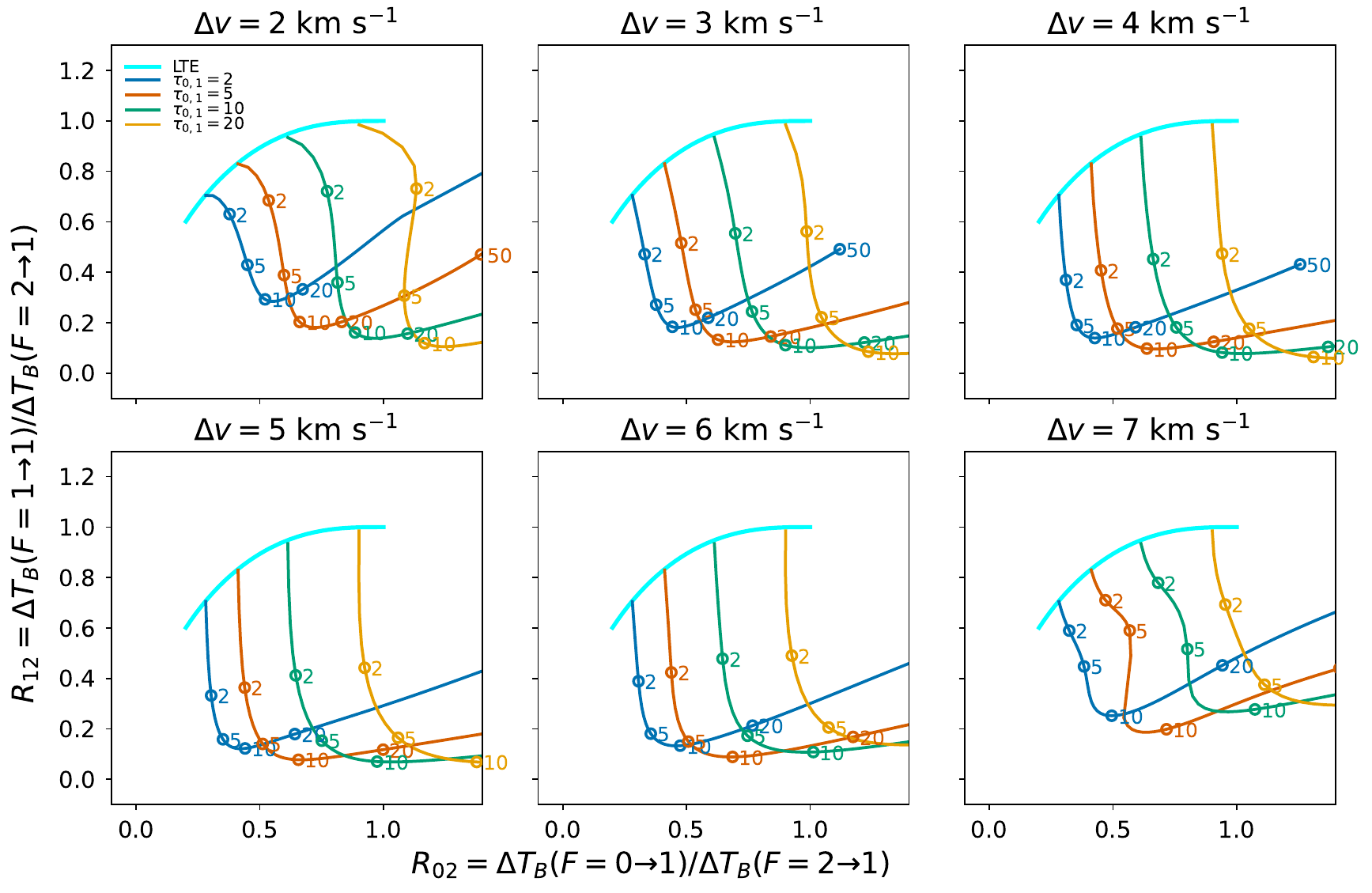} 
    \caption{The fitted hyperfine intensity ratios for \hcn\, for a two cloud model, with a foreground cloud and background cloud with different parameters: excitation temperature $T_{ex}$, velocity dispersion $\sigma$, LSR velocity $V$, and optical depth $\tau$.  The models shown here use $\sigma_1$ = $\sigma_2$ = 1.5 \kms, $T_{ex,1} = 50$ K, and $T_{ex,2} = 5$ K. The fitted values assume a single HCN line with hyperfine structure, with the fit applied to the emergent two-cloud HCN line profile.  The amplitude of each hyperfine component is fitted independently.  The cyan line represents the locus of LTE \hcn\, hyperfine line ratios for various optical depths, with optically thin ratios (0.2,0.6) at the lower left, and optically thick ratios (1.0, 1.0) at the upper right.  The other lines represent the locus of the fitted points with varying optical depths of the foreground cloud, from $\tau_{0,2}= 0$ to 50.  Each colored line indicates a different value for the optical depth of the background cloud $\tau_{0,1}$, ranging from 2 to 50, as indicted in the legend.  Here $\tau_0$ indicates the peak optical depth at line center for a theoretical, unsplit line.  The peak optical depth of the main $F= 2-1$ hyperfine line is $5/9~ \tau_0$.  The smallest optical depth for the foreground cloud occurs at the point nearest the LTE curve, as the foreground cloud will not attenuate the background cloud when the optical depth is zero. Fiducial values of $\tau_{0,2}$ are indicated by larger points and labels of their values (2, 5, 10, 20, 50).  Each panel shows the models for different velocity differences $\Delta V = V_2 - V_1$, as indicated above each panel.  The optical depth of the background cloud is indicated in the legend.  When the optical depth of the foreground cloud becomes large enough, the main $F = 2 \to 1$ hyperfine component of the foreground cloud absorbs the $F = 1 \to 1$ component of the background cloud and suppresses the $R_{12}$ ratio.}
    \label{fig:model_hcn_hf_ratios}
\end{figure}

Figure \ref{fig:model_hcn_hf_ratios} shows some representative results.
These plots of $R_{12}$ vs. $R_{02}$ show the locus of LTE ratios from Equation (4) in cyan.  The other lines on these plots show the locus of points of the fitted hyperfine ratios, for various values of $\tau_{0,1}$, and letting $\tau_{0,2}$ vary from 0 to 50.  Here $\tau_{0,1}$ and $\tau_{0,2}$ are the optical depth at line center for a hypothetical unsplit \hcn\, line.  The optical depths of the center of each of the hyperfine lines are then 0.1111, 0.5555, and 0.3333 times $\tau_0$ for the $F = 0 \to 1, F=2 \to 1$, and $F=1 \to 1$ hyperfine lines, respectively. 
Small values of $\tau_{0,2}$ will lie close to the LTE curves because the foreground cloud is barely attenuating the background cloud.  For larger values of $\tau_{0,2}$, however, self-absorption can greatly change the line ratios from the LTE values.  In Fig. \ref{fig:model_hcn_hf_ratios}, fiducial values of $\tau_{0.2} = 2, 5, 10, 20,$ and 50 are marked with circles and labeled.

The models shown here use the following parameters.  The background continuum temperature was taken to be $T_{bg} = 2.7$ K.  For Cloud 1, the background cloud, the velocity is $V_1 = 0.0$ \kms, the dispersion $\sigma_1 = 1.5$ \kms, and the excitation temperature $T_{ex,1} = 50$ K.  For Cloud 2, the foreground cloud, the dispersion $\sigma_1 = 1.5$ \kms, and the excitation temperature $T_{ex,2} = 5$ K.  Figure \ref{fig:model_hcn_hf_ratios} shows models for values of $V_2 - V_1$ ranging from 2 to 7 \kms.  The exact choice of parameters will of course change the fitted hyperfine line ratios, but these parameters are chosen as reasonable values for dense molecular clumps.  The suppression of the background $F = 1 \to 1$ hyperfine line increases as $T_{ex,1} - T_{ex,2}$ increases, and as $\tau_{0,2}$ increases.

The two-cloud model can broadly reproduce the observed \hcn\, hyperfine intensity anomalies, specifically very low values of $R_{12}$, if certain conditions are met. 
First, the foreground cloud must be redshifted with respect to the background cloud.  Obviously, the maximum absorption will occur when the main hyperfine line of the foreground cloud exactly coincides with the velocity of the $F = 1 \to 1$ hyperfine line of the background cloud, or $V_2 - V_1 = 4.8$
\kms.  However, the $F = 1 \to 1$ hyperfine line can also be suppressed even for smaller redshifts (as low as $V_2 - V_1 \approx 2$
\kms) or larger redshifts (as high as $V_2 - V_1 \approx 7$
\kms).  Second, the foreground cloud must have a velocity dispersion large enough to cover most of the hyperfine line from the background cloud ($\sigma_2 \gtrapprox \sigma_1$).  Third, the foreground cloud must have a large optical depth $\tau_{0,2} >> 1$.  Finally, the foreground cloud must have a much smaller excitation temperature than the background cloud ($T_{ex,2} << T_{ex,1}$).

These simple models demonstrate that self-absorption can indeed suppress the flux of the \hcn\, $F = 1 \to 1$ hyperfine line to small values while retaining normal ratios of the other two hyperfine lines.  Thus, the unusually low values of $R_{12}$ seen toward our sample can be explained as the result of self-absorption in a collapsing clump.

Finally, we use the two-cloud model to support the idea that the infall velocities for clumps in this sample are typically much larger than those found in previous measurements.  
Unfortunately, many parameters determined by the two-cloud model, especially those related to the line amplitudes, are poorly constrained due to degeneracies in the fit parameters.  Indeed, estimates of the excitation temperature and the beam filling factor, and to some extent the optical depth of the background cloud are highly degenerate.  However, since the shape of the line profile is very sensitive to the relative velocity $\Delta V = (V_2 - V_1)$, the infall velocity is well constrained.  As expected for collapsing clumps, the modeled velocity offsets $\Delta V = (V_2 - V_1)$ are typically positive. For the two-cloud model,  the mean value for  $\Delta V =V_2 - V_1$ is $4.77 \pm 0.79$ \kms.  Since the infall speed is $\sim \Delta V/2$, the infall speeds for this sample are $V_{inf} \sim 2.4$ \kms.  This velocity is larger than those typically observed, $V_{inf} \sim 0.1$ to 1 \kms\,  \citep{Fuller2005, Csengeri2011, Peretto2013}, although \cite{Wyrowski2016} find three clumps with $V_{inf} = 1.6$ to 2.8 \kms.  This sample may represent clumps with among the largest known collapse velocities.  In fact, three clumps in our sample have $V_{inf} > 5$ \kms.

Although it is tempting to extend the two-cloud analysis by trying to model three or more lines simultaneously, attempts to do so were unsuccessful.  For instance, when jointly modeling \hcn, \hcop, and \hnc\, simultaneously, the two-cloud model was unable to match simultaneously all three line profiles very well.  We surmise that the optical depths among the lines may vary considerably.  If so, the $\tau=1$ surface for each line may probe regions of the clump with very different physical conditions, especially the excitation temperature.  If that is the case, it would be difficult for the model to reproduce all three line profiles simultaneously.  Perhaps the two-cloud model worked so well for \hcn\, and \hcop\, because their optical depths have similar, large values.

\subsection{Consistency with Collapse}

This paper examines a sample of sources with potentially large blue asymmetries in their \hcop\ lines in order to verify the asymmetries and to determine whether these asymmetries are consistent with collapse.  Several predictions of the collapse scenario are consistent with the observations.  First, the large blue asymmetries in the \hcop\, line are confirmed.  Second, the blue asymmetries decrease in lines of decreasing optical depth.  \hcn and \hnc\,  also show typically show positive, but smaller blue asymmetries than \hcop\, and the optically thinner \htcop\, line shows smaller blue asymmetries still.  Third, the unusual \hcn\, hyperfine line intensity ratios are also consistent with a colder, optically thick, redshifted clump exterior absorbing \hcn\, line emission from a warmer, blueshifted clump interior.  Finally, estimates of the infall velocities from the two-cloud model suggest large collapse velocities, $\sim 2.4$ \kms.  All of these observations are consistent with the clumps in this sample undergoing gravitational collapse and having a gradient in excitation temperature, with smaller values in the exterior and larger values in the interior.

Other possibilities, however, might explain the observed blue asymmetries.  For example, this sample may happen to include clumps with bright molecular outflows in which the blueshifted outflow lobe is unusually brighter than the redshifted lobe and produces excess emission on the blueshifted side of the line profile.
The selection of this sample would in fact be biased to include such sources.  It is conceivable that such sources, though rare, may have been selected from the MALT90 sample as having the most extreme blue asymmetries.  Indeed, only 27 candidates were selected for this study from over 3000 MALT90 clumps. 

Asymmetric molecular outflows with a brighter blue lobe compared to the red lobe have indeed been detected and imaged by interferometers \citep{Zhang2014, Aso2018, Stephens2015, Stephens2019}.  Such asymmetries might reflect viewing angles.  Since the blueshifted lobe is launched in front of the parental cloud but the redshifted lobe behind it, the redshifted lobe is therefore more likely to be self-absorbed by the parental cloud.

Still, two arguments favor the gravitational collapse interpretation over the asymmetric outflow interpretation for the blue asymmetries. First, the spatial extent of the Mopra beam is large, about 0.6 pc across at a distance of 3 kpc, and one might expect tens to hundreds of outflows in an active star forming region over this size scale.  Any asymmetries in individual sources may be averaged away if several outflows are blended.  For the asymmetric outflow interpretation to work, a single, luminous outflow would need to dominate blueshifted emission.  Second, outflows typically have broad wings in their line emission.  Although a few sources studied here show wing emission to the blueshifted side in \hcop, most sources do not.  The Mopra data alone cannot conclusively establish or refute the presence of outflows as the cause of the asymmetry, and future interferometric studies will be necessary.  We can conclude, though, that the blue asymmetries observed here do match expectations for a sample of rapidly collapsing clumps.

\section{Summary}
We present spectra taken with the Mopra telescope of several molecular lines toward dense clumps for which MALT90 measurements show evidence for collapse via a large blue asymmetry.  These new spectra have longer integration times and better signal-to-noise ratios than those from the MALT90 Survey.

From these data, we draw the following conclusions:

1. The initial collapsing clump candidates were selected from the MALT90 sample as having large blue asymmetries, as quantified by their asymmetry parameters: \ahcop $\ge 0.8$.  The deeper pointed observations presented here confirm the highly asymmetric profiles for this sample, with a mean value of
$\bar{A}_{HCO^+} = 0.69\pm0.01$, much higher than the MALT90 sample mean value of $\bar{A}_{HCO^+} = 0.083\pm0.010$.  Positive values of \ahcop\, for individual sources are found for 92\% of the sample.

2. The asymmetry parameter for a collapsing clump is expected to decrease with tracers of decreasing optical depth.  This expectation is confirmed.  The highest optical depth line, \hcop, has the highest mean asymmetry parameter, $\bar{A}_{HCO^+} = 0.69\pm0.01$.  The \hcn\, and \hnc\, lines, with smaller, yet still optically thick, optical depths have smaller asymmetry parameters, $\bar{A}_{HCN} = 0.35\pm0.01$ and $\bar{A}_{HNC} = 0.28\pm0.01$. Finally, lines with the smaller optical depths, \cch\ and \htcop, have smaller asymmetry parameters, $\bar{A}_{C_2H} = 0.15\pm0.02$ and $\bar{A}_{H^{13}CO^+} = 0.18\pm0.03$.

3. The observed hyperfine intensity ratios for \nthp\, match the expected values for optically thin, LTE emission.  The hyperfine intensity ratios for \hcn, however, do not match their optically thin, LTE values.  Specifically, the $F = 1 \to 1$ hyperfine lines are much fainter than expected.  A simple two-cloud model meant to represent a collapsing cloud was developed to test whether self-absorption by cold, foreground gas can account for these anomalous hyperfine line ratios.  This two-cloud model with a cold, redshifted, foreground cloud and a warm background cloud can indeed reproduce the observed intensity ratios.  In this model, extreme self-absorption arises as the main $F = 2 \to 1$ hyperfine line in the foreground cloud significantly obscures the $F = 1 \to 1$ hyperfine line from the background cloud. 

4.  Collapse speeds of at least 2 \kms\, are required to explain the line profiles and the \hcn\, hyperfine ratios, and the two-cloud model implies average collapse speeds of $\sim 2.4$ \kms.  Such speeds can easily be generated by free-fall collapse within 1 pc of a clump with the average mass in this sample. 

5.  All of these results are consistent with an optically thick, collapsing clump model with a high-excitation interior and a low-excitation exterior.  This sample may represent collapsing clumps with extreme infall velocities, excitation temperature gradients, and optical depths.

\section{Acknowledgments} 
J. M. Jackson gratefully acknowledges start up funding from the University of Newcastle, Australia that enabled these observations. The National Radio Astronomy Observatory and Green Bank Observatory are facilities of the U.S. National Science Foundation operated under cooperative agreement by Associated Universities, Inc. The Mopra telescope is part of the Australia Telescope National Facility (https://ror.org/05qajvd42) which is funded by the Australian Government for operation as a National Facility managed by CSIRO.  We acknowledge the Gomeroi people as the Traditional Owners of the Observatory site.  The University of New South Wales Digital Filter Bank (MOPS) used for the observations with the Mopra Telescope was provided with support from the Australian Research Council.  We thank Nigel Maxted for his assistance and support for the Mopra observations.  R. Simon gratefully acknowledges support within the Collaborative Research Centers 1601 (SFB 1601 sub-project B2) and 956 funded by the Deutsche Forschungsgemeinschaft (DFG) – project IDs 500700252 and 184018867, respectively. P. Sanhueza was partially supported by a Grant-in-Aid for Scientific Research (KAKENHI Number JP22H01271 and JP23H01221) of JSPS.

\facilities{Mopra}

\software{Astropy \citep{Astropy2022},  Miriad \citep{Sault1995}}

\movetabledown=5.5cm
\begin{rotatetable}
\begin{deluxetable*}{lcccccccccc}
    \tablecaption{Source List\label{tab:parameters}}
    \tabletypesize{\footnotesize}
    \tablecolumns{12}
    \tablehead{
        \colhead{Name} &
        \colhead{Gal. Long.} &
        \colhead{Gal. Lat.} &
        \colhead{Classif.} & 
        \colhead{Peak AGAL Flux} &  
        \colhead{Distance} &
        \colhead{T$_{dust}$} & 
        \colhead{log \textless N\textgreater} & 
        \colhead{R$_{eff}$} & 
        \colhead{Virial} & 
        \colhead{M$_{dust}$} \\
        \colhead{} & 
        \colhead{(deg)} & 
        \colhead{(deg)} &
        \colhead{} & 
        \colhead{(Jy/beam)} & 
        \colhead{(kpc)} & 
        \colhead{(K)} & 
        \colhead{(log cm$^{-2}$)}& 
        \colhead{(pc)} & 
        \colhead{Ratio}  & 
        \colhead{(\Msun)} \\
        }              
    \startdata
        AGAL305.589+00.462 & 305.589 & 0.462 & Unknown       &  0.76 &  4.95 & 16.0 & 22.2 & 0.3 & 0.460 & 7.20E+02 \\  
        AGAL310.879+00.004 & 310.879 & 0.004 & HII region    &  2.46 &  5.56 & 22.0 & 22.3 & 0.5 & 0.329 & 1.30E+03 \\
        AGAL327.109+00.539 & 327.109 & 0.539 & CHII          &  1.02 &  5.27 & 15.3 & 22.4 & 0.4 & 0.666 & 1.35E+03 \\
        AGAL331.491$-$00.116 & 331.491 & -0.116 & Protostellar         &  5.06 &  11.57 & 20.3 & 22.6 & 4.3 & \ldots & 3.96e+04 \\        
        AGAL332.251+00.184 & 332.251 & 0.184 & Protostellar  &  0.36 &  3.49 & 12.8 & 22.2 & 0.4 & 1.276 & 1.24E+02 \\
        AGAL332.276$-$00.071 & 332.276 & -0.071 & Protostellar &  0.71 &  3.42 & 18.0 & 22.2 & 0.2 & 0.375 & 4.02E+02 \\
        AGAL333.449$-$00.182 & 333.449 & -0.182 & Quiescent    &  1.18 &  3.28 & 17.0 & 22.3 & 0.5 & 3.916 & 3.87E+02 \\
        AGAL333.466$-$00.164 & 333.466 & -0.164 & HIIregion    &  9.01 & 11.99 & 25.0 & 22.5 & 3.8 & 0.208 & 3.50E+04 \\
        AGAL333.604$-$00.212 & 333.604 & -0.212 & HIIregion    & 36.15 &  3.52 & 32.0 & 22.3 & 1.9 & 2.109 & 4.34E+03 \\
        AGAL335.231$-$00.316 & 335.231 & -0.316 & Protostellar &  0.53 &  3.07 & 11.6 & 22.5 & 0.4 & 0.966 & 2.60E+02 \\
        AGAL335.284$-$00.154 & 335.284 & -0.154 & HIIregion    & 0.57  & 12.18 & 23.2 & 21.9 & 2.1 & \ldots & 1.79E+03 \\
        AGAL327.152$-$00.062 & 327.152 & -0.062 & Uncertain          &  0.64 &  11.38 & 22.2 & 22.2 & 2.0 & \ldots & 2.89E+03 \\  
        AGAL337.261$-$00.146 & 337.261 & -0.146 & HIIregion    & 0.56  & 11.86 & 28.0 & 21.7 & 1.6 & 0.888 & 8.49E+02 \\
        AGAL337.284$-$00.159 & 337.284 & -0.159 & HIIregion    & 1.03  & 11.84 & 25.2 & 22.1 & 0.8 & 0.333 & 3.12E+03 \\
        AGAL337.771$-$00.534 & 337.771 & -0.534 & Quiescent    & 0.48  & 12.55 & 14.0 & 22.3 & 1.8 & 0.299 & 3.73E+03 \\
        AGAL337.891$-$00.491 & 337.891 & -0.491 & Quiescent    & 0.68  &  3.20 & 21.0 & 22.3 & 1.4 & 0.187 & 4.98E+02 \\
        AGAL337.916$-$00.477 & 337.916 & -0.477 & Protostellar & 24.75 &  3.23 & 30.0 & 22.6 & 0.8 & 0.659 & 1.99E+03 \\
        AGAL337.927$-$00.432 & 337.927 & -0.432 & Unknown      & 0.88  &  3.20 & 22.0 & 21.9 & 3.6 & 0.827 & 5.13E+02 \\
        AGAL337.934$-$00.507 & 337.934 & -0.507 & PDR          & 3.75  &  3.20 & 21.7 & 22.4 & 3.0 & 0.184 & 1.16E+03 \\
        AGAL338.327$-$00.409 & 338.327 & -0.409 & Protostellar & 1.05  &  3.20 & 20.0 & 22.2 & 1.5 & 0.201 & 4.42E+02 \\
        AGAL338.394$-$00.406 & 338.394 & -0.406 & Protostellar & 1.05  &  3.20 & 21.1 & 22.2 & 2.1 & 0.209 & 5.26E+02 \\
        AGAL338.869$-$00.479 & 338.869 & -0.479 & Quiescent    & 0.67  &  3.04 & 11.1 & 22.6 & 0.2 & 0.418 & 7.10E+02 \\
        AGAL340.942$-$00.364 & 340.942 & -0.364 & Quiescent    & 1.37  &  3.78 & 17.7 & 22.4 & 0.9 & 0.971 & 5.06E+02 \\
        AGAL341.211$-$00.271 & 341.211 & -0.271 & Unknown      & 0.33  &  3.75 & 14.0 & 22.2 & 0.3 & 4.179 & 5.73E+02 \\
        AGAL345.058$-$00.224 & 345.058 & -0.224 & Protostellar & 18.51 &  3.12 & 25.8 & 22.7 & 1.4 & 0.554 & 2.51E+02 \\
        AGAL346.076$-$00.056 & 346.076 & -0.056 & HIIregion    & 0.66  & 15.21 & 26.0 & 21.8 & 1.7 & 0.078 & 1.51E+04 \\
        AGAL349.774+00.019 & 349.774 & 0.019  & Quiescent    & 1.21  & 23.29 & 28.0 & 21.9 & 2.8 & 0.067 & 2.80E+04 \\                     
    \enddata
\end{deluxetable*}
\end{rotatetable}

\begin{deluxetable*}{ll}

\tablecaption{Rest Frequencies}
\label{tab:restfreqs}
\tablecolumns{2}
    \tablehead{
        \colhead{Transition} &
        \colhead{Rest Frequency} \\
        \colhead { } & {(MHz)} \\
        }

    \startdata
        \nthp           & 93 173.711\\
        \tcs            & 92 494.308\\
        H41$\alpha$     & 92 034.434\\
        \chtcn          & 91 985.314\\
        \hctn           & 90 979.023\\
        \tctfs          & 90 926.026\\
        \hnc            & 90 663.568\\
        \hctccn         & 90 593.059\\
        \hcop           & 89 188.525\\
        \hcn            & 88 631.847\\
        \hncofo         & 88 239.020\\
        \hncofz         & 87 925.237\\
        \cch            & 87 316.925\\
        \hntc           & 87 090.850\\
        \sio            & 86 846.960\\
        \htcop          & 86 754.288\\
    \enddata

    \tablecomments{Except where noted below, the rest frequencies are taken from the splatalogue website \citep[\url{https://www.splatalogue.online},   ][]{Remijan2007} and references therein.  The \htcopnt\, rest frequency is the mean of two blended hyperfine lines, weighted by their statistical weights \citep{SchmidBurgk2004}. For \nthpnt, \cite{Jackson2019} recommend a rest frequency of 93173.703 MHz, based on the multiple, split line frequencies presented by \cite{Pagani2009}.  The adoption of this rest frequency would change all \nthp\, velocities by 0.003 \kms, less than 3\% of the width of a single spectral channel ($\Delta V_{chan} = 0.11$ \kms), and would not significantly change any of the derived parameters.}

\end{deluxetable*}

\begin{deluxetable*}{lcccc}

\tablecaption{\nthp\, Fit Parameters. }
\label{tab:nthp_fit}
\tablecolumns{5}
    \tablehead{
        \colhead{Name} &
        \colhead{Amplitude\tablenotemark{a}} &
        \colhead{Velocity} &
        \colhead{$\sigma$} &
        \colhead{RMS} \\
        \colhead{} &
        \colhead{(K)} &
        \colhead{(\kms\,)} &
        \colhead{(\kms\,)} &
        \colhead{(K)} \\
        }

\startdata
    AGAL305.589$+$00.462 & 0.335$\pm$0.005 & -35.71$\pm$0.02 & 1.09$\pm$0.02 & 0.022 \\ 
    AGAL310.879$+$00.004 & 0.380$\pm$0.006 & -57.73$\pm$0.02 & 1.12$\pm$0.02 & 0.024 \\
    AGAL327.109$+$00.539 & 0.294$\pm$0.005 & -79.92$\pm$0.03 & 1.46$\pm$0.03 & 0.021 \\
    AGAL332.251$+$00.184 & 0.313$\pm$0.006 & -49.33$\pm$0.02 & 0.90$\pm$0.02 & 0.019 \\
    AGAL332.276$-$00.071 & 0.442$\pm$0.006 & -48.8$\pm$0.02 & 1.02$\pm$0.02 & 0.021 \\
    AGAL333.449$-$00.182 & 0.243$\pm$0.005 & -43.45$\pm$0.04 & 1.51$\pm$0.03 & 0.023 \\
    AGAL333.466$-$00.164 & 1.425$\pm$0.021 & -43.02$\pm$0.03 & 1.49$\pm$0.02 & 0.043 \\
    AGAL333.604$-$00.212 & 0.395$\pm$0.007 & -47.63$\pm$0.06 & 2.37$\pm$0.05 & 0.033 \\
    AGAL335.231$-$00.316 & 0.556$\pm$0.009 & -40.07$\pm$0.02 & 0.93$\pm$0.02 & 0.030 \\
    AGAL337.261$-$00.146 & 0.162$\pm$0.006 & -52.11$\pm$0.05 & 1.11$\pm$0.05 & 0.025 \\
    AGAL337.284$-$00.159 & 0.111$\pm$0.006 & -52.24$\pm$0.09 & 1.35$\pm$0.08 & 0.026 \\
    AGAL337.771$-$00.534 & 0.222$\pm$0.021 & -39.41$\pm$0.07 & 0.67$\pm$0.07 & 0.066 \\
    AGAL337.891$-$00.491 & 1.608$\pm$0.091 & -39.77$\pm$0.06 & 0.94$\pm$0.06 & 0.071 \\
    AGAL337.916$-$00.477 & 0.451$\pm$0.006 & -38.92$\pm$0.03 & 1.64$\pm$0.02 & 0.022 \\
    AGAL337.927$-$00.432 & 1.845$\pm$0.009 & -38.61$\pm$0.01 & 1.33$\pm$0.01 & 0.023 \\
    AGAL337.934$-$00.507 & 2.292$\pm$0.008 & -38.58$\pm$0.01 & 1.33$\pm$0.01 & 0.024 \\
    AGAL338.327$-$00.409 & 1.122$\pm$0.007 & -38.18$\pm$0.01 & 1.20$\pm$0.01 & 0.022 \\
    AGAL338.394$-$00.406 & 0.748$\pm$0.006 & -37.62$\pm$0.01 & 1.07$\pm$0.01 & 0.020 \\
    AGAL338.869$-$00.479 & 0.577$\pm$0.006 & -36.10$\pm$0.01 & 1.19$\pm$0.01 & 0.022 \\
    AGAL340.942$-$00.364 & 0.331$\pm$0.007 & -43.98$\pm$0.02 & 0.94$\pm$0.02 & 0.020 \\
    AGAL341.211$-$00.271 & 1.056$\pm$0.010 & -43.17$\pm$0.01 & 0.91$\pm$0.01 & 0.027 \\
    AGAL345.058$-$00.224 & 0.599$\pm$0.008 & -26.68$\pm$0.02 & 1.25$\pm$0.02 & 0.028 \\
    AGAL346.076$-$00.056 & 0.592$\pm$0.006 & -84.38$\pm$0.02 & 1.33$\pm$0.01 & 0.020 \\
    AGAL349.774+00.019 & 0.193$\pm$0.005 & -106.25$\pm$0.04 & 1.30$\pm$0.04 & 0.023 \\
\enddata
    
\tablenotetext{a}{The amplitude is that of the $F_1 = 2 \to 1$ main hyperfine component.}

\end{deluxetable*}

\begin{deluxetable*}{lccccc}

\tablecaption{Asymmetry Parameters, using \nthp\, as the optically thin reference line.  }
\label{tab:Asymm_all}
\tablecolumns{6}
    \tablehead{
        \colhead{Name} &
        \colhead{Asymmetry} &
        \colhead{Asymmetry} &
        \colhead{Asymmetry} &
        \colhead{Asymmetry} &
        \colhead{Asymmetry} \\ 
        \colhead{} &
        \colhead{Parameter} &
        \colhead{Parameter\tablenotemark{a}} &
        \colhead{Parameter} &
        \colhead{Parameter} &
        \colhead{Parameter} \\
        \colhead{} &
        \colhead{\hcop} &
        \colhead{\hcn} &
        \colhead{\hnc}  &
        \colhead{\cchnt\, (1-0)} &
        \colhead{\htcop} \\
        }

    \startdata
       AGAL305.589+00.462 & -0.056$\pm$0.016 & -0.153$\pm$0.034 & 0.056$\pm$0.029& 0.412$\pm$ 0.076 & 0.051$\pm$0.067 \\
        AGAL310.879+00.004 & 0.623$\pm$0.026 & 0.261$\pm$0.029 & 0.085$\pm$0.019& 0.079$\pm$0.054 & 0.011$\pm$0.102 \\
        AGAL327.109+00.539 & 0.784$\pm$0.023 & 0.306$\pm$0.031 & 0.481$\pm$0.019& 0.174$\pm$ 0.071 & 0.534$\pm$0.060\\
        AGAL332.251+00.184 & 0.548$\pm$0.027 & 0.137$\pm$ 0.021 & 0.197$\pm$0.025& 0.216$\pm$0.086 & 0.051$\pm$0.074\\
        AGAL332.276$-$00.071 & 0.634$\pm$0.016 & 0.301$\pm$ 0.021 & 0.266$\pm$0.014& 0.217$\pm$ 0.045 & 0.509$\pm$0.102\\
        AGAL333.449$-$00.182 & 0.978$\pm$0.040 & 0.508$\pm$0.036 & 0.380$\pm$0.024& 0.591$\pm$0.059 & 0.452$\pm$0.081\\
        AGAL333.466$-$00.164 & 0.841$\pm$0.053 & 1.023$\pm$0.105 & 0.411$\pm$0.035& 0.308$\pm$0.045 & 0.173$\pm$0.066\\
        AGAL333.604$-$00.212 & 0.932$\pm$0.030 & 0.693$\pm$0.030 & 0.148$\pm$0.020& 0.099$\pm$0.032 & -0.012$\pm$0.025\\
        AGAL335.231$-$00.316 & 0.311$\pm$0.046 & 0.111$\pm$0.039 & 0.096$\pm$0.035& -0.503$\pm$0.220 & 0.068$\pm$0.078\\
        AGAL337.261$-$00.146 & 0.244$\pm$0.023 & 0.179$\pm$0.023 & 0.183$\pm$0.037& 0.366$\pm$0.053 & 0.094$\pm$0.076 \\
        AGAL337.284$-$00.159 & 0.016$\pm$0.031 & 0.070$\pm$0.026 & 0.053$\pm$0.045& 0.438$\pm$0.078 & 0.140$\pm$0.266 \\
        AGAL337.771$-$00.534 & 0.808$\pm$0.095 & 0.359$\pm$0.119 & 0.469$\pm$0.139& 0.190$\pm$0.161 & 0.294$\pm$0.212\\
        AGAL337.891$-$00.491 & 0.718$\pm$0.141 & 0.374$\pm$0.149 & 0.192$\pm$0.083& 0.025$\pm$0.246 & 0.047$\pm$0.412\\
        AGAL337.916$-$00.477 & 1.101$\pm$0.017 & 0.825$\pm$0.025 & 0.640$\pm$0.021& 0.332$\pm$0.107 & 0.332$\pm$0.021\\
        AGAL337.927$-$00.432 & 0.957$\pm$0.011 & 0.675$\pm$0.011 & 0.475$\pm$0.030& 0.302$\pm$0.028 &  0.143$\pm$0.033\\
        AGAL337.934$-$00.507 & 1.842$\pm$0.059 & 0.966$\pm$0.021 & 0.756$\pm$0.012& 0.210$\pm$0.015 & 0.116$\pm$0.039\\
        AGAL338.327$-$00.409 & 0.946$\pm$0.021 & 0.364$\pm$0.024 & 0.181$\pm$0.014& 0.069$\pm$0.048 & 0.195$\pm$0.053\\
        AGAL338.394$-$00.406 & 0.815$\pm$0.023 & 0.057$\pm$0.019 & -0.083$\pm$0.017& 0.127$\pm$0.037 & -0.010$\pm$0.050\\
        AGAL338.869$-$00.479 & 0.271$\pm$0.035 & 0.031$\pm$0.028 & 0.019$\pm$0.011& 0.002$\pm$0.072 & 0.230$\pm$0.048\\
        AGAL340.942$-$00.364 & 0.657$\pm$0.018 & 0.165$\pm$0.077 & 0.456$\pm$0.030& -0.036$\pm$0.054 & -0.037$\pm$0.047 \\
        AGAL341.211$-$00.271 & 0.666$\pm$0.020 & 0.204$\pm$0.025 & 0.425$\pm$0.038&  0.078$\pm$0.027 & 0.019$\pm$0.069\\
        AGAL345.058$-$00.224 & 0.962$\pm$0.055 & 0.502$\pm$0.034 & 0.530$\pm$0.022& -0.004$\pm$0.084 & 0.425$\pm$0.154 \\
        AGAL346.076$-$00.056 & 0.653$\pm$0.012 & 0.390$\pm$0.024 & 0.220$\pm$0.020& -0.100$\pm$0.046 & -0.078$\pm$0.063\\
        AGAL349.774+00.019 & 0.283$\pm$0.032 & 0.136$\pm$0.022 & 0.126$\pm$0.038& -0.008$\pm$0.109 & 0.434$\pm$0.199\\
        \hline
        Mean\tablenotemark{b}& 0.69$\pm$0.01 & 0.35$\pm$0.01 & 0.28$\pm$0.01 & 0.15$\pm$0.02 & 0.18$\pm$0.03 \\
        Fraction with $A>3\sigma$ & 0.92 & 0.71 & 0.75 & 0.5 & 0.33 \\
    \enddata

    \tablenotetext{a}{The asymmetry parameter for \hcn\, is for the main $F = 2 \to 1$ hyperfine line.}
    \tablenotetext{b}{Errors for the mean values are calculated by $\sigma = \sqrt{\sum{\sigma_i}^2}/N$, where $\sigma_i$ are the errors on the individual values and $N$ is the size of the sample.}

\end{deluxetable*}

\begin{deluxetable*}{lcc}

\tablecaption{A Comparison of the \hcop\, Asymmetry Parameters using \nthp\, and \htcop\, as the optically thin reference lines}
\label{tab:A_hcop_compare_ref}
\tablecolumns{3}
    \tablehead{
        \colhead{Name} &
        \colhead{\hcop} &
        \colhead{\hcop} \\ 
        \colhead{} &
        \colhead{Asymmetry Parameter} &
        \colhead{Asymmetry Parameter} \\ 
        \colhead{} &
        \colhead{\nthp\, Reference} &
        \colhead{\htcop\, Reference} \\
        }

    \startdata
        AGAL305.589+00.462 & -0.056$\pm$0.016 & -0.056$\pm$0.019\\
        AGAL310.879+00.004 & 0.623$\pm$0.026 & 0.623$\pm$0.032\\
        AGAL327.109+00.539 & 0.784$\pm$0.023 & 0.683$\pm$0.026\\
        AGAL332.251+00.184 & 0.548$\pm$0.027 & 0.450$\pm$0.031\\
        AGAL332.276$-$00.071 & 0.634$\pm$0.016 & 0.409$\pm$0.061\\
        AGAL333.449$-$00.182 & 0.978$\pm$0.040 & 0.945$\pm$0.038\\
        AGAL333.466$-$00.164 & 0.841$\pm$0.053 & 0.833$\pm$0.023\\
        AGAL333.604$-$00.212 & 0.932$\pm$0.030 & 0.882$\pm$0.025\\
        AGAL335.231$-$00.316 & 0.311$\pm$0.046 & 0.309$\pm$0.065\\
        AGAL337.261$-$00.146 & 0.244$\pm$0.023 & -0.586$\pm$0.131\\
        AGAL337.284$-$00.159 & 0.016$\pm$0.031 & 0.047$\pm$0.089\\
        AGAL337.771$-$00.534 & 0.808$\pm$0.095 & 0.808$\pm$0.043\\
        AGAL337.891$-$00.491 & 0.718$\pm$0.141 & 0.647$\pm$0.083\\
        AGAL337.916$-$00.477 & 1.101$\pm$0.017 & 0.748$\pm$0.021\\
        AGAL337.927$-$00.432 & 0.946$\pm$0.021 & 0.927$\pm$0.013\\
        AGAL337.934$-$00.507 & 1.842$\pm$0.059 & 1.819$\pm$0.057\\
        AGAL338.327$-$00.409 & 0.946$\pm$0.021 & 0.902$\pm$0.019\\
        AGAL338.394$-$00.406 & 0.815$\pm$0.023 & 0.815$\pm$0.023\\
        AGAL338.869$-$00.479 & 0.271$\pm$0.035 & 0.247$\pm$0.032\\
        AGAL340.942$-$00.364 & 0.657$\pm$0.018 & 0.657$\pm$0.021\\
        AGAL341.211$-$00.271 & 0.666$\pm$0.020 & 0.639$\pm$0.027\\
        AGAL345.058$-$00.224 & 0.962$\pm$0.055 & 0.945$\pm$0.059\\
        AGAL346.076$-$00.056 & 0.653$\pm$0.012 & 0.693$\pm$0.026\\
        AGAL349.774+00.019 & 0.283$\pm$0.032 & 0.117$\pm$0.090\\
        \hline
        Mean\tablenotemark{a}& 0.69$\pm$0.01 & 0.60$\pm$0.01\\
        Fraction with \mbox{$A>3\sigma$} & 0.92 & 0.83\\
    \enddata

    \tablenotetext{a}{Errors for the mean values are calculated by $\sigma = \sqrt{\sum{\sigma_i}^2}/N$, where $\sigma_i$ are the errors on the individual values and $N$ is the size of the sample.}

\end{deluxetable*}

\begin{deluxetable*}{lccccccccccc}
\tablecaption{Mean Asymmetry Parameters for Selected Lines}
\label{tab:Asymm_mean}
\tablecolumns{3}
    \tablehead{
        \colhead{Line} &
        \colhead{\hcopnt} &
        \colhead{\hcnnt} &
        \colhead{\hncnt} &
        \colhead{\cchnt} &
        \colhead{\htcopnt} &
        \colhead{\nthpnt} &
        \colhead{\hncofznt\tablenotemark{a}} &
        \colhead{\hctnnt} &
        \colhead{\siont} &
        \colhead{\tcsnt} &
        \colhead{\hntcnt} \\ 
        }

    \startdata
        \underline{All Values} & ~ & ~ & ~ & ~ & ~ & ~ & ~ & ~ & ~ & ~ & ~ \\
        Mean $A$ & 0.69 & 0.35 & 0.28 & 0.15 & 0.18 & -0.04 & 0.01 & -0.70 & 1.05 & -0.18 & -0.72 \\ 
        $\sigma_A$ \tablenotemark{b}& 0.01 & 0.01 & 0.01 & 0.02 & 0.03 & 0.01 & 0.16 & 2.76 & 11.54 & 0.18 & 1.61 \\
        Fraction with $A>3\sigma$ & 0.92 & 0.71 & 0.75 & 0.46 & 0.33 & 0.00 & 0.13 & 0.08 & 0.04 & 0.08 & 0.04 \\ \hline
        \underline{Excluding $I<3\sigma$}\tablenotemark{c} & ~ & ~ & ~ & ~ & ~ & ~ & ~ & ~ & ~ & ~ & ~ \\
        Mean $A$ & 0.69 & 0.35 & 0.28 & 0.15 & 0.18 & -0.04 & 0.16 & 0.09 & -0.01 & 0.06 & -0.09 \\
        $\sigma_A$ \tablenotemark{b}& 0.01 & 0.01 & 0.01 & 0.02 & 0.02 & 0.01 & 0.04 & 0.04 & 0.04 & 0.04 & 0.03 \\
        Number with $I>3\sigma$ & 24 & 24 & 24 & 24 & 23 & 24 & 19 & 17 & 19 & 16 & 18 \\   
    \enddata
    \tablenotetext{a}{This line is the \hncofz\, transition.}  
    \tablenotetext{b}{Errors for the mean values are calculated by $\sigma = \sqrt{\sum{\sigma_i}^2}/N$, where $\sigma_i$ are the errors on the individual values and $N$ is the size of the sample.}
    \tablenotetext{c}{The values for this section of the Table exclude sources where the total integral $I = I_{blue} + I_{red} < 3\sigma$.  This eliminates sources with unusually large errors.}

\end{deluxetable*}

\begin{deluxetable*}{lccrcrccrrcrrc}
{
\tabletypesize{\scriptsize}
\tablecaption{Best fit parameters for the two-cloud model applied simultaneously to the \hcn\ and \hcop\ data.\label{tab:two-cloud}}
\tablecolumns{14}
\tablehead{
  \colhead{Name} &
  \colhead{$T_{\rm ex,1}$} &
  \colhead{$T_{\rm ex,2}$} &
  \colhead{$V_1$} &
  \colhead{$\epsilon(V_1)$} &
  \colhead{$\Delta V$\tablenotemark{*}} &
  \colhead{$\epsilon(\Delta V)$} &
  \colhead{$\phi$} &
  \colhead{$\sigma_1$} &
  \colhead{$\sigma_2$} &
  \colhead{$\tau_{0,1}$} &
  \colhead{$\tau_{0,2}$} &
  \colhead{$T_{\rm bg}$} &
  \colhead{$\tau(\mathrm{HCO}^+)/\tau(\mathrm{HCN})$} \\
  \colhead{} &
  \colhead{(K)} &
  \colhead{(K)} &
  \colhead{(km s$^{-1}$)} &
  \colhead{(km s$^{-1}$)} &
  \colhead{(km s$^{-1}$)} &
  \colhead{(km s$^{-1}$)} &
  \colhead{} &
  \colhead{(km s$^{-1}$)} &
  \colhead{(km s$^{-1}$)} &
  \colhead{} &
  \colhead{} &
  \colhead{(K)} &
  \colhead{}
}
\startdata
AGAL305.589+00.462 & 40.0 & 2.5 & -35.46 & 0.07 & 0.40 & 0.28 & 0.19 & 2.94 & 10.00 & 1.0 & 2.9 & 2.8 & 0.1 \\
AGAL310.879+00.004 & 43.9 & 2.9 & -57.82 & 0.07 & 6.16 & 0.50 & 0.49 & 1.55 & 3.42 & 0.1 & 4.3 & 2.9 & 0.8 \\
AGAL327.109+00.539 & 29.1 & 6.0 & -82.37 & 0.11 & 5.84 & 0.17 & 0.02 & 1.55 & 1.34 & 1.8 & 4.8 & 2.8 & 2.9 \\
AGAL332.251+00.184 & 42.0 & 3.8 & -49.77 & 0.02 & -1.18 & 0.57 & 0.27 & 0.90 & 6.60 & 0.1 & 0.3 & 2.8 & 1.2 \\
AGAL332.276$-$00.071 & 54.7 & 2.9 & -49.06 & 0.01 & 11.18 & 0.40 & 0.66 & 0.91 & 3.05 & 0.1 & 14.6 & 2.9 & 0.7 \\
AGAL333.449$-$00.182 & 45.3 & 1.4 & -46.37 & 0.50 & 6.35 & 0.50 & 0.01 & 3.72 & 2.40 & 4.2 & 3.9 & 2.9 & 0.7 \\
AGAL333.466$-$00.164 & 48.0 & 4.9 & -45.24 & 0.09 & 4.30 & 0.09 & 0.04 & 4.91 & 1.52 & 2.8 & 14.6 & 4.0 & 0.4 \\
AGAL333.604$-$00.212 & 54.1 & 5.0 & -43.01 & 0.11 & 3.60 & 0.10 & 0.28 & 4.93 & 3.60 & 3.1 & 3.6 & 19.1 & 1.1 \\
AGAL335.231$-$00.316 & 53.6 & 3.8 & -40.31 & 0.03 & 0.69 & 0.30 & 0.69 & 0.61 & 3.73 & 0.1 & 1.9 & 3.7 & 1.8 \\
AGAL337.261$-$00.146 & 47.9 & 7.4 & -52.26 & 0.04 & 11.79 & 0.17 & 0.45 & 2.00 & 1.94 & 0.1 & 12.4 & 7.6 & 0.7 \\
AGAL337.284$-$00.159 & 35.4 & 6.3 & -53.86 & 1.11 & 14.83 & 0.96 & 0.02 & 10.00 & 3.95 & 1.7 & 1.9 & 15.9 & 1.1 \\
AGAL337.771$-$00.534 & 41.5 & 5.5 & -38.82 & 0.36 & 2.46 & 0.35 & 0.16 & 2.62 & 0.78 & 0.1 & 50.0 & 5.3 & 1.1 \\
AGAL337.891$-$00.491 & 78.2 & 6.0 & -39.81 & 0.09 & 4.48 & 0.33 & 0.96 & 1.58 & 1.99 & 0.1 & 13.0 & 6.0 & 0.3 \\
AGAL337.916$-$00.477 & 51.3 & 9.3 & -38.97 & 0.04 & 2.88 & 0.04 & 0.09 & 2.24 & 1.11 & 8.8 & 23.2 & 10.4 & 0.2 \\
AGAL337.927$-$00.432 & 60.6 & 2.3 & -38.25 & 0.05 & 2.73 & 0.06 & 0.06 & 2.14 & 1.41 & 2.2 & 12.5 & 2.7 & 0.4 \\
AGAL337.934$-$00.507 & 53.5 & 5.5 & -39.66 & 0.08 & 3.50 & 0.08 & 0.07 & 3.97 & 1.32 & 1.1 & 17.2 & 6.5 & 0.2 \\
AGAL338.327$-$00.409 & 60.9 & 2.9 & -37.16 & 0.16 & 2.61 & 0.18 & 0.76 & 2.33 & 1.85 & 0.1 & 17.8 & 2.8 & 0.5 \\
AGAL338.394$-$00.406 & 42.0 & 2.9 & -35.16 & 0.17 & 2.82 & 0.91 & 0.62 & 2.66 & 3.05 & 0.1 & 4.9 & 2.8 & 4.0 \\
AGAL338.869$-$00.479 & 39.6 & 6.4 & -36.06 & 0.08 & 0.90 & 0.27 & 0.01 & 1.59 & 2.11 & 5.7 & 2.5 & 2.7 & 0.3 \\
AGAL340.942$-$00.364 & 32.4 & 3.5 & -44.70 & 0.69 & 6.09 & 0.65 & 0.06 & 3.16 & 1.81 & 0.4 & 23.7 & 2.9 & 1.1 \\
AGAL341.211$-$00.271 & 48.9 & 3.5 & -43.20 & 0.31 & 4.36 & 0.39 & 0.22 & 3.47 & 1.87 & 0.3 & 20.1 & 3.1 & 0.3 \\
AGAL345.058$-$00.224 & 47.9 & 4.4 & -26.31 & 0.15 & 2.28 & 0.14 & 0.46 & 3.97 & 1.70 & 0.1 & 18.5 & 4.2 & 0.3 \\
AGAL346.076$-$00.056 & 55.0 & 2.9 & -84.87 & 0.05 & 6.09 & 0.19 & 0.69 & 1.41 & 2.94 & 0.1 & 6.5 & 2.8 & 0.6 \\
AGAL349.774+00.019 & 42.6 & 2.9 & -106.46 & 0.02 & 9.38 & 0.12 & 0.51 & 1.26 & 1.75 & 0.1 & 7.6 & 2.8 & 0.7 \\
\enddata
\tablenotetext{*}{Here $\Delta V = V_2 - V_1$.  $\Delta v$ approximates the infall velocity.}
\tablecomments{Beam-filling factors $\phi$ are corrected for Mopra’s main-beam efficiency ($\eta_{\rm mb}=0.49$). The column $\epsilon(V_1)$ is the formal 1$\sigma$ uncertainty on $V_1$, and $\epsilon(\Delta V)$ is the formal 1$\sigma$ uncertainty on $\Delta V$.}
}
\end{deluxetable*}
\begin{deluxetable*}{lcccccc}
{\bf
\tabletypesize{\scriptsize}
\tablecaption{Asymmetry parameters: data vs.\ two-cloud model and residuals.\label{tab:two-cloud-asym}}
\tablecolumns{7}
\tablehead{
  \colhead{Name} &
  \colhead{$A(\mathrm{HCN})$} &
  \colhead{$A(\mathrm{HCO}^+)$} &
  \colhead{$A(\mathrm{HCN})$} &
  \colhead{$A(\mathrm{HCO}^+)$} &
  \colhead{$\Delta A(\mathrm{HCN})$} &
  \colhead{$\Delta A(\mathrm{HCO}^+)$} \\
    \colhead{} &
  \colhead{Data} &
  \colhead{Data} &
  \colhead{Model} &
  \colhead{Model} &
  \colhead{Residual} &
  \colhead{Residual}
}
\startdata
AGAL305.589+00.462 & -0.15 & -0.06 & -0.09 & -0.04 & -0.06 & -0.01 \\
AGAL310.879+00.004 & 0.26  & 0.62  & 0.18  & 0.41  & 0.08  & 0.21 \\
AGAL327.109+00.539 & 0.31  & 0.78  & 0.36  & 0.73  & -0.06 & 0.06 \\
AGAL332.251+00.184 & 0.14  & 0.55  & 0.00  & 0.29  & 0.14  & 0.25 \\
AGAL332.276$-$00.071 & 0.30  & 0.63  & 0.34  & 0.27  & -0.04 & 0.37 \\
AGAL333.449$-$00.182 & 0.51  & 0.98  & 0.63  & 0.83  & -0.12 & 0.15 \\
AGAL333.466$-$00.164 & 1.02  & 0.84  & 0.87  & 0.88  & 0.16  & -0.04 \\
AGAL333.604$-$00.212 & 0.69  & 0.93  & 0.56  & 0.70  & 0.14  & 0.23 \\
AGAL335.231$-$00.316 & 0.11  & 0.31  & -0.04 & 0.06  & 0.15  & 0.25 \\
AGAL337.261$-$00.146 & 0.18  & 0.24  & 0.18  & 0.07  & -0.01 & 0.17 \\
AGAL337.284$-$00.159 & 0.07  & 0.02  & 0.12  & 0.06  & -0.05 & -0.05 \\
AGAL337.771$-$00.534 & 0.36  & 0.81  & 0.21  & 0.46  & 0.15  & 0.34 \\
AGAL337.891$-$00.491 & 0.37  & 0.72  & 0.33  & 0.37  & 0.04  & 0.35 \\
AGAL337.916$-$00.477 & 0.83  & 1.10  & 0.92  & 0.73  & -0.09 & 0.38 \\
AGAL337.927$-$00.432 & 0.68  & 0.95  & 0.72  & 0.73  & -0.04 & 0.21 \\
AGAL337.934$-$00.507 & 0.97  & 1.84  & 1.12  & 0.99  & -0.15 & 0.85 \\
AGAL338.327$-$00.409 & 0.36  & 0.95  & 0.46  & 0.67  & -0.09 & 0.28 \\
AGAL338.394$-$00.406 & 0.06  & 0.82  & 0.08  & 0.87  & -0.02 & -0.06 \\
AGAL338.869$-$00.479 & 0.03  & 0.27  & 0.06  & 0.06  & -0.03 & 0.21 \\
AGAL340.942$-$00.364 & 0.17  & 0.66  & 0.09  & 0.68  & 0.07  & -0.02 \\
AGAL341.211$-$00.271 & 0.20  & 0.67  & 0.32  & 0.56  & -0.12 & 0.10 \\
AGAL345.058$-$00.224 & 0.50  & 0.96  & 0.64  & 0.66  & -0.14 & 0.30 \\
AGAL346.076$-$00.056 & 0.39  & 0.65  & 0.36  & 0.54  & 0.03  & 0.12 \\
AGAL349.774+00.019 & 0.14  & 0.28  & 0.08  & 0.16  & 0.05  & 0.12 \\
\enddata
\tablecomments{$\Delta A = A_{\rm Data} - A_{\rm Model}$. Data values are from the Mopra analysis using \nthp\ as the reference; model values are from the two-cloud fits.}
}
\end{deluxetable*}
\clearpage
\clearpage

\newpage
\newpage

\clearpage

\clearpage

\end{document}